\documentclass[aps,prl,preprint,groupedaddress]{revtex4-1}

\usepackage{psfrag,xcolor}
\usepackage{graphicx,amssymb,textcomp,stmaryrd}
\usepackage{epstopdf,epsfig}
\usepackage{latexsym,amsmath}

\providecommand\bnabla{\mathbf{\nabla}}
\providecommand\bcdot{\mathbf{\cdot}}

\newcommand\etal{\mbox{\textit{et al.~}}}

\newcommand\diff{\mbox{d}}            

\newcommand\p{\ensuremath{\partial}}

\providecommand\bnabla{\mathbf{\nabla}}
\providecommand\bcdot{\mathbf{\cdot}}

\newcommand{\vetu}[0]{\mathbf{u}}
\newcommand{\vetx}[0]{\mathbf{x}}

\newcommand{\vete}[0]{\mathbf{e}}

\begin{document}

\title{Statistics and structure of spanwise rotating turbulent channel flow
at moderate Reynolds numbers}

\author{Geert Brethouwer$^1$}  %
\affiliation{
$^1$Linn\'e FLOW Centre, KTH Mechanics, SE-10044 Stockholm, Sweden
}

\date{\today}


\begin{abstract}
A study of fully developed plane turbulent channel flow subject to spanwise system rotation
through direct numerical simulations is presented.
In order to study both the influence of the Reynolds number and spanwise rotation on
channel flow,
the Reynolds number $Re = U_b h/\nu$ is varied from a low 3000 
to a moderate $31\,600$ and
the rotation number $Ro = 2 \Omega  h/U_b$ is varied from 0 to 2.7,
where $U_b$ is the mean bulk velocity, $h$ the channel half gap and $\Omega$ the
system rotation rate.
The mean streamwise velocity profile displays also at higher $Re$ the characteristic
linear part with a slope near to $2 \Omega$ and
a corresponding linear part in the profiles of the production and 
dissipation rate of turbulent kinetic energy appears.
With increasing $Ro$ a distinct unstable side with 
large spanwise and wall-normal Reynolds stresses
and a stable side with much weaker turbulence develops in the channel.
The flow starts to relaminarize on the stable side of the channel
and persisting turbulent-laminar patterns appear at higher $Re$.
If $Ro$ is further increased the flow on the stable side becomes laminar-like
while at yet higher $Ro$ the whole flow relaminarizes, although
the calm periods might be disrupted by repeating bursts of turbulence, as
explained by Brethouwer (2016).
The influence of the Reynolds number is considerable, in particular on the stable
side of the channel where velocity fluctuations are stronger 
and the flow relaminarizes less quickly at higher $Re$.
Visualizations and statistics show that at $Ro=0.15$ and 0.45
large-scale structures and large counter rotating streamwise roll cells
develop on the unstable side. These become less noticeable and eventually vanish
when $Ro$ raises, especially at higher $Re$. 
At high $Ro$, the largest energetic structures are larger at lower $Re$.
\end{abstract}

\maketitle

\section{Introduction}
\label{intro}

Turbulent plane channel flow subject to rotation about the spanwise direction
displays several phenomena of interest to engineering applications,
turbulence modelling (Jakirli{\'c} \etal 2002, 
Smirnov \& Menter 2009, Arolla \& Durbin 2013, Hsieh \etal 2016)
and subgrid-scale modelling in large-eddy simulation
(Marstorp \etal 2009, Yang \etal 2012).
Effects of rotation on flow and turbulence are in this case non-trivial. 
Channel flow is unaffected by rotation if fluid motions
are two-dimensional and perpendicular to the rotation axis
(Tritton 1992),
for example, laminar Poiseuille flow and waves perpendicular
to the rotation axis are not influenced
(Brethouwer \etal 2014).
Yet, it is well-established that spanwise rotation strongly influences Reynolds stresses,
anisotropy and structures, and the mean velocity profile
in plane turbulent channel flow (Johnston \etal 1972).
Rotation effects are therefore obviously related to 
three-dimensional turbulent processes.

A useful parameter when discussing rotation effects on turbulent shear flows
is the ratio of background and mean shear vorticity.
For a unidirectional shear flow with mean velocity $U(y)$ in the $x$-direction 
and shear vorticity $-\mbox{d} U / \mbox{d} y$ rotating
with angular velocity $\Omega$ about the $z$-axis, this ratio is given by
\begin{equation}
S = - \frac{2 \Omega}{\mbox{d} U / \mbox{d} y}.
\label{ratio}
\end{equation}
When $S > 0$ the mean shear and background vorticity have the same sense of rotation
and rotation is cyclonic and when $S < 0$ they have the opposite sense
and rotation is anti-cyclonic.
Displaced particle analysis (Tritton 1992), rapid distortion theory,
large-eddy and direct numerical simulation (DNS) of rotating homogeneous turbulent shear
flow demonstrate that turbulence is damped if $S>0$ and $S< -1$
and augmented if $-1 < S < 0$ (Salhi \& Cambon 1997, Brethouwer 2005).

Experiments of fully developed turbulent plane channel flow subject to spanwise rotation
by Johnston \etal (1972) at $Re = U_b h /\nu = 5750$, $Ro = 2 \Omega h / U_b \leq 0.21$
and $Re = 17500$, $Ro \leq 0.081$, and by Nakabayashi \& Kitoh (2005) at 
$Re \leq 2750$ and $Ro \leq 0.056$,
where $U_b$ is the bulk mean velocity, $h$ channel half-width, $\nu$ viscosity and $\Omega$ rotation rate,
have shown that turbulence is suppressed
on one side of the channel where $S>0$ and augmented on the other side where
$-1 < S < 0$, in accordance with the discussion above. These sides are from now on
called the stable and unstable side, respectively.

DNS of spanwise rotating channel flow have been carried out by Kristoffersen \& Andersson (1993)
at $Re_\tau = 194$ and $Ro \leq 0.5$, Lamballais \etal (1996) at $Re=2500$
and $Ro \leq 1.5$, Nagano \& Hattori (2003) at $Re_\tau = 150$ and $Ro_\tau \leq 5$
corresponding to $Ro \lesssim 0.3$, and
Liu \& Lu (2006) at $Re_\tau = 194$ and $Ro_\tau \leq 7.5$
corresponding to $Ro \lesssim 0.5$.
Here, $Re_\tau$ and $Ro_\tau$ are based on the friction velocity instead of $U_b$. 
Yang \etal (2012) have performed DNSs of rotating channel flow at
$Re=7000$ and $Ro \leq 0.6$ to validate their large-eddy simulations.
These DNSs were broadly consistent with the experimental observations
and show that at sufficiently high $Ro$ 
the flow becomes laminar-like on the stable side
owing to the strong suppression of turbulence.
The asymmetry of the Reynolds stresses induces a skewed mean velocity profile
and differences in the shear stresses on the two walls.
Another notable feature is the appearance of a region in the channel
where the mean velocity profile
is linear with a slope $\diff U/\diff y \approx 2\Omega$, i.e. $S \approx -1$, 
implying that the absolute mean
vorticity is nearly zero.
Spanwise rotation affects turbulence anisotropy too since wall-normal
fluctuations are typically strongly augmented on the unstable side 
(Kristoffersen \& Andersson 1993), 
again in line with DNS of rotating homogeneous shear flow (Brethouwer 2005).
Another consequence of spanwise system rotation is the emergence of large
streamwise roll cells at certain $Ro$ induced by the Coriolis force, as shown by
experiments and DNS (Johnston \etal 1972, Kristoffersen \& Andersson 1993,
Dai \etal 2016).

Higher $Ro$ was considered by
Grundestam \etal (2008), who have performed DNS of spanwise rotating channel flow at
$Re_\tau = 180$ and $Ro \leq 2.49$. 
Besides one-point statistics, they studied the effect of rotation on turbulent
structures and Reynolds stress budgets.
They observed that at higher $Ro$ 
turbulence was weak as well on the
unstable side, suggesting that the flow fully relaminarizes 
at sufficiently high $Ro$.
This can be understood from the fact that in Poiseuille flow
$S \geq 0$ and $S \leq -1$
everywhere in the channel if $Ro \geq 3.0$.
A more rigorous stability analysis shows that all modes with spanwise
wavenumber $\beta \neq 0$ are linearly stable for $Ro > Ro_c$ in plane Poiseuille flow
with spanwise rotation (Wallin \etal 2013). The critical rotation number $Ro_c$
is a monotonic function of $Re$, i.e. $Ro_c = 2.80$ for $Re=10000$
and $Ro_c \rightarrow 3.0$ for $Re \rightarrow \infty$. 
DNS confirms that the flow relaminarizes when $Ro \rightarrow Ro_c$
(Wallin \etal 2013).

However, Tollmien-Schlichting (TS) waves with a wave vector normal 
to the rotation axis are unaffected by spanwise rotation
and become linearly unstable in plane Poiseuille flow if $Re \geq 3848$
(Schmid \& Henningson 2001). DNS indeed reveal TS wave
instabilities resulting in a continuous cycle of
turbulent bursts if $Ro \simeq Ro_c$ 
and the flow is basically laminar (Wallin \etal 2013).
Absence of turbulence though is not a prerequisite for TS instabilities, as shown by
Brethouwer \etal (2014) who have observed cyclic bursts of intense turbulence
triggered by an unstable TS wave in DNS at $Re=20000$ and $Ro=1.2$
when turbulence is strong on the unstable side.
An extensive study of DNSs of spanwise rotating channel flow reveals
that these cyclic turbulent bursts and TS wave instabilities 
show up in a quite wide range of $Re$ and $Ro$ including cases
with overall weak turbulence as well as strong continuous 
turbulence on the unstable side (Brethouwer 2016). 
The observed TS instability growth was compared with linear stability theory predictions
and analyzed in detail.

DNSs at $Re_\tau = 180$ and a wide range of $Ro$ were also performed by Xia \etal (2016).
They studied various one-point statistics and observed a linear part in the
profile of the streamwise Reynolds stress production 
at sufficiently high $Ro$.
Another study of spanwise rotating channel flow was performed by
Yang \& Wu (2012) who have carried out DNSs 
at $Re_\tau = 180$ and performed a helical wave decomposition. This shows
that at low $Ro$ energy concentrates in large-scale modes, presumably
streamwise roll cells, while at higher $Ro$ energy accumulates at smaller scales. 
In DNSs with periodic boundary conditions these roll cells appear as pairs of counter-rotating vortices.
Dai \etal (2016) have detected roll cells,
sometimes called Taylor-G{\"o}rtler vortices, in DNS at $Re=2800$ and
$0.1 \leq Ro \leq 0.5$ and $Re=7000$ and $Ro=0.3$, and studied their effect on the turbulence.
They found that, somewhat counter intuitively, turbulence is enhanced 
on the unstable side in the low-wall-shear-stress region where
the fluid is pumped away from the wall by the counter-rotating roll cells.
Hsieh \& Biringen have carried out DNSs of rotating channel flow at $Re_\tau \approx 200$ with $0 \leq Ro \leq0.5$ 
and $Re_\tau \approx 400$ with $Ro=0.2$ with varying domain sizes. At low $Re_\tau$ they observed
that when the spanwise domain was too small to capture a full pair of counter-rotating
roll cells the mean velocity profiles and Reynolds stresses were incorrect,
illustrating that the roll cells have a significant impact on the momentum transfer
and turbulence.
At higher $Re_\tau$ the mean velocity and Reynolds stresses changed significantly when the
spanwise domain size was reduced from $2 \pi h$ to $\pi h$.

These previous studies of spanwise rotating channel flow were mostly limited to
low Reynolds numbers, $Re_\tau \leq 194$, 
meaning that the influence of the Reynolds number on the statistics, turbulent
structures and relaminarization on the stable side can be significant.
Exceptions are the experiments of Johnston \etal (1972)
and DNSs of Dai \etal (2016) and Hsieh \& Biringen (2016)
at somewhat higher $Re$ albeit limited to moderate $Ro$.
It is therefore not completely clear if the previous
observations have been influenced by the low $Re$.
I have carried out DNSs of
plane turbulent channel flow subject to spanwise rotation
at higher Reynolds numbers than in previous studies
with $Re$ up to $31\,600$ and a wide range of $Ro$.
My aim is to assess the influence of spanwise rotation
on the mean velocity, one-point statistics and Reynolds stress 
budgets at these higher $Re$.
With these simulations I will investigate 
whether the effects of rotation on the mean flow and turbulence discussed above are 
either generic or influenced by the Reynolds number.
Another goal is to study 
the relaminarization of the flow on the stable channel side at high $Ro$ 
and turbulence structures, and examine if roll cells exist
in rotating channel flow at these higher $Re$.
Some previous observations of the flow structures and relaminarization
may have been affected by the limited computational domains
that were often used in the numerical studies.
In the present study, I therefore use larger
computational domains and study the structures through
visualizations, two-point correlations and spectra.
Finally, the effect of $Re$ on the flow
statistics and structures is studied at a fixed $Ro$.
The present study is also motivated by the need for higher $Re$ data 
of rotating channel flow for turbulence modelling.

\section{Numerical method and parameters}
\label{sec_num}

%
%
The velocity $\vetu$ in the DNSs is governed by the incompressible Navier-Stokes equations
\begin{equation}
\frac{\p \vetu}{\p t} + \vetu \bcdot \bnabla \vetu = - \bnabla p + \frac{1}{Re} \nabla^2 \vetu - Ro (\vete_z \times \vetu),~~~
\bnabla \bcdot \vetu = 0,
\label{gov_eq}
\end{equation}
where $\vete_z$ is the unit vector in the $z$-direction and $p$ the pressure including the centrifugal acceleration.
The equations are nondimensionalized by
the mean bulk velocity $U_b$ and channel half gap $h$; 
time $t$ is thus given in terms of a convection time $h/U_b$.
Streamwise, wall-normal, spanwise coordinates are denoted by $x$, $y$, $z$, respectively,
and boundary conditions are periodic in the streamwise and spanwise directions and 
no-slip at the walls. A sketch of the geometry is shown in Brethouwer (2016).
In the present DNSs $y=-1$ and 1 correspond to the wall on the unstable and 
stable channel side, respectively.

Equations (\ref{gov_eq})
are solved with a pseudo-spectral code with Fourier expansions in the homogeneous
$x$- and $z$-direction and Chebyshev polynomials in the $y$-direction (Chevalier \etal 2007)
and the spatial resolution is similar as in previous channel flow DNSs (Lee \& Moser 2015).
In all runs the flow rate and thus $Re$ was kept constant by adapting the mean pressure gradient.
$Re$ is varied from 3000 up to 31\,600 and $Ro$ from 
0 (nonrotating) to about $Ro_c$ which is between 2.67 to 2.87 for this range of $Re$. 
In most DNSs, the streamwise and spanwise domain size are either $12 \pi h \times 10.5 h$
or $8 \pi h \times 3 \pi h$, but in some DNSs the domain size was chosen
differently to accommodate TS instabilities, as 
explained in Brethouwer (2016), but this variation has little effect
on results presented here.
The runs are sufficiently long to reach a statistically stationary state in all DNSs. 
Parameters of the DNSs at $Re=3000$ to $31\,600$ are listed in table \ref{sim_par}.
The friction velocity is calculated as 
$u_\tau = [u^2_{\tau u}/2 + u^2_{\tau s}/2 ]^{1/2}$, where
$u_{\tau u}$ and $u_{\tau s}$ are the friction velocity of unstable
and stable channel side, respectively (Grundestam \etal 2008).
With this definition the mean dimensional pressure gradient $\p P/\p x =
\rho u^2_\tau/h$, where $\rho$ is the fluid density.
$Re^u_\tau$ and $Re^s_\tau$ are
Reynolds numbers based on 
$u_{\tau u}$ and $u_{\tau s}$, respectively, and $Ro_\tau = 2 \Omega h / u_\tau$.
The DNSs studied here are basically the same as those reported in Brethouwer (2016).
\begin{table}
\begin{center}
\def~{\hphantom{0}}
\begin{tabular}{llrrrrrr}
$Re_b$ & $Ro$ & $Re_\tau$ & $Re^u_\tau$ & $Re^s_\tau$ & $Ro_\tau$ & $L_x/h \times L_z/h$ & $N_x \times N_y \times N_z$\\[3pt]
31\,600	&	0	& 1505 & 1505 & 1505 & 0 & $12\pi\times 10.5$ & $6144 \times 577 \times 3456$\\	
31\,600	&	0.45	& 1213 & 1445 & 925 & 11.7 & $12\pi\times 10.5$ & $4608 \times 481 \times 2560$\\	
31\,600	&	0.9	& 822 & 988 & 613 & 34.6 & $12\pi\times 10.5$ & $3072 \times 385 \times 1728$\\
31\,600	&	1.2	& 562 & 670 & 428 & 67.4 & $12\pi\times 10.5$ & $2048 \times 257 \times 1152$\\[3pt]
30\,000	&	1.5	& 415 & 462 & 363 & 108 & $8\pi\times 3\pi$ & $1024 \times 193 \times 768$\\	
30\,000	&	2.1	& 319 & 318 & 319 & 198 & $8\pi\times 4.8\pi$ & $864 \times 161 \times 864$\\
30\,000	&	2.4	& 302 & 306 & 298 & 238 & $8\pi\times 3\pi$ & $640 \times 193 \times 512$\\
30\,000	& 	2.7     & 301 & 302 & 300 & 269 & $7.5\pi \times 3\pi$ & $432 \times 161 \times 384$\\[3pt]
20\,000	&	0	& 1000 & 1000 & 1000 & 0 & $8\pi\times 3\pi$ & $2560 \times 385 \times 1920$\\	
20\,000	&	0.15	& 976 & 1107& 825 & 3.1 & $8\pi\times 3\pi$ & $2304 \times 385 \times 1728$\\	
20\,000	&	0.45	& 800 & 964 & 594 & 11.2 & $8\pi\times 3\pi$ & $2048 \times 361 \times 1536$\\	
20\,000	&	0.65	& 700 & 851 & 505 & 18.6 & $8\pi\times 3\pi$ & $1920 \times 321 \times 1440$\\	
20\,000	&	0.9	& 544 & 677 & 365 & 33.1 & $8\pi\times 3\pi$ & $1536 \times 257 \times 1152$\\	
20\,000	&	1.2	& 423 & 501 & 326 & 56.7 & $8\pi\times 3\pi$ & $1152 \times 217 \times 864$\\
20\,000	&	1.5	& 333 & 370 & 292 & 90.0 & $29.4 \times 4\pi$ & $864 \times 193 \times 768$\\
20\,000	&	2.1	& 259 & 265 & 252 & 162 & $8\pi\times 3\pi$ & $512 \times 161 \times 432$\\[3pt]
10\,000	&	0	& 544 & 544 & 544 & 0 & $8\pi\times 3\pi$ & $1152 \times 193 \times 864$\\	
10\,000	&	0.45	& 435 & 535 & 304 & 10.3 & $8\pi\times 3\pi$ & $1024 \times 193 \times 768$\\	
10\,000	&	0.9	& 339 & 416 & 240 & 26.5 & $8\pi\times 3\pi$ & $768 \times 161 \times 576$\\	
10\,000	&	1.2	& 277 & 325 & 219 & 43.3 & $8\pi\times 3\pi$ & $576 \times 129 \times 432$\\	
10\,000	&	1.5	& 226 & 249 & 199 & 66.5 & $16\pi\times 6\pi$ & $1280 \times 161 \times 960$\\	
10\,000	&	1.8	& 196 & 207 & 185 & 91.6 & $16\pi\times 6\pi$ & $1024 \times 129 \times 768$\\	
10\,000	&	2.1	& 182 & 186 & 178 & 115 & $16\pi\times 6\pi$ & $640 \times 129 \times 512$\\[3pt]	
5000 & 0	& 297 & 297 & 297 & 0	& $8\pi \times 3\pi$ & $576 \times 109 \times 432$\\
5000 & 0.15	& 277 & 326 & 217 & 2.7 & $8\pi \times 3\pi$ & $576 \times 109 \times 432$\\
5000 & 0.45	& 251 & 310 & 174 & 8.9 & $8\pi \times 3\pi$ & $512 \times 109 \times 384$\\
5000 & 0.9	& 214 & 258 & 159 & 21.0 & $8\pi \times 3\pi$ & $432 \times 109 \times 320$\\
5000 & 1.2	& 182 & 211 & 148 & 32.9 & $8\pi \times 3\pi$ & $384 \times 97 \times 288$\\
5000 & 1.5	& 154 & 170 & 137 & 48.6 & $8\pi \times 3\pi$ & $320 \times 97 \times 240$\\
5000 & 1.8	& 137 & 144 & 129 & 65.8 & $27 \times 3\pi$ & $288 \times 97 \times 216$\\
5000 & 2.1	& 128 & 130 & 125 & 82.2 & $8\pi \times 3\pi$ & $256 \times 97 \times 192$\\[3pt]
3000 & 0	& 190 & 190 & 190 & 0	& $8\pi \times 3\pi$ & $320 \times 97 \times 256$\\
3000 & 0.15	& 179 & 213 & 138 & 2.5 & $8\pi \times 3\pi$ & $320 \times 97 \times 256$\\
3000 & 0.45	& 174 & 211 & 127 & 7.8 & $8\pi \times 3\pi$ & $320 \times 97 \times 256$\\
3000 & 0.9	& 153 & 181 & 118 & 17.7 & $8\pi \times 3\pi$ & $256 \times 97 \times 216$\\
3000 & 1.2	& 134 & 154 & 112 & 26.8 & $8\pi \times 3\pi$ & $216 \times 97 \times 180$\\
3000 & 1.5	& 117 & 128 & 105 & 38.6 & $8\pi \times 3\pi$ & $192 \times 97 \times 144$\\
\end{tabular}
\caption{DNS parameters: $N_x$, $N_y$ and $N_z$ are the number of modes in the streamwise,
wall-normal and spanwise direction, and $L_x$ and $L_z$, are the streamwise
and spanwise computational domain size, respectively.
}
\label{sim_par}
\end{center}
\end{table}
%

\section{Flow visualizations}
\label{sec_viz}

\begin{figure}
\begin{center}
\setlength{\unitlength}{1cm}
\includegraphics[width=120mm]{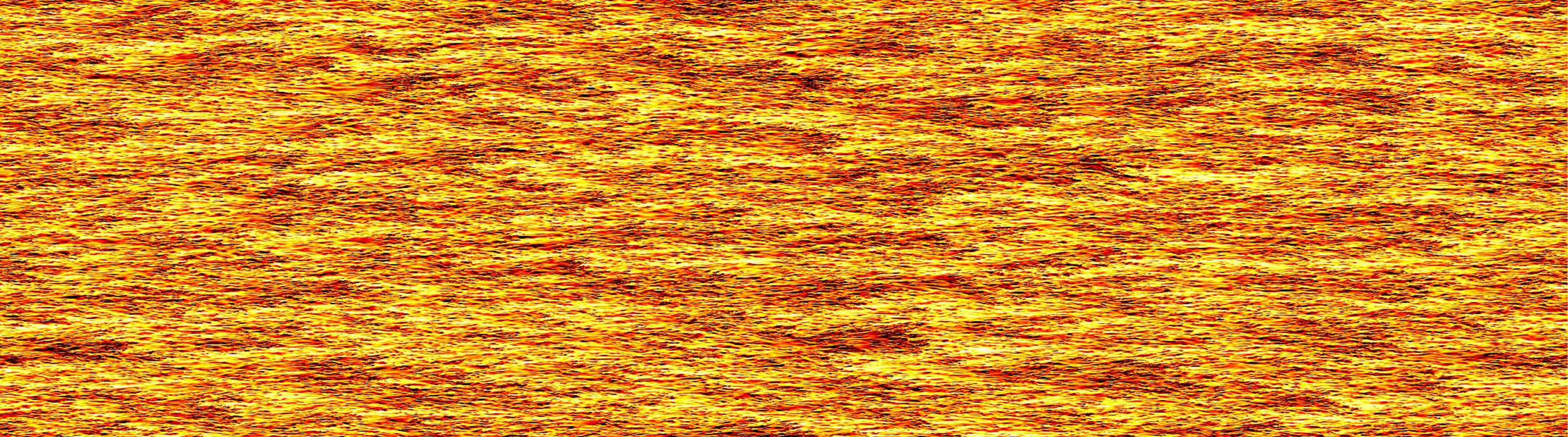}
\put(0.1,3.0){$\displaystyle (a)$}
\vskip2mm
\includegraphics[width=120mm]{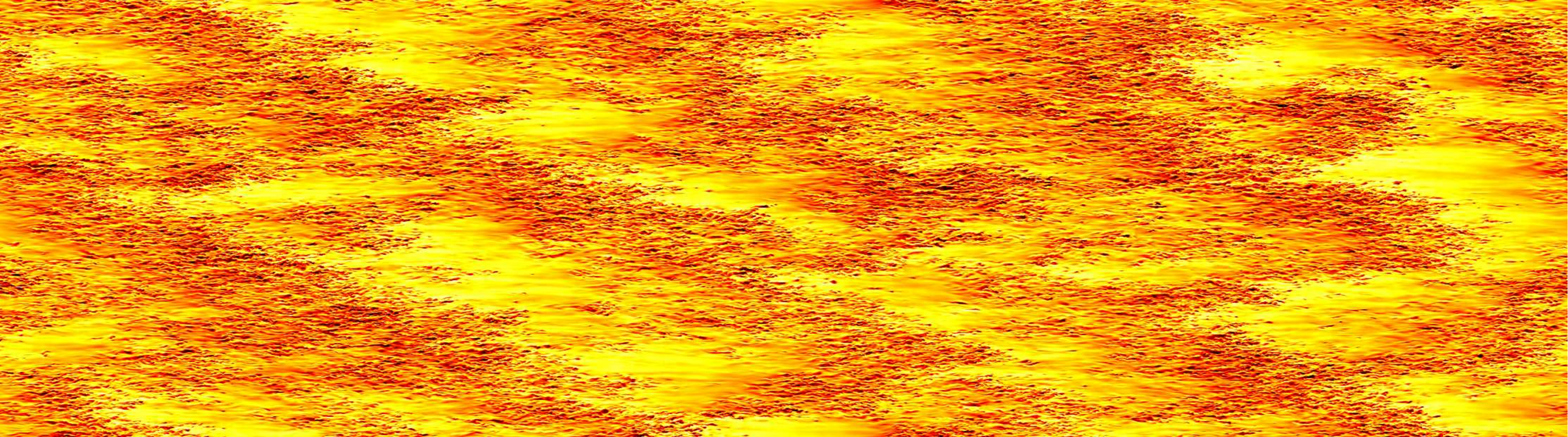}
\put(0.1,3.0){$\displaystyle (b)$}
\vskip2mm
\includegraphics[width=120mm]{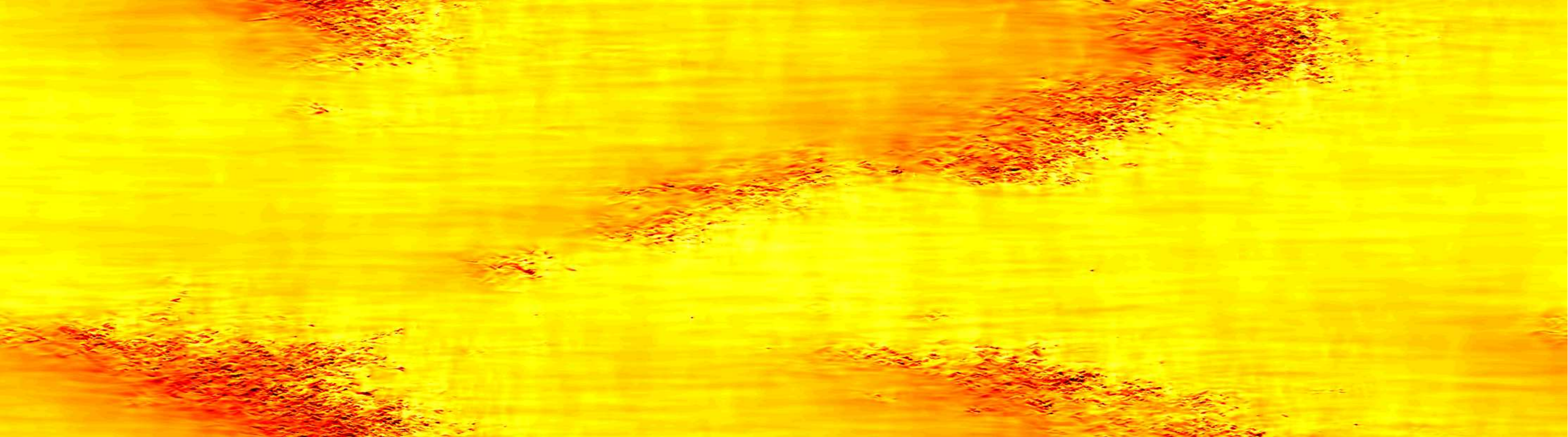}
\put(0.1,3.0){$\displaystyle (c)$}
\vskip2mm
\includegraphics[width=120mm]{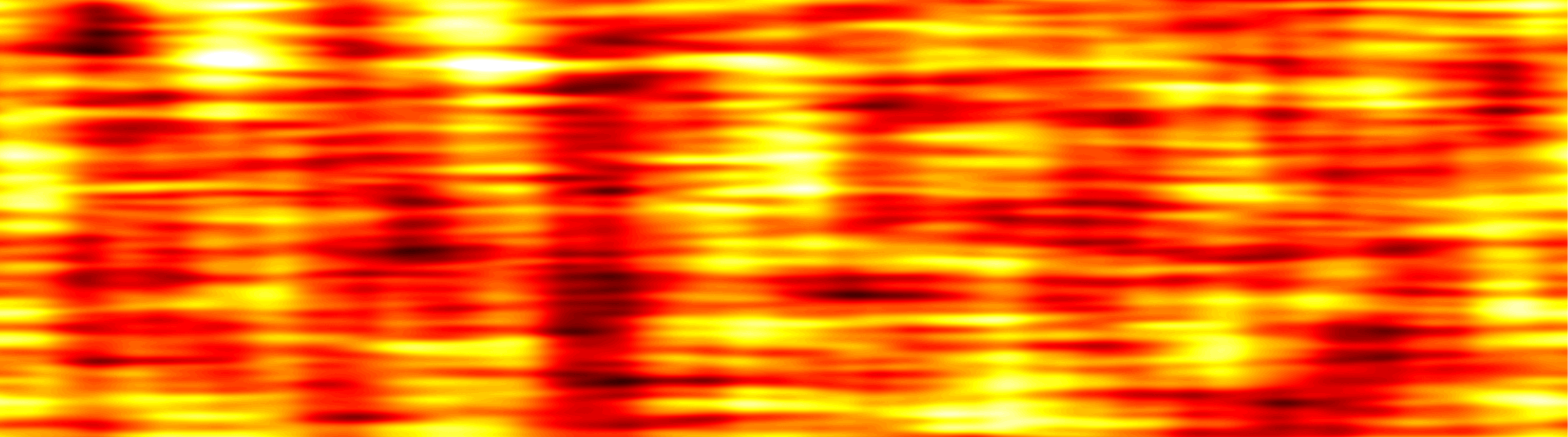}
\put(0.1,3.0){$\displaystyle (d)$}
\vskip2mm
\includegraphics[width=80mm]{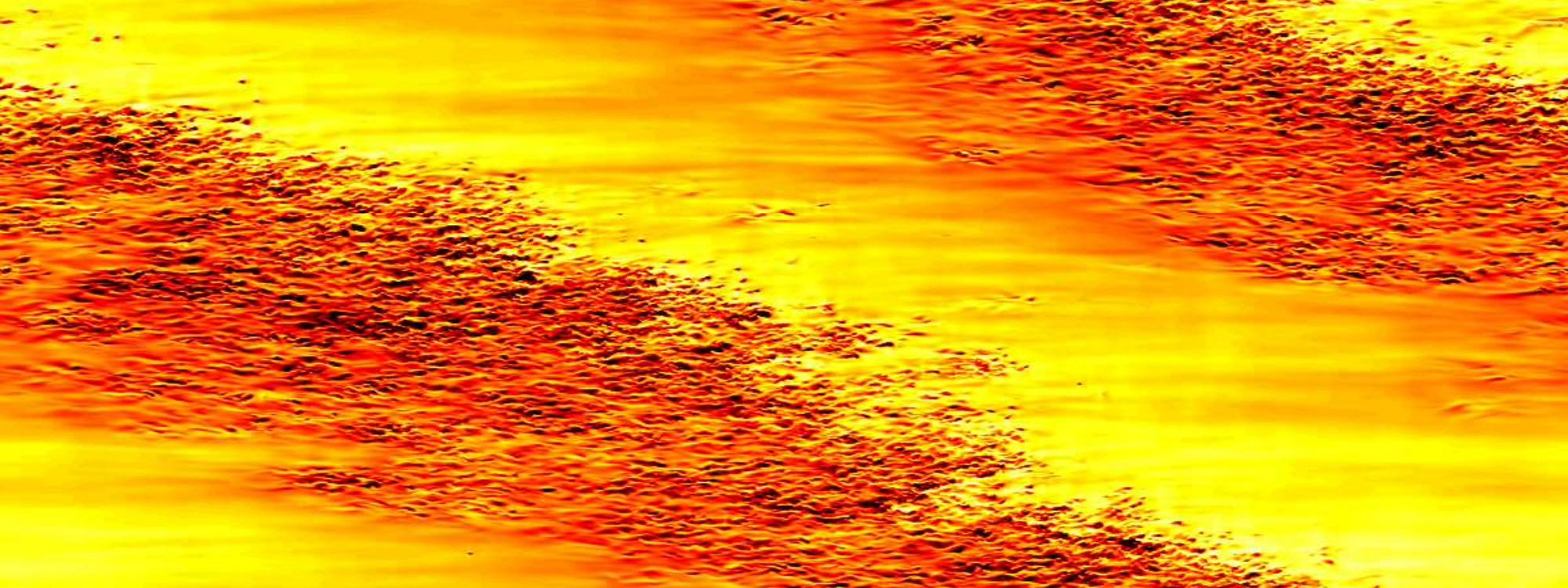}
\put(0.1,2.65){$\displaystyle (e)$}
\end{center}
\caption{(Colour online) Visualizations of the instantaneous streamwise velocity in an $x$-$z$ plane 
at $y^+ \approx 5$ on the stable channel side. 
Dark colours signify high velocities.
$Re=31\,600$ and 
({\it a}) $Ro=0$, 
({\it b}) $Ro=0.45$, 
({\it c}) $Ro=0.9$, 
({\it d}) $Ro=1.2$. 
({\it e}) $Re=20\,000$ and $Ro=0.45$.}
\label{vis_xz}
\end{figure}

Before discussing flow statistics and spectra, visualizations are presented
to get an understanding of the main flow characteristics.
In all following two-dimensional visualizations the complete domain is shown.
Figure \ref{vis_xz}.({\it a-d}) shows plots of the instantaneous streamwise velocity
in an $x$-$z$ plane close to wall on the stable side 
at $Re=31\,600$ and $Ro \leq 1.2$.
Structures with a width of $O(h)$ are vaguely visible at $Ro=0$ 
(figure \ref{vis_xz}.{\it a}) indicating 
near-wall turbulence modulation by large-scale outer structures (Mathis \etal 2009).
In rotating channel flow, turbulence becomes progressively weaker on the stable side
with increasing rotation speeds. 
At low rotation rates, the flow is fully turbulent on the stable side, but 
at $Ro=0.45$ it is not fully turbulent anymore and regions with small-scale
turbulence as well as regions where turbulence is mostly absent 
can be seen in figure \ref{vis_xz}.({\it b}).
The regions with small-scale turbulence have a weakly visible oblique band-like structure
with an angle of about $30^o$ degree to the flow direction.
Similar oblique patterns have been observed in e.g. transitional plane Couette flow
(Barkley \& Tuckerman 2007, Duguet \etal 2010) as well as in other flows
with some stabilizing force like stratified open channel flow,
MHD channel flow and rotating Couette flows (Brethouwer \etal 2012) and
stratified Ekman layers (Deusebio \etal 2014).
In rotating channel flow the turbulent-laminar patterns only exist on the 
stable side since the other side is fully turbulent. 
The turbulent-laminar patterns are not transitional but persists in time, like in the 
aforementioned studies.

When $Ro$ is raised to 0.9 the turbulent fraction becomes smaller and the turbulent regions
appear as spots and bounded band-like oblique structures 
(figure \ref{vis_xz}.{\it c}) while at $Ro=1.2$ and higher
no small-scale turbulence is present and only weak larger scale fluctuations are seen
on the stable channel side (figure \ref{vis_xz}.{\it d}).
In the latter case, a continuous cycle of strong turbulent bursts with a long
period of $O(1000 h/U_b)$ on the stable side
occurs caused by a linear TS wave instability (Brethouwer 2016) and a vague imprint
of the TS wave is visible in figure \ref{vis_xz}. At other times the TS wave is often more
prominent. 

At a lower $Re=20\,000$ the flow is 
fully turbulent on the stable channel side at $Ro=0.15$ (shown later), whereas
at $Ro=0.45$ one distinct oblique turbulent and laminar 
banded pattern can be observed (figure \ref{vis_xz}.{\it e}).
At the same $Ro$ but a higher $Re=31\,600$ 
in a DNS with a larger domain
turbulent-laminar patterns are more numerous but less explicit (figure \ref{vis_xz}.{\it b}).
Although the angle of the oblique pattern in figure figure \ref{vis_xz}.{\it e} is
determined by the periodic boundary conditions and the domain size,
it is similar as the angle of the patterns seen in figure \ref{vis_xz}.{\it b} and {\it c}
and the oblique patterns in other flow cases (Duguet \etal 2010, Brethouwer \etal 2012).
Duguet \& Schlatter (2013) argue that the obliqueness of the patterns is caused
by a large-scale flow with a non-zero spanwise velocity.
In previous DNSs of rotating channel flow discussed in the Introduction 
such patterns have not been observed, which is likely owing to the use of fairly
limited computational domains.

However, Johnston \etal (1972) observed
in a certain $Re$-$Ro$ range 
laminar flows interluded with turbulent spots on the stable channel side 
in their experiments confirming that such transitional flows are found at lower $Ro$
and $Re$.
Dai \etal (2016) observed in DNS at $Re=7000$ and $Ro=0.3$ quasi-periodic behaviour 
of the wall shear stress and turbulence intensity on the stable side because the flow
continuously alternated between laminar-like and intermittent with large-scale streamwise bands 
with either turbulent or laminar features.
They attributed this quasi-periodic behaviour to the dynamics of the streamwise roll cells 
in their DNS. 
Streamwise turbulent-laminar bands, as observed by Dai \etal (2016), are not
seen in the present DNSs which may be related to the size of the computational domain. 
On the other hand, strong quasi-periodic variations of the wall shear stress and turbulence
intensity are observed in some of the present DNSs, but these were caused
by a linear instability of a TS-like wave, as explained in Brethouwer (2016). 
This quasi-periodic behaviour is not observed 
in all but one DNS without this linear instability. 
The exception is the DNS at
$Re=10\,000$ and $Ro=0.45$ where large variations of about $30\%$
are seen in the wall shear stress on the stable channel side.
Visualizations (not shown here) reveal that these variations are related to
quasi-periodically growing and decaying turbulent spots on this side.
When the wall shear stress reaches a minimum the spots almost disappear.
Large cyclic variations in the wall shear stress have also been observed
in DNSs of transitional strongly stratified channel and plane Couette flows
(Garcia-Villalba \& del {\'A}lamo 2011, Deusebio \etal 2015),
but these variations disappeared when the computational domain was enlarged,
indicating that this behaviour is strongly affected by the size of domain.
DNSs of spanwise rotating channel flow by Hsieh \& Biringen (2016) confirm that 
the intermittency on the stable channel side can be strongly influenced by the computational
domain size when the flow is transitional there.

Continuing with the present cases,
if $Ro\geq 0.9$ and $Re=20\,000$ or lower, 
no turbulent spots or band-like structures are seen like in 
figure \ref{vis_xz}.{\it c},
illustrating that $Re$ effects can be appreciable, as discussed in more detail later.
Only weak larger-scale fluctuations are observed on the stable side at high $Ro$, 
as illustrated in figure \ref{vis_xz}.{\it d}.
However, in a number of cases the calm periods on the stable side are interrupted
by violent bursts of turbulence triggered by a linear instability, as explained before.

\begin{figure}
\begin{center}
\setlength{\unitlength}{1cm}
\includegraphics[height=20mm]{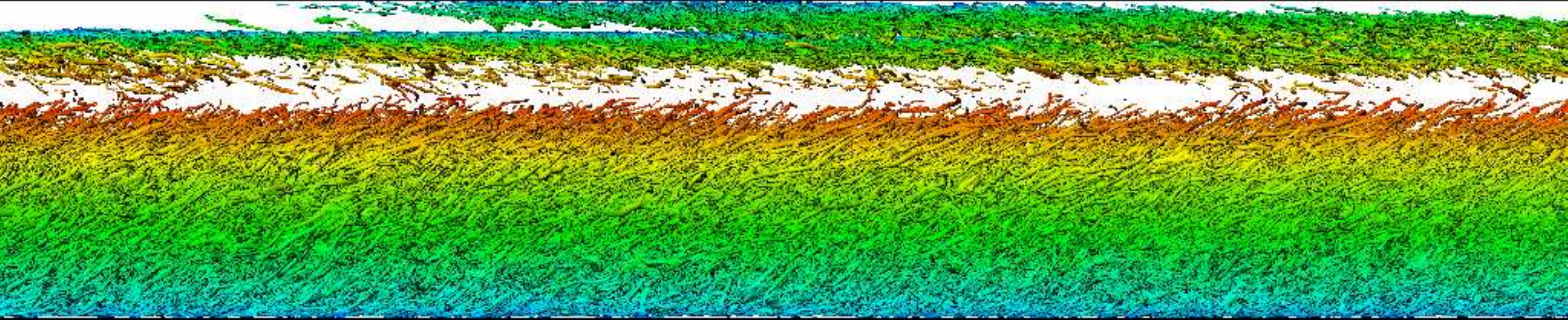}
\put(0.2,1.5){$\displaystyle (a)$}
\vskip4mm
\includegraphics[height=20mm]{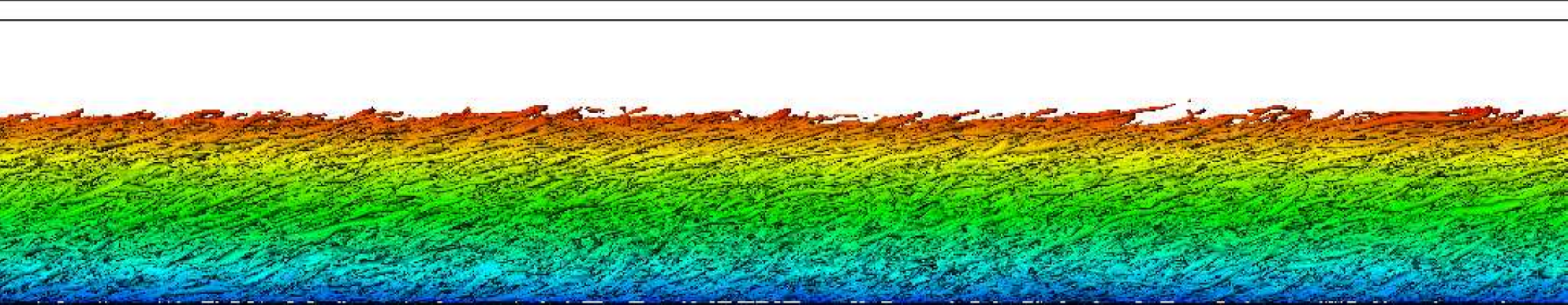}
\put(0.2,1.5){$\displaystyle (b)$}
\vskip4mm
\includegraphics[height=20mm]{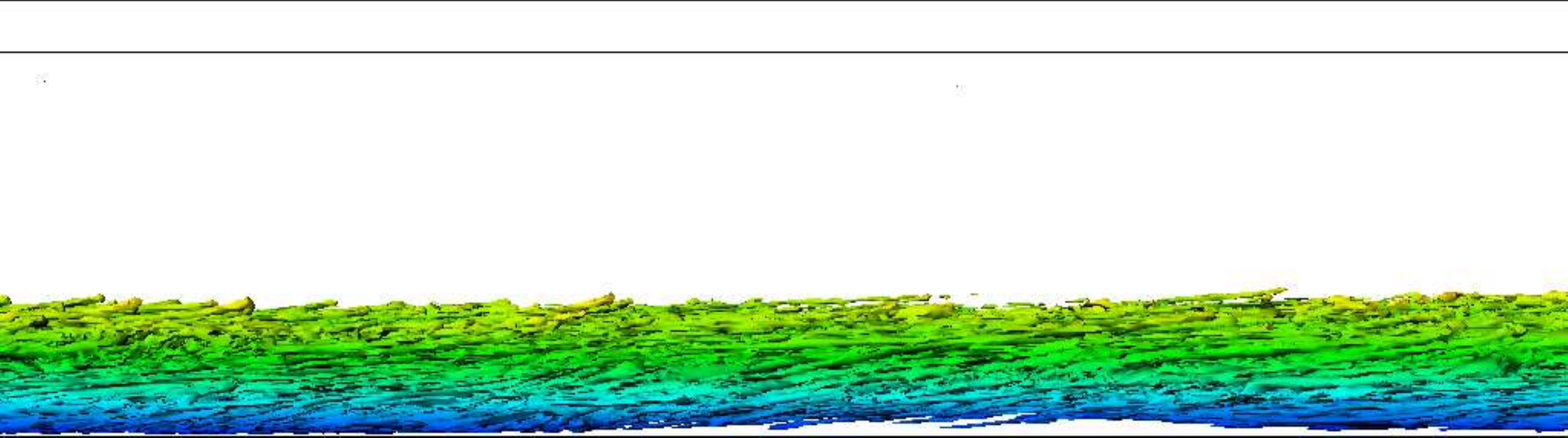}
\put(0.2,1.5){$\displaystyle (c)$}
\vskip4mm
\includegraphics[height=35mm]{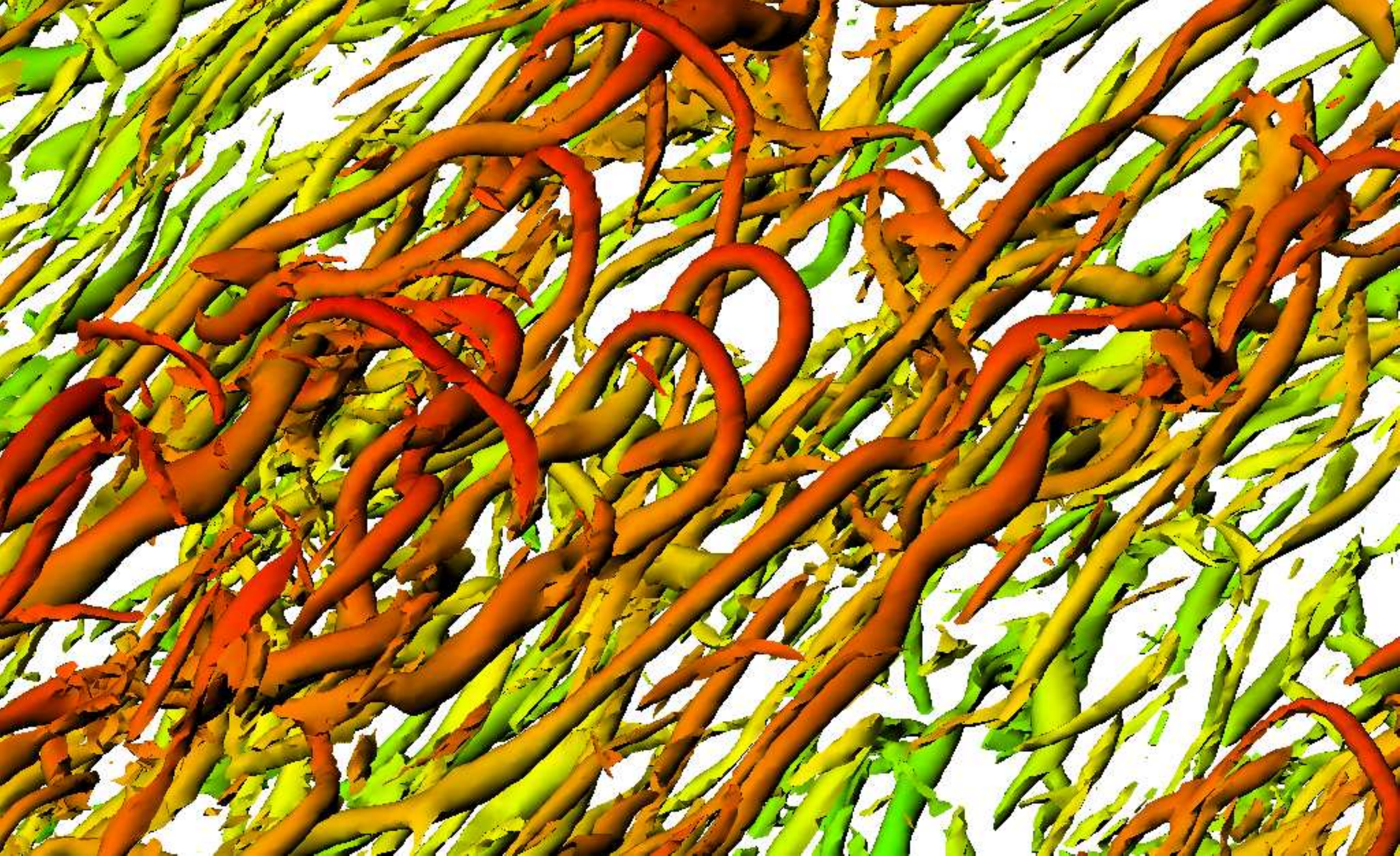}
\put(0.1,3.2){$\displaystyle (d)$}
\hskip7mm
\includegraphics[height=35mm]{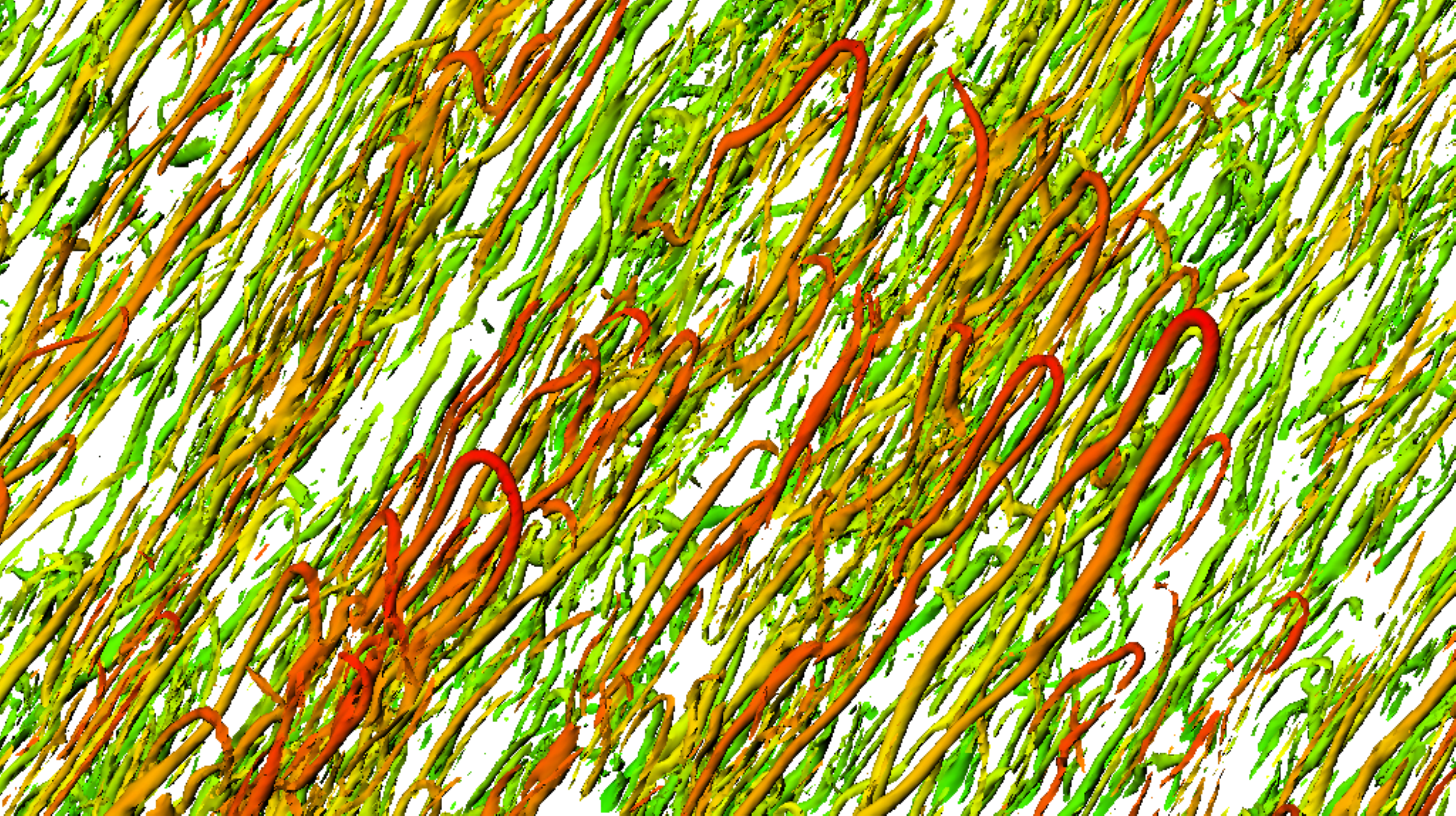}
\put(0.1,3.2){$\displaystyle (e)$}
\end{center}
\caption{(Colour online) Iso-contours of 
({\it a}) $\lambda_2 = -9$ at $Re=31\,600$ and $Ro=0.9$,
({\it b}) $\lambda_2 = -2.25$ at $Re=31\,600$ and $Ro=1.2$,
({\it c}) $\lambda_2 = -0.45$ at $Re=30\,000$ and $Ro=2.1$ (side views, stable
channel side is at the top)
and close up at iso-contours of
({\it d}) $\lambda_2 = -5.6$ at $Re=31\,600$ and $Ro=0.9$,
({\it e}) $\lambda_2 = -2.25$ at $Re=31\,600$ and $Ro=1.2$.
The isosurfaces are coloured with streamwise velocity. The
velocity increases from blue to red.}
\label{vis_lam}
\end{figure}

Figure \ref{vis_lam} shows at three different $Ro$
$\lambda_2$ isocontours (Jeong \& Hussain 1995) coloured
with the streamwise velocity to identify vortices.
The strongly turbulent unstable side of the channel with intense
vortices obviously shrinks with $Ro$. Between the unstable side with
strong turbulence and clearly
identifiable vortices and stable side with weak turbulence a seemingly sharp and flat
border exists
(figure \ref{vis_lam}.{\it a}-{\it c}), as was noted by Johnston \etal (1972),
although when $Ro=0.9$ some areas with vortices 
corresponding to the patterns in figure \ref{vis_xz}.{\it c}
can still be seen on the stable side.
Rotation promotes the formation of streamwise vortices on the unstable side
(Dai \etal 2016), especially in the region where the absolute mean
vorticity is about zero (Lamballais \etal 1996).
Indeed, close ups of the vortices on the
border between the unstable and stable channel side at $Ro=0.9$ and 1.2
in figure \ref{vis_lam}.{\it d} and {\it e} reveal elongated streamwise
vortices and, remarkably, packages of hairpin vortices 
which are detached from the wall.
The red colour signifies that hairpin vortices are mostly found near
the streamwise velocity maximum on the border between the unstable and stable channel side.
At higher $Ro$ vortices including the head of the hairpin vortices become more and more
aligned with the streamwise direction
and hairpin vortices,
as discussed by Adrian (2007), become less explicitly visible (not shown here).
Yang \& Wu (2016) argued that the Coriolis force reduces the inclination angle
of the vortices and favours their streamwise elongation on the unstable side
whereas on the stable channel side this force impedes this streamwise elongation. 
Lamballais \etal (1996) observed that vortices become aligned with the flow direction
with $Ro$ and remarked that especially in the region where the absolute mean
vorticity is about zero streamwise vortex stretching is promoted.
These findings are broadly consistent with the present visualizations.

Large, steady streamwise roll cells induced by the Coriolis force 
have been observed on the unstable side
in several experimental and numerical studies of rotating channel flow (Dai \etal 2016).
Steady means here that they
have a relatively long life time, although in the experiments of Johnston \etal (1972)
the roll cells changed in time whereas
in a DNS by Kristoffersen \& Andersson (1993) they were more coherent and
stationary at $Ro=0.15$ and spanned the whole channel. 

\begin{figure}
\begin{center}
\setlength{\unitlength}{1cm}
\includegraphics[width=80mm]{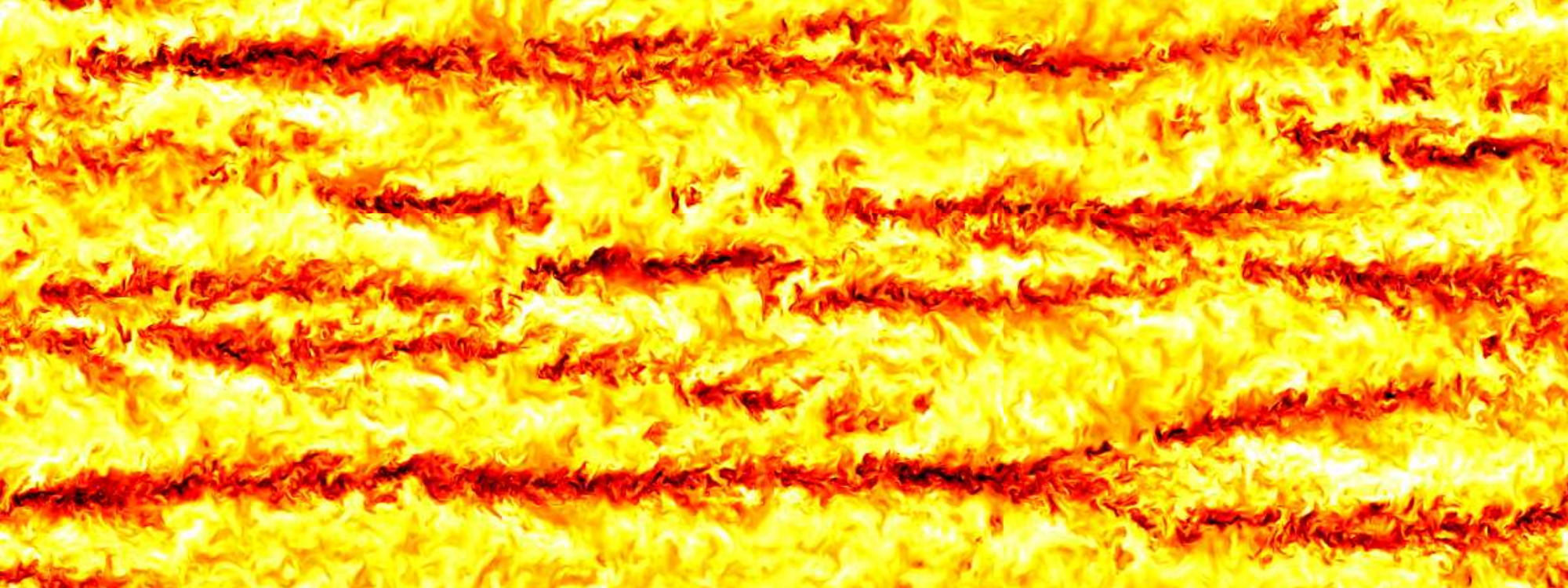}
\put(0.1,2.65){$\displaystyle (a)$}
\vskip2mm
\includegraphics[width=120mm]{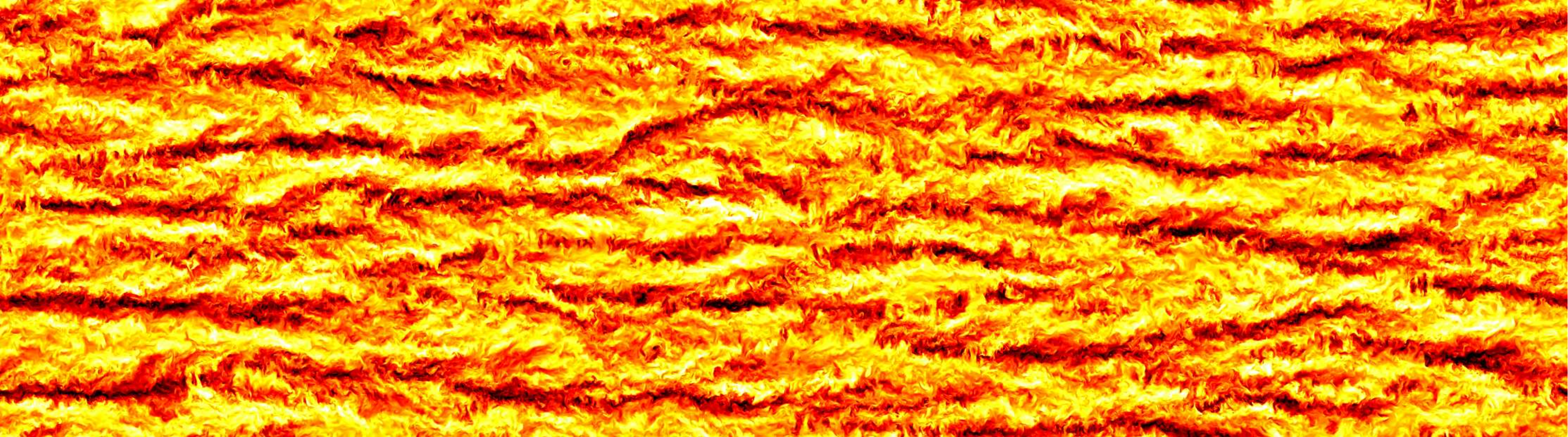}
\put(0.1,3.0){$\displaystyle (b)$}
\vskip2mm
\includegraphics[width=120mm]{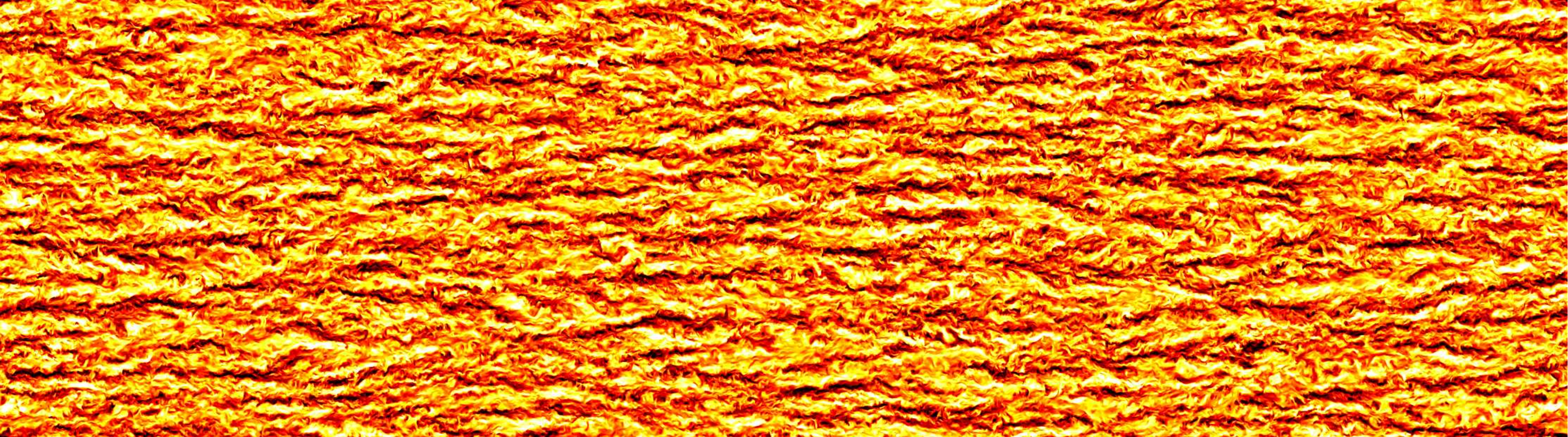}
\put(0.1,3.0){$\displaystyle (c)$}
\vskip2mm
\includegraphics[height=16mm]{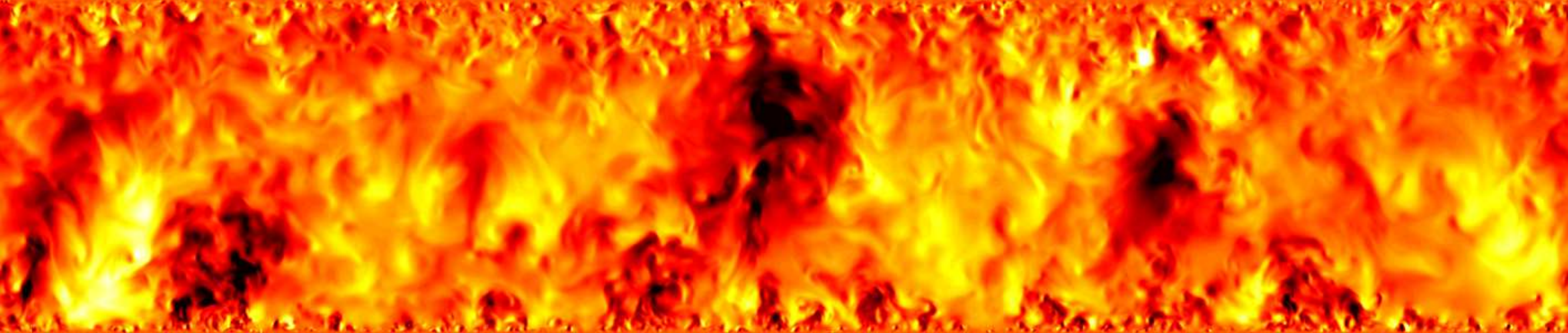}
\put(0.1,1.3){$\displaystyle (d)$}
\vskip2mm
\includegraphics[height=16mm]{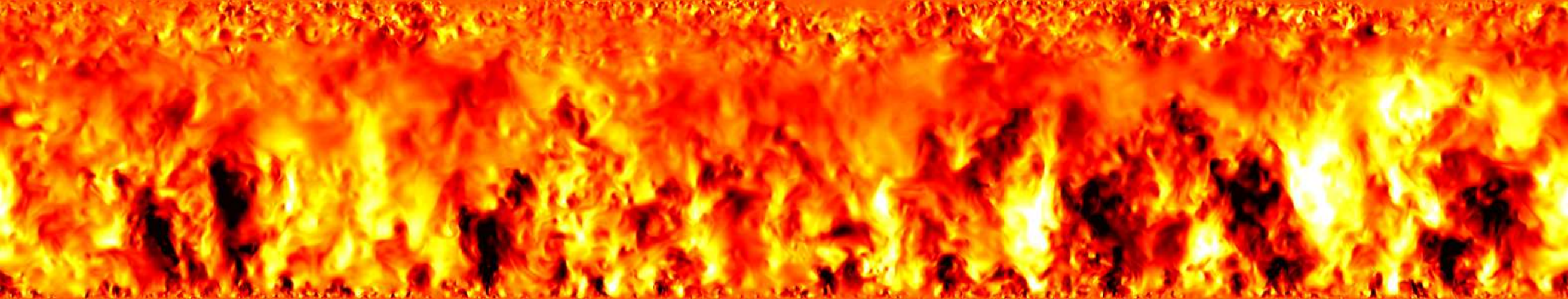}
\put(0.1,1.3){$\displaystyle (e)$}
\vskip2mm
\includegraphics[height=16mm]{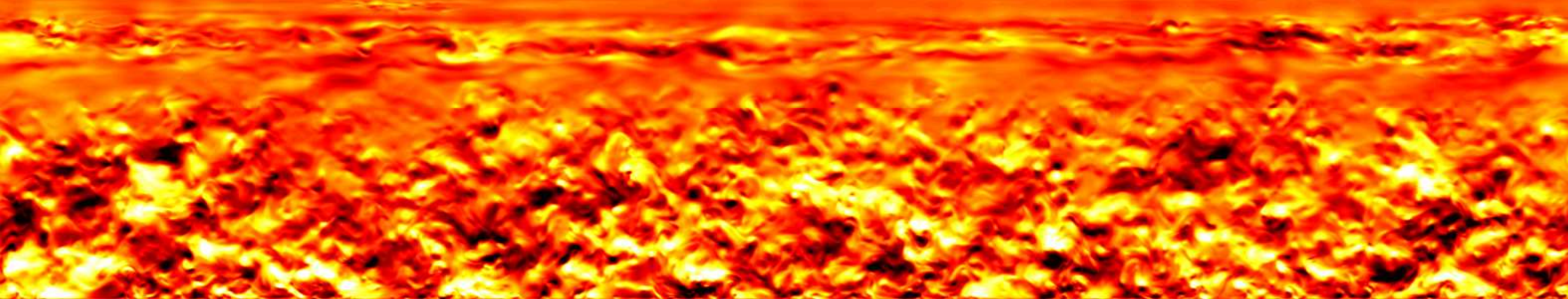}
\put(0.1,1.3){$\displaystyle (f)$}
\vskip2mm
\includegraphics[height=30mm]{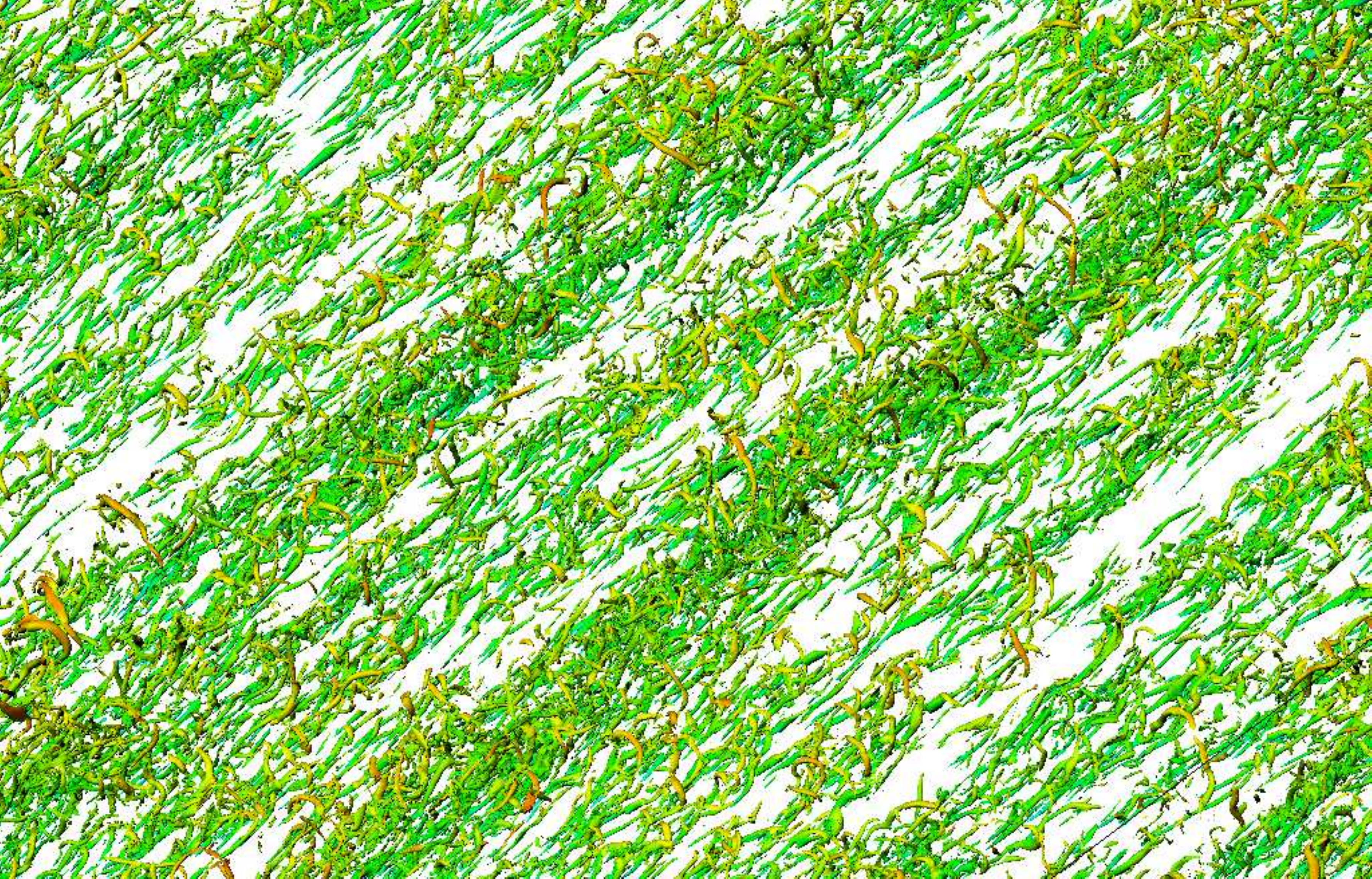}
\put(0.1,2.65){$\displaystyle (g)$}
\end{center}
\caption{(Colour online) Visualizations of the instantaneous wall-normal velocity in an $x$-$z$ plane 
({\it a}) at $y=-0.5$ and $Re=20\,000$, $Ro=0.15$,
({\it b}) at $y=-0.5$ and $Re=31\,600$, $Ro=0.45$,
({\it c}) at $y=-0.75$ and $Re=31\,600$, $Ro=0.9$
and in a $y$-$z$ plane at 
({\it d}) $Re=20\,000$, $Ro=0.15$,
({\it e}) $Re=31\,600$, $Ro=0.45$,
({\it f}) $Re=31\,600$, $Ro=0.9$.
Dark colours signify positive velocities.
({\it g}) Isocontours of $\lambda_2 = -36$ at $Re=20\,000$ and $Ro=0.15$ near the wall
on the unstable side. The flow is from bottom left to upper right corner.
}
\label{vis_roll}
\end{figure}

Visualizations of the instantaneous wall-normal velocity field in the present DNSs
are shown in figure \ref{vis_roll}.
Narrow elongated streaks with positive wall-normal velocity away from the wall in a
wall parallel $x$-$z$ plane on the unstable side (figure \ref{vis_roll}.{\it a}) and 
regions with alternating positive and negative wall-normal velocity in a
cross stream $y$-$z$ plane at $Ro=0.15$ and $Re=20\,000$ (figure \ref{vis_roll}.{\it d})
indicate the presence of large streamwise roll cells
on the unstable side that extend up to the border between the unstable
and the stable side.
The structures are long but not as coherent as in the DNS by Kristoffersen \& Andersson (1993) 
since they appear to break up or split at some places.
Possible reasons for the lesser coherency can be the higher $Re$ and the larger computational domain
in the present study which puts less constrains on the dynamics
of the structures through the periodic boundary conditions.

The clustering of intense vortices in streamwise near-wall streaks
seen in figure \ref{vis_roll}.{\it g} 
show that the roll cells modulate the near-wall dynamics on the unstable side.
Dai \etal (2016) observed in their DNSs of spanwise rotating channel flow that
turbulence and vortices on the unstable side are stronger in 
the regions where the fluid is pumped away from the wall by the counter-rotating
roll cells and the local mean wall shear stress has a minimum,
but this augmentation was weaker at a higher $Re$.
The higher vorticity was found to be caused by the Coriolis term
and strong vortex stretching.
This apparently contrasts the effect of large-scale motions in non-rotating wall flows.
Experiments by Talluru \etal (2014) 
suggest the presence of large-scale
counter-rotating vortices in turbulent boundary layer flow. 
They show that turbulence is weaker in the near-wall regions where the fluid is pumped
away from the wall by these vortices and the local wall shear stress has a minimum
and stronger where the large-scale motions induce a high wall shear stress,
in agreement with Agostini \& Leschziner (2016).
The different effect of large-scale motions in non-rotating vs. rotating
wall flows is presumably a result of the Coriolis term which affects both the
Reynolds stresses and vorticity.
Roll cells in rotating flows are likely also 
more coherent and steady and induce stronger wall-normal velocities.

Streaks with positive wall-normal velocity
(figure \ref{vis_roll}.{\it b})
and large-scale regions with positive and negative wall-normal velocity
(figure \ref{vis_roll}.{\it e})
at $Ro=0.45$ and $Re=31\,600$ indicate roll cells, 
but they seem to be smaller and less coherent than at $Ro=0.15$. 
Streaky structures with a positive wall-normal velocity are seen too
in figure \ref{vis_roll}.{\it c} 
in the DNS at $Ro=0.9$ 
but it is not clear if these upward and downward motions seen in
figure \ref{vis_roll}.{\it f} 
can be interpreted as signs of roll cells.
The spacing and form of the streaks indicate that the roll cells, if they exist, are 
smaller and less coherent than at lower $Ro$. 
Roll cells became smaller too in the DNS of rotating channel flow at $Re = 2800$ by 
Dai \etal (2016) when $Ro$ was raised from 0.1 to 0.5.
Signs of roll cells are visible in the DNSs at $Ro=0.45$ and 0.9
but lower $Re$ (not shown here).
When $Ro=1.2$ or higher no visible
signs of roll cells are found at $Re=20\,000$ and $31\,600$. However, at $Re=5000$
visualizations hint at the existence of roll cells and Grundestam
\etal (2008) observed them at $Ro=1.27$ and $Re_\tau=180$.
These observations suggest that at lower $Re$ roll cells exist up to
higher $Ro$. 
A more quantitative study of the structures in rotating channel flow is
presented in Section \ref{sec_spec}.

\begin{figure}
\begin{center}
\setlength{\unitlength}{1cm}
\includegraphics[width=80mm]{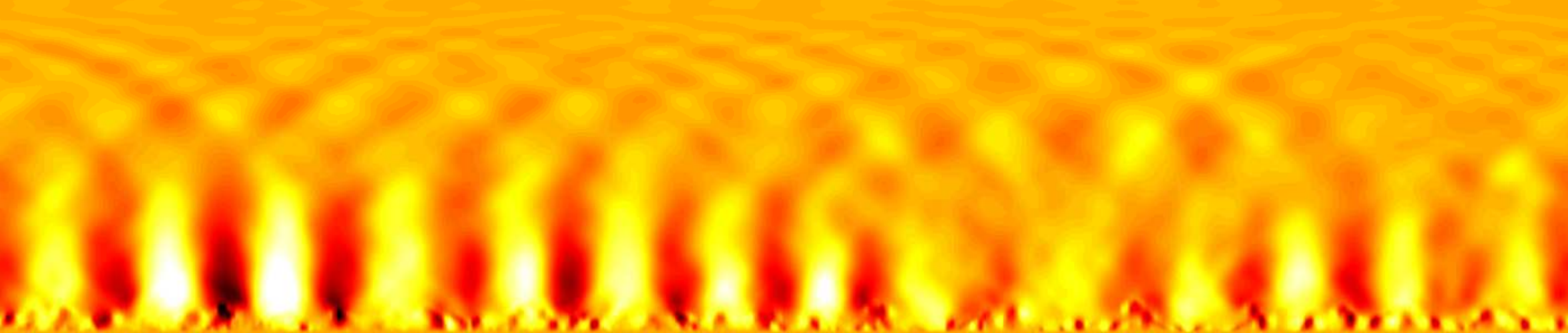}
\put(0.1,1.4){$\displaystyle (a)$}
\vskip2mm
\includegraphics[width=75mm]{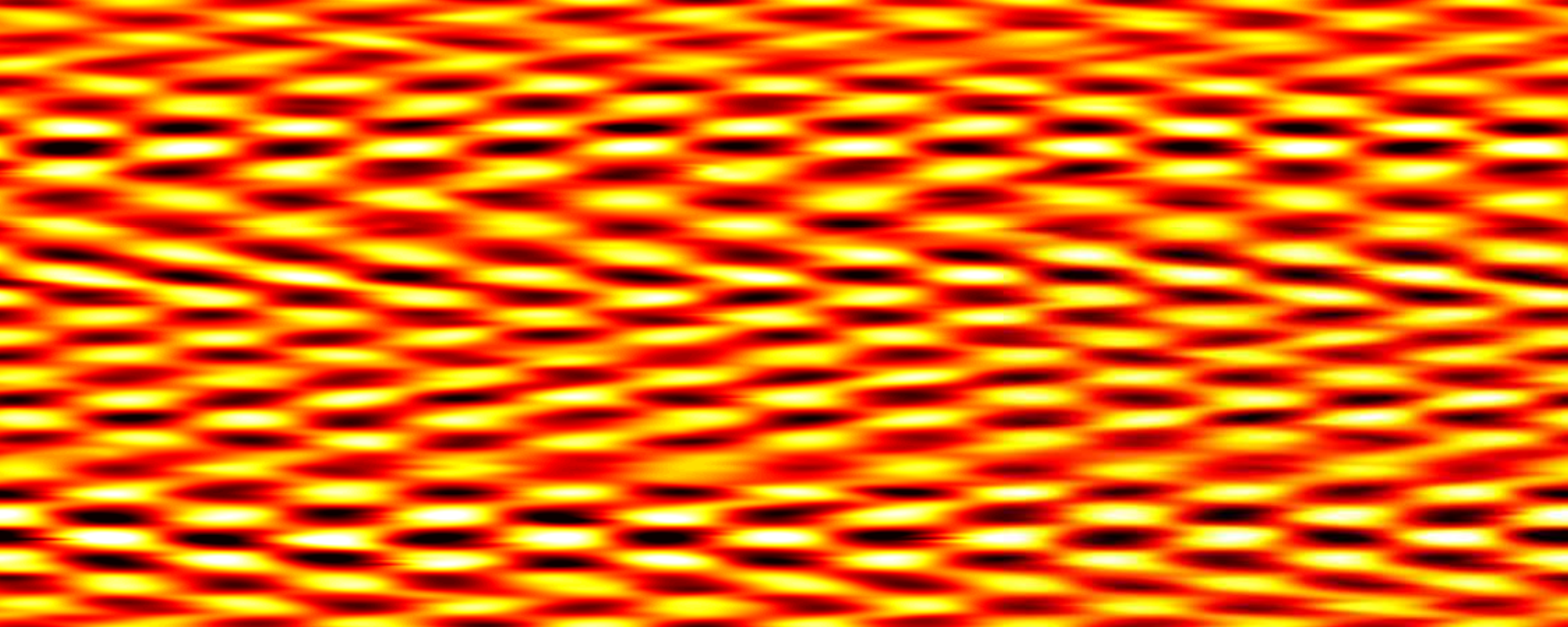}
\put(0.1,2.7){$\displaystyle (b)$}
\end{center}
\caption{(Colour online) Visualization of the instantaneous wall-normal velocity 
for $Re=30\,000$ and $Ro=2.7$ in 
a ({\it a}) $y$-$z$ plane 
and in an ({\it b}) $x$-$z$ plane at $y=-0.8$.
Colour scale ranges from $v=-0.021$ to 0.021 in {\it b}.}
\label{vis_obl}
\end{figure}

At higher $Ro$ the unstable turbulent side becomes smaller and smaller
and the flow tends to fully laminarize if $Ro$ approaches $Ro_c$ (Wallin \etal 2013).
Yet, even if $Ro \rightarrow Ro_c$, streamwise and oblique modes 
are still
unstable as a result of rotation according
to linear stability theory (Brethouwer 2016) and some of the largest linearly
unstable modes become noticeable in the DNSs
when turbulence gets weak on the unstable channel side. 
Figure \ref{vis_obl}
shows the resulting typical oblique waves on the unstable side
in a DNS at $Re=30\,000$ and $Ro=2.7$, close to $Ro_c=2.87$ for this $Re$.
The waves are quite weak, i.e., the wall-normal velocity is about 2 to 3\% of $U_b$
in this case.

\section{Flow statistics}
\label{sec_stat}

In this section, one-point statistics of the flow
are presented and the effects of rotation are discussed.
As mentioned before, linearly unstable TS waves cause strong recurring bursts of turbulence
on a long time scale in some DNSs. The instabilities and bursts are the topic of another
study (Brethouwer 2016) and therefore not discussed in detail here. However, it should
be noted that they significantly affect flow structures and turbulence,
especially around the bursting moment and on the stable side where the bursts are most intense.
Time series of, for instance, the volume-averaged turbulent kinetic energy $K_m$ show distinct sharp
peaks as a result of the bursts while in the calm periods between the bursts 
the same quantity only shows small to moderate variations, see e.g. figure 3.({\it b})
in Brethouwer \etal (2014).
I have excluded the burst periods 
when computing the statistics by excluding the periods with distinct peaks in $K_m$
caused by the linear instability. 
The statistics are thus 
based on the long relatively calm periods of $O(100 h/U_b)$ between the bursts
when $K_m$ only shows small to moderate variations and the turbulence does not appear
to be strongly influenced by the instability. 
The reason for excluding these bursts periods from the statistics is that
the bursts are not the subject of this study 
and that the bursts would obscure the direct effect of rotation on the turbulence.
Besides, the bursts occur on a very long time scale 
of $O(1000 h/U_b)$,
which makes it practically impossible to obtain reasonable converged statistics
if they are included, and they cannot be predicted by Reynolds stress and two-equation
turbulence models.
Brethouwer (2016) lists all the cases when bursts happen. At $Re = 30\,000$ to
$31\,600$ linear instabilities and bursts develop when $Ro\geq 1.2$.
In the next parts, $U$ is the mean streamwise velocity and $u$, $v$, $w$
are the streamwise, wall-normal and spanwise velocity fluctuation, respectively.
An overline implies temporal and spatial averaging in the homogeneous directions.

\begin{figure}
\begin{center}
\setlength{\unitlength}{1cm}
\includegraphics[width=62mm]{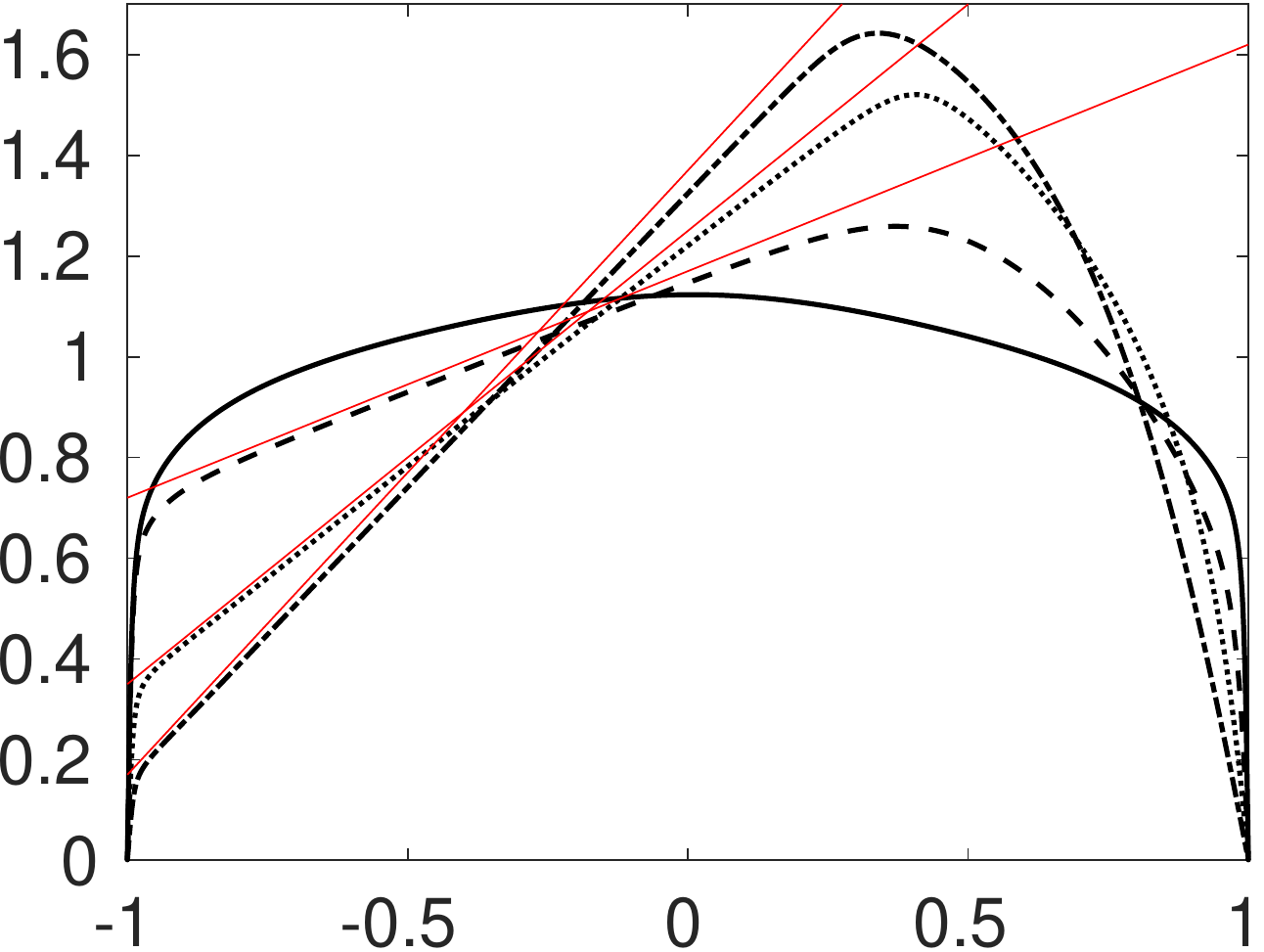}
\put(-0.1,4.3){$\displaystyle (a)$}
\put(-3.1,-0.3){$y$}
\put(-6.9,2.3){$\frac{U}{U_b}$}
\hskip9mm
\includegraphics[width=62mm]{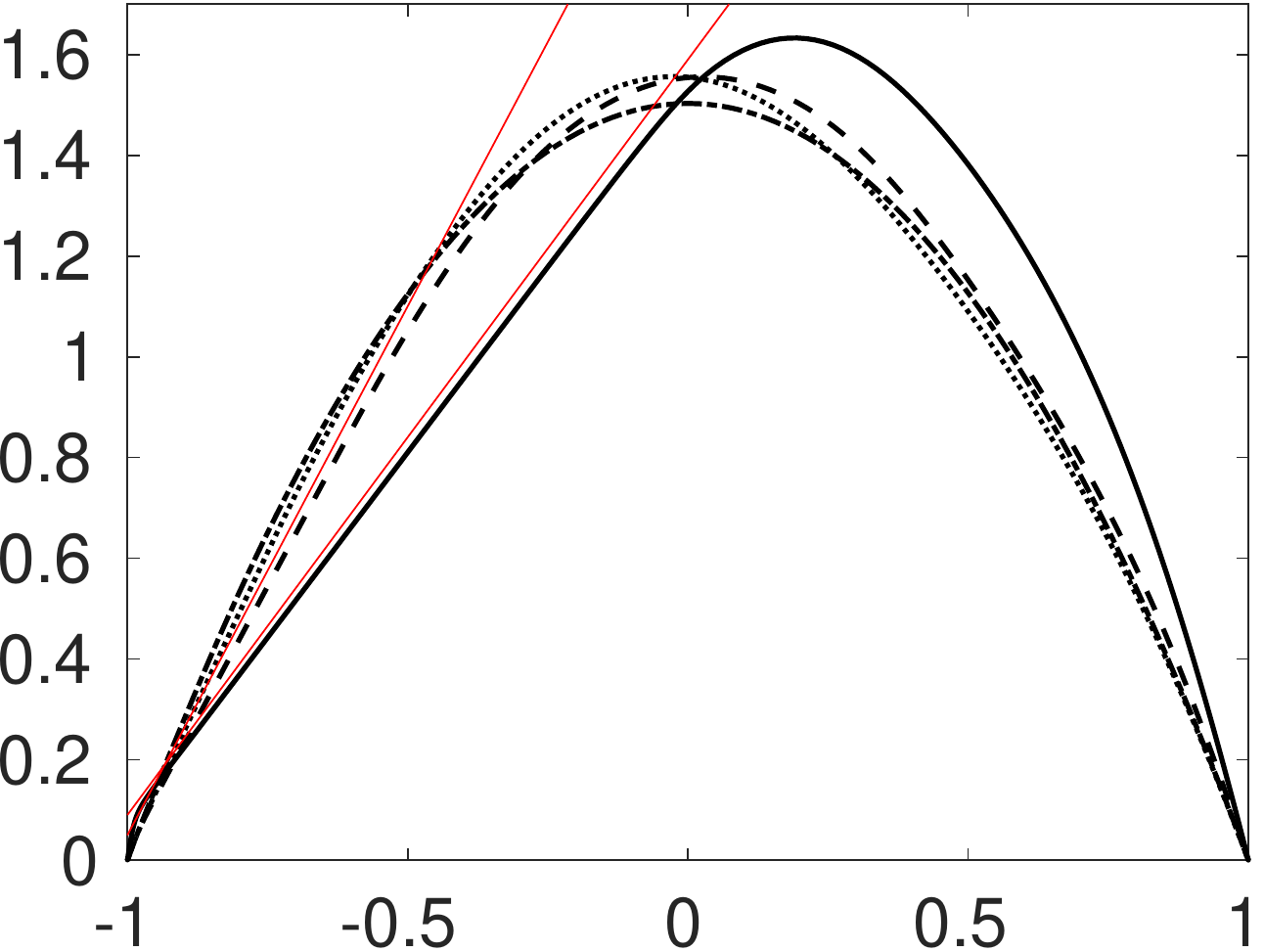}
\put(-0.1,4.3){$\displaystyle (b)$}
\put(-3.1,-0.3){$y$}
\put(-6.9,2.3){$\frac{U}{U_b}$}
\end{center}
\caption{(Colour online) Mean velocity profiles scaled by $U_b$.
({\it a}) $Re=31\,600$.
$~^{\line(1,0){20}}$, $Ro=0$;
$---$, $Ro=0.45$;
$\cdot\cdot\cdot$, $Ro=0.9$;
$-\cdot-\cdot-$, $Ro=1.2$.
({\it b}) $Re=30\,000$.
$~^{\line(1,0){20}}$, $Ro=1.5$;
$---$, $Ro=2.1$;
$\cdot\cdot\cdot$, $Ro=2.4$;
$-\cdot-\cdot-$, $Ro=2.7$.
Slopes with $S=-1$ are shown by straight red lines.}
\label{umean}
\end{figure}

Figure \ref{umean} shows mean streamwise velocity profiles scaled
by $U_b$ at the highest $Re$ considered for $Ro$ up to 2.7.
As in previous studies of turbulent channel flow subject to spanwise rotation
(e.g. Kristoffersen \& Andersson 1993,
Grundestam \etal 2008, Xia \etal 2016), 
the mean velocity profile becomes asymmetric and
develops an extended linear region where the slope 
$\mbox{d}U/\mbox{d}y \simeq 2 \Omega$, i.e. 
$S\simeq -1$, implying an absolute mean vorticity close to zero.
If $Ro\leq 1.5$ the velocity on the unstable side goes down while on the stable side
it goes up since the turbulence becomes stronger and weaker on
the unstable and stable side, respectively, as will be shown later.
At higher $Ro$, the profile becomes more and more parabolic-like and
beyond $Ro=2.4$ the linear slope region disappears and the velocity profile
approaches a laminar Poiseuille profile, 
as in the DNSs by Grundestam \etal (2008) and Xia \etal (2016).

\begin{figure}
\begin{center}
\setlength{\unitlength}{1cm}
\includegraphics[width=62mm]{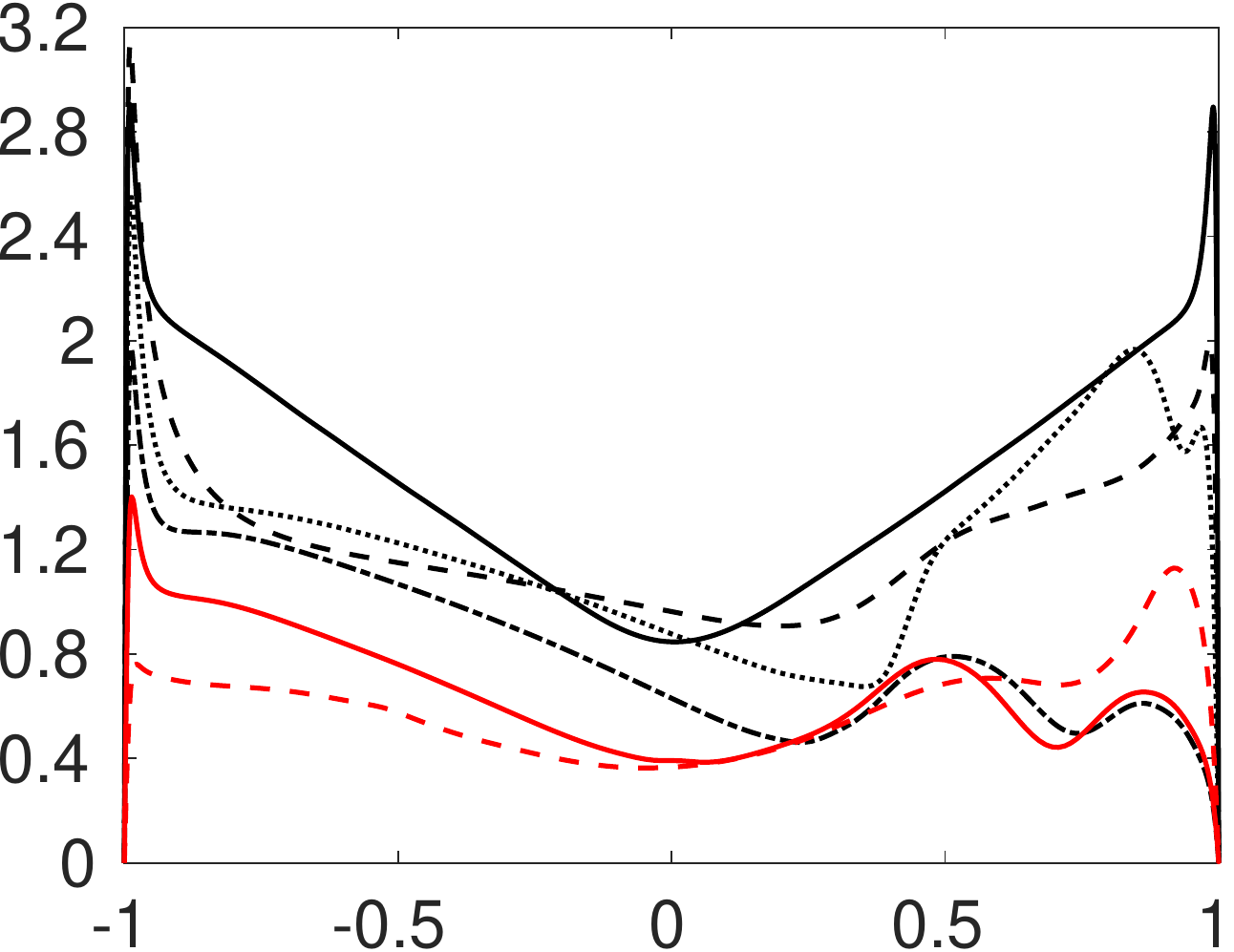}
\put(-0.1,4.3){$\displaystyle (a)$}
\put(-3.1,-0.3){$y$}
\put(-6.9,2.3){$u^+$}
\hskip9mm
\includegraphics[width=62mm]{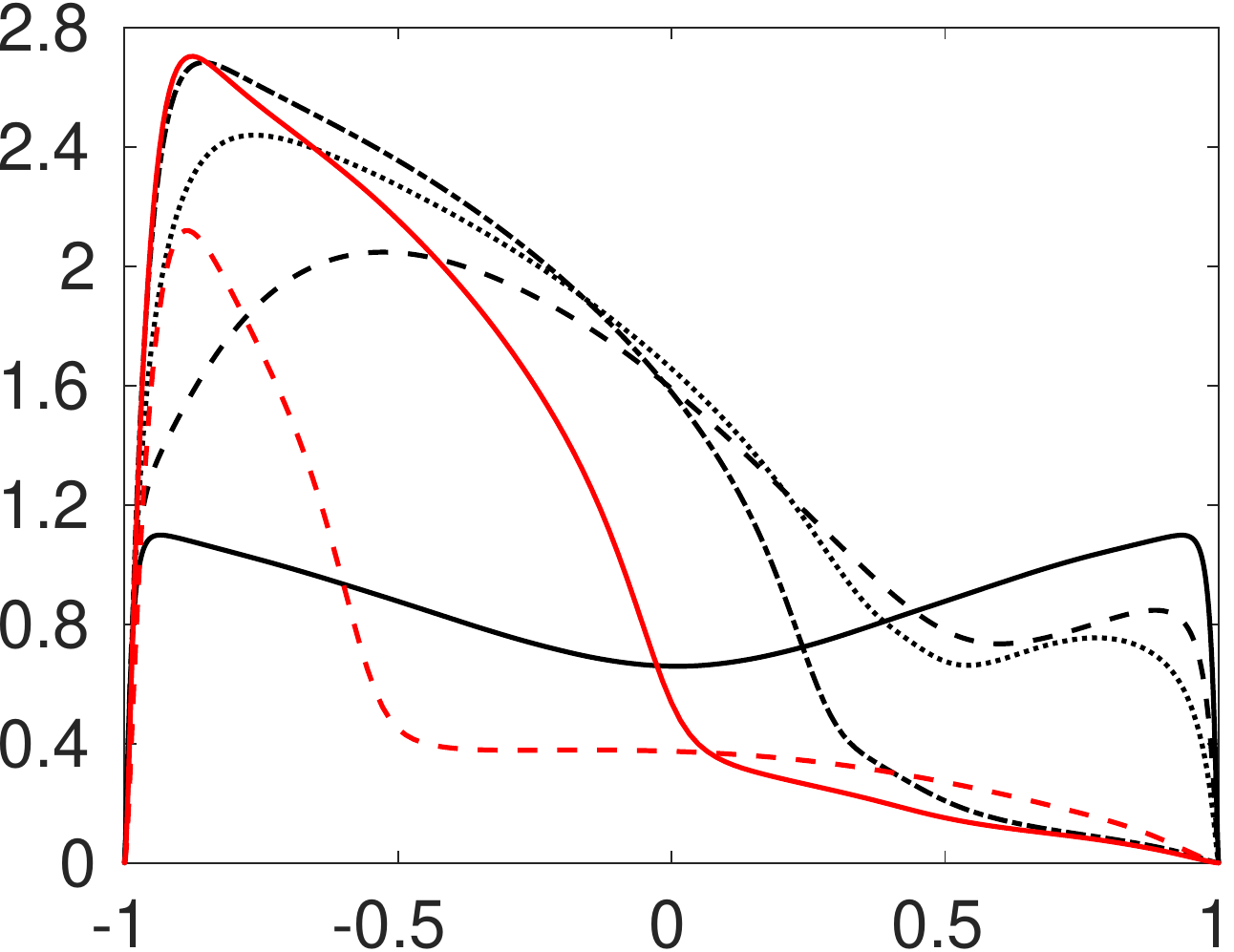}
\put(-0.1,4.3){$\displaystyle (b)$}
\put(-3.1,-0.3){$y$}
\put(-6.9,2.3){$v^+$}

\includegraphics[width=62mm]{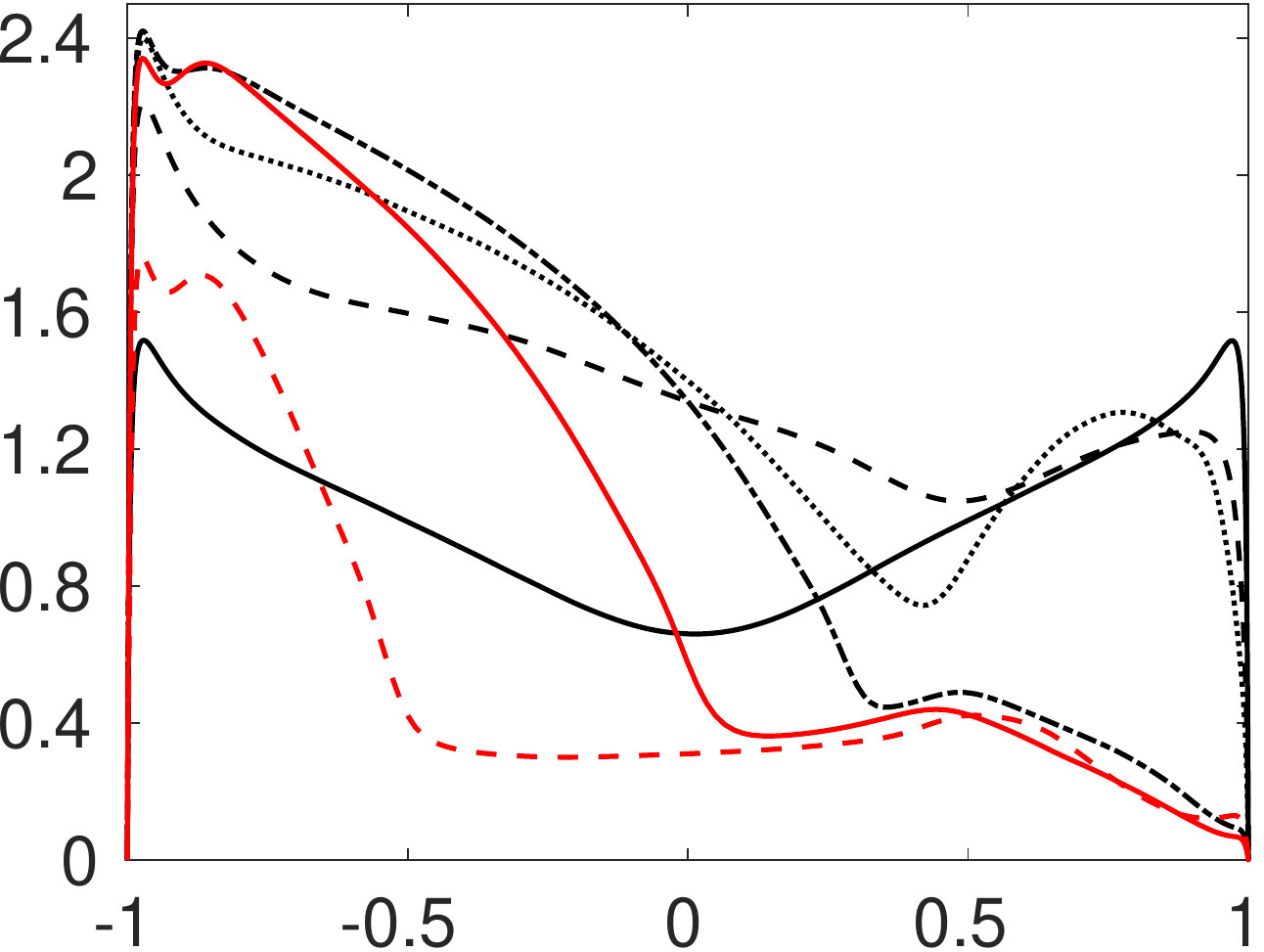}
\put(-0.1,4.3){$\displaystyle (c)$}
\put(-3.1,-0.3){$y$}
\put(-6.9,2.3){$w^+$}
\hskip9mm
\includegraphics[width=62mm]{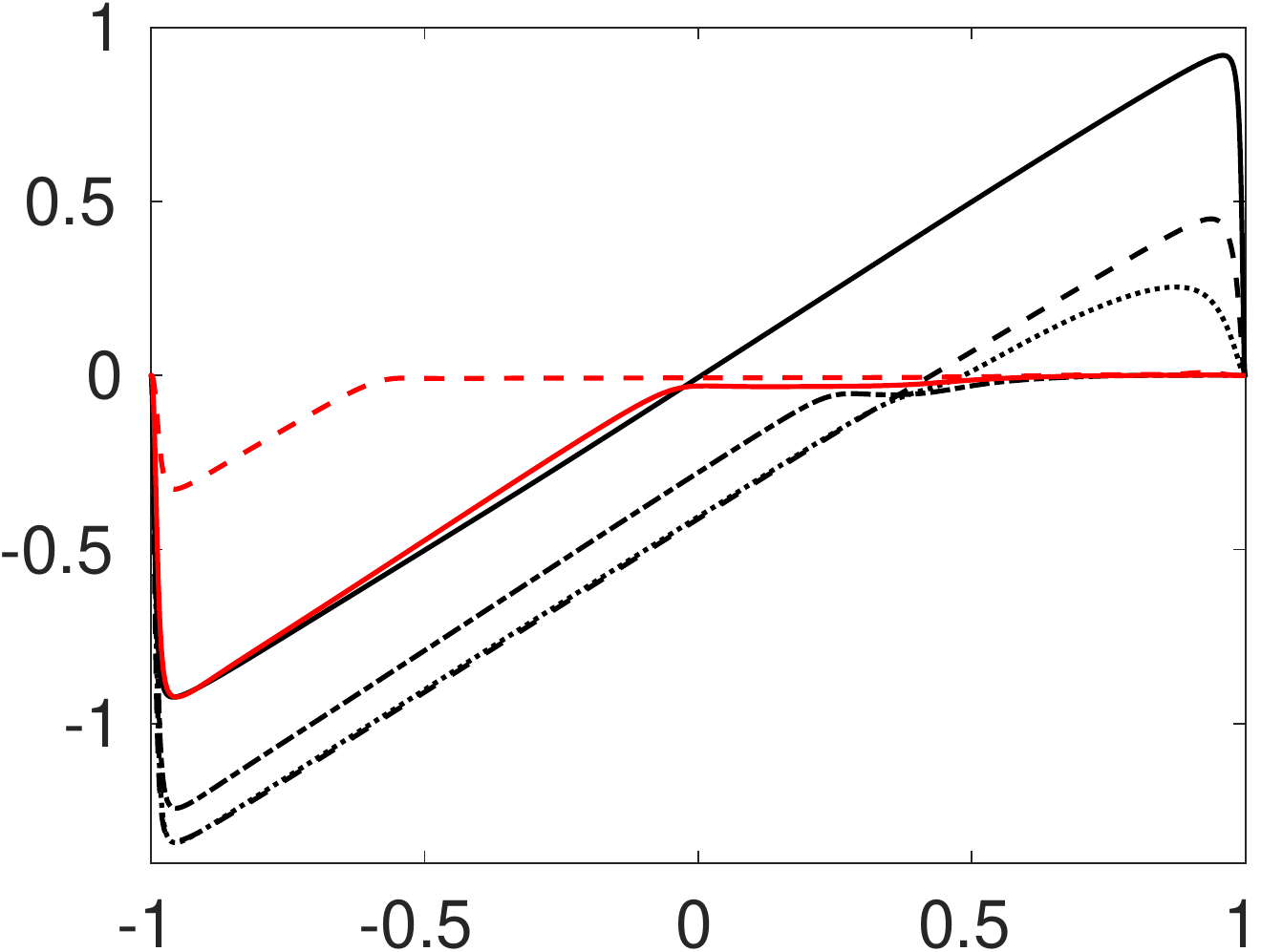}
\put(-0.1,4.3){$\displaystyle (d)$}
\put(-3.1,-0.3){$y$}
\put(-6.9,2.3){$uv^+$}
\end{center}
\caption{(Colour online) Rms profiles of 
({\it a}) streamwise,
({\it b}) wall-normal
({\it c}) spanwise
velocity fluctuations and ({\it d}) $\overline{uv}$ profiles in wall units.
$Re=31\,600$ and
($~^{\line(1,0){20}}$) $Ro=0$,
($---$) $Ro=0.45$,
($\cdot\cdot\cdot$) $Ro=0.9$,
($-\cdot-\cdot-$) $Ro=1.2$.
$Re=30\,000$ and
($\textcolor{red}{~^{\line(1,0){20}}}$) $Ro=1.5$,
($\textcolor{red}{---}$) $Ro=2.1$.}
\label{urms}
\end{figure}

Profiles of the root-mean-square (rms) of the 
streamwise, wall-normal and spanwise velocity fluctuations,
$u^+$, $v^+$ and $w^+$, respectively, and the Reynolds
shear stress $\overline{uv}^+$
normalized by $u_\tau [u^2_{\tau u} /2 + u^2_{\tau s} /2 ]^{1/2}$ 
for $Ro$ up to 2.1 are shown in figure \ref{urms}.
The results are for the highest $Re$ considered here, $31\,600$ and $30\,000$.
Note that for $Ro \geq 1.2$ the flow is subject to a
TS-wave instability resulting in intense bursts of turbulence
at this $Re$ (Brethouwer 2016), 
but these periods with turbulent bursts are excluded as much as possible
when computing the statistics as mentioned before. The statistics are thus 
based on the long calm periods between the bursts.

Figure \ref{urms} shows that rotation causes a reduction of the turbulence intensity
on the stable side, as expected (Johnston \etal 1972). 
Especially for the wall-normal component (figure \ref{urms}.{\it b})
this reduction is apparent while for the other two velocity components it becomes most notable
for $Ro \geq 1.2$. The turbulence on the unstable channel side displays a more complex
behaviour. The peak of $u^+$ first raises with $Ro$ but then declines if $Ro > 0.45$
whereas $w^+$ and especially $v^+$ strongly grow with $Ro$ and only start to
notably decline when $Ro > 1.5$. At higher $Ro$ turbulence is progressively suppressed
(not shown here)
and the flow approaches more and more a laminar Poiseuille flow, as observed for
the mean flow. Rotation has thus not only
a marked influence on the turbulence intensity but also
on its anisotropy, as in rotating homogeneous
shear flows (Brethouwer 2005). 
The observed trends are in qualitative agreement with DNSs 
of rotating channel flow at lower $Re$ (Grundestam \etal 2008).
The maximum of $v^+$ is near the wall at $Ro=0$ but remarkably far from
the wall on the unstable side
at $Ro=0.45$, which may be due to the presence of roll cells, 
while at higher $Ro$ it approaches the wall again.

From the mean momentum balance for rotating channel flow follows 
\begin{equation}
-\overline{uv} + \nu \frac{\diff U}{\diff y} 
= u^2_{\tau u}
- u^2_\tau ( y +1),
\label{mom_bal}
\end{equation}
where the velocities are dimensional.
The sum of the viscous and turbulent shear stresses is thus linear in $y$
but is shifted owing to the difference in the wall shear stresses
on the unstable and stable channel sides in the rotating flow cases.
These stresses are naturally higher on the unstable side owing to the more intense turbulence.
When viscous stresses are negligible 
the $\overline{uv}^+$ profile is also linear in $y$ according to equation (\ref{mom_bal}).
From this equation follows that in the part of the
channel where $\nu\diff U/\diff y \simeq 2 \nu \Omega= U^2_b Ro/Re $
the turbulent shear stress approximates
\begin{equation}
\overline{uv} \simeq
u^2_\tau ( y +1) - u^2_{\tau u}
 + U^2_b \frac{Ro}{Re}. 
\label{stress_approx}
\end{equation}
This shows that 
the $\overline{uv}^+$ profile is linear in $y$ and has a unit slope
even if viscous stresses are not negligible (Xia \etal 2016).
Figure \ref{urms}.({\it d}) confirms that on the unstable
side the $\overline{uv}^+$ profiles have a unit slope.
The turbulent momentum transfer shifts progressively towards
the unstable side with $Ro$ and for $Ro \geq 1.2$
it is in fact negligible on the stable side 
where viscous stresses dominate.
On the unstable side the magnitude of $\overline{uv}^+$
starts to decline when $Ro > 0.9$ and is small for $Ro \geq 2.1$.
Viscous shear stresses are significant on the strongly turbulent
unstable channel side at higher $Ro$ because of the steep mean 
velocity gradient. If the total
shear stress is estimated as $u^2_\tau$, it follows that on the unstable side
where $\diff U/\diff y \simeq 2 \Omega$ the ratio between viscous
and total stresses is approximately $Ro\,Re / Re^2_\tau$. From that
follows that the viscous contribution grows with $Ro$ and
is about 12\% and 26\% at
$Re=31\,600$ and $Ro=1.2$ and 1.5, respectively, and even higher at
the same $Ro$ but lower $Re$. Once $Ro \gtrsim 2.0$ viscous stresses dominate also
on the unstable side.

\begin{figure}
\begin{center}
\setlength{\unitlength}{1cm}
\includegraphics[width=62mm]{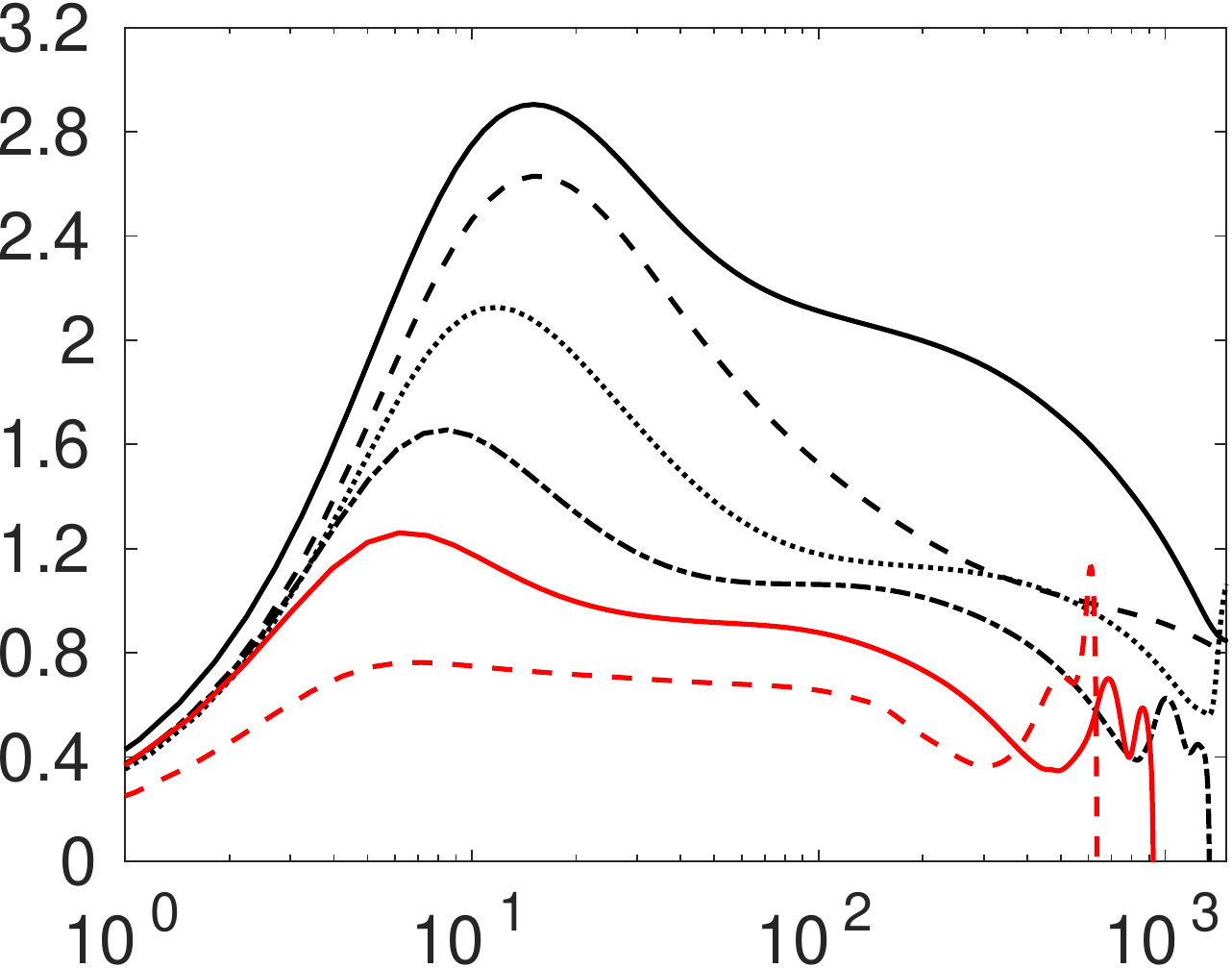}
\put(-0.1,4.3){$\displaystyle (a)$}
\put(-3.1,-0.3){$y^*$}
\put(-6.9,2.3){$u^*$}
\hskip9mm
\includegraphics[width=62mm]{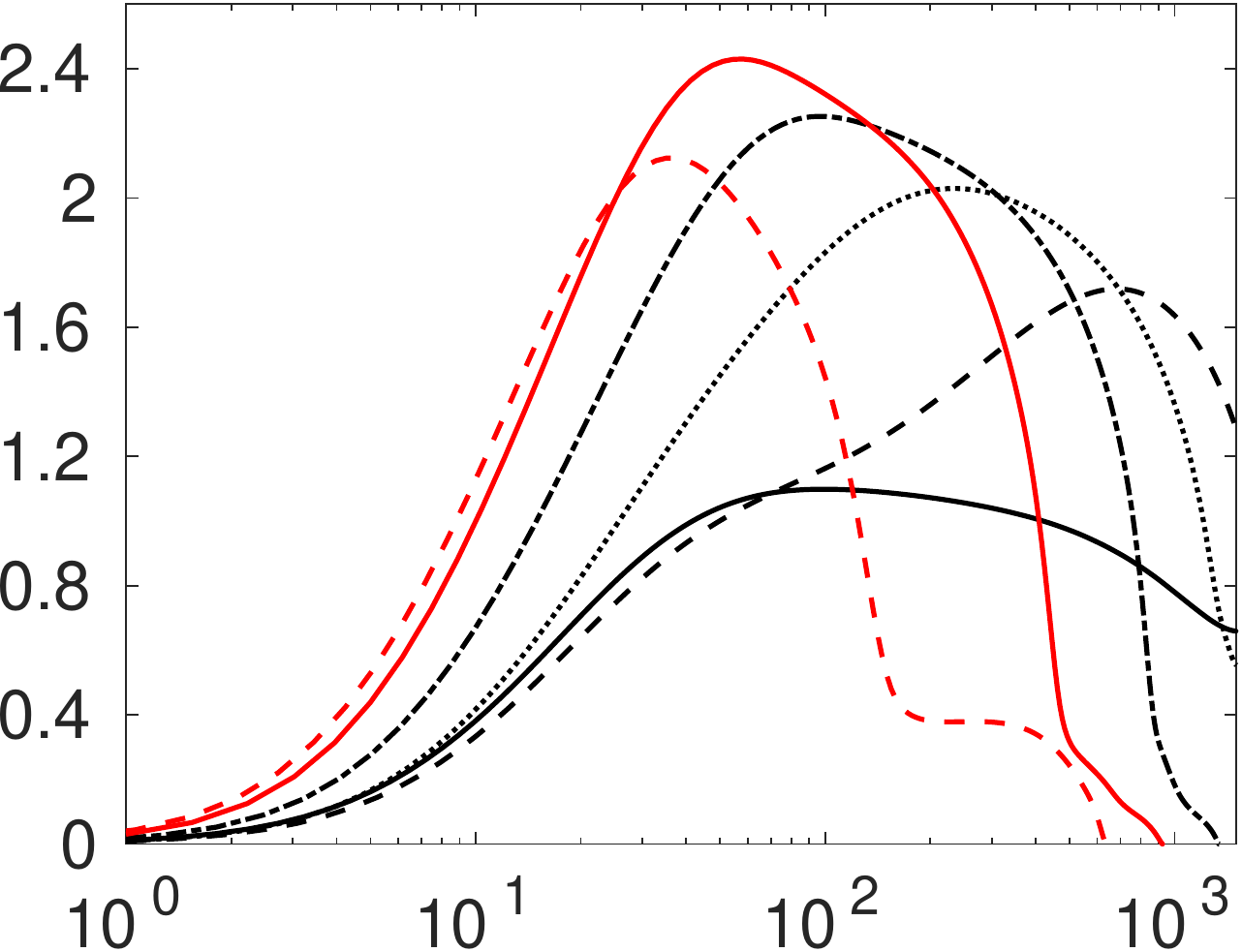}
\put(-0.1,4.3){$\displaystyle (b)$}
\put(-3.1,-0.3){$y^*$}
\put(-6.9,2.3){$v^*$}

\includegraphics[width=62mm]{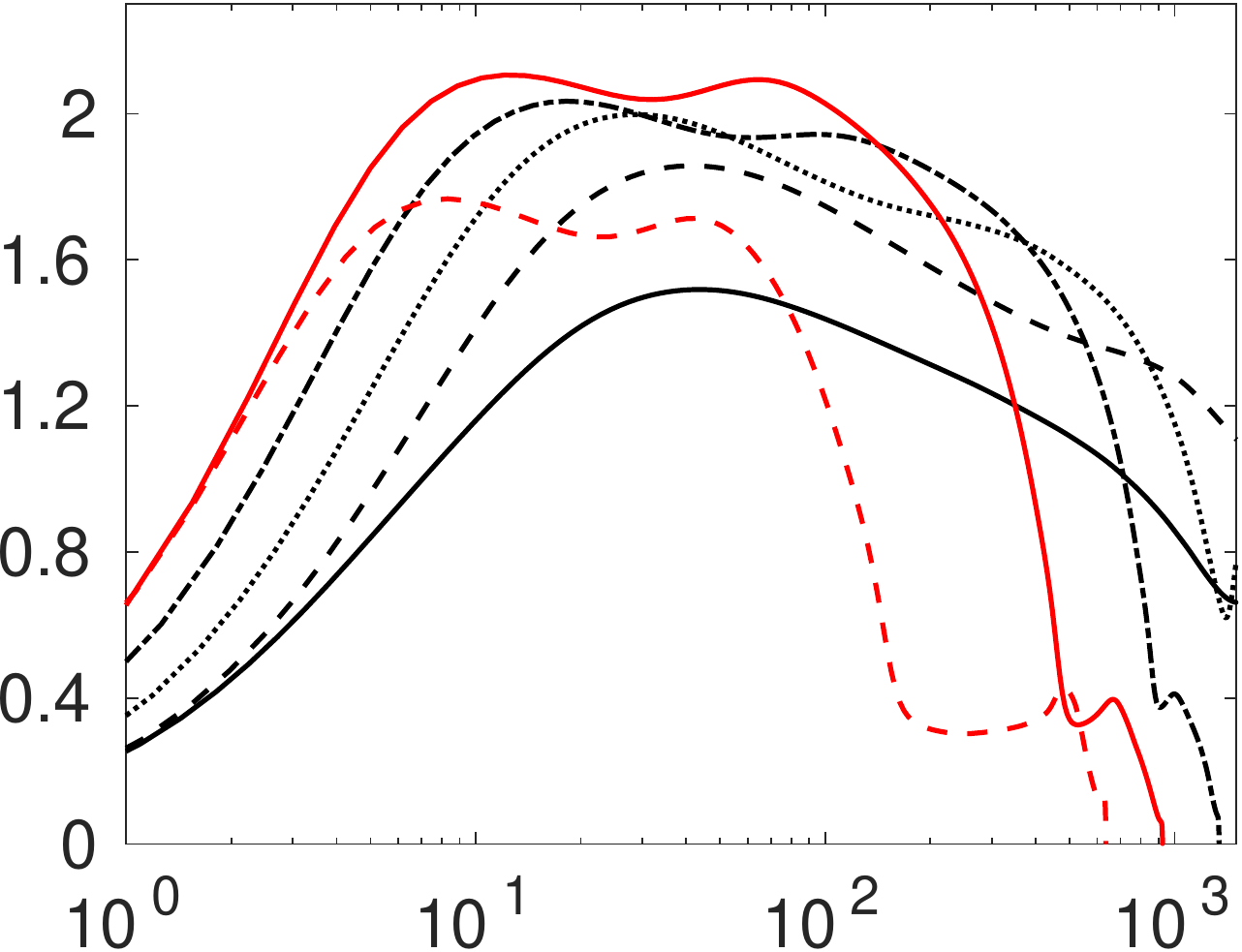}
\put(-0.1,4.3){$\displaystyle (c)$}
\put(-3.1,-0.3){$y^*$}
\put(-6.9,2.3){$w^*$}
\end{center}
\caption{(Colour online) Rms profiles of 
({\it a}) streamwise,
({\it b}) wall-normal and
({\it c}) spanwise
velocity fluctuations in wall units of the unstable side.
$Re=31\,600$ and
($~^{\line(1,0){20}}$) $Ro=0$,
($---$) $Ro=0.45$,
($\cdot\cdot\cdot$) $Ro=0.9$,
($-\cdot-\cdot-$) $Ro=1.2$.
$Re=30\,000$ and
($\textcolor{red}{~^{\line(1,0){20}}}$) $Ro=1.5$,
($\textcolor{red}{---}$) $Ro=2.1$.}
\label{rms_rot}
\end{figure}

To investigate in more detail the turbulence near the wall
on the unstable channel side, I present in figure \ref{rms_rot}
rms-profiles of the streamwise, wall-normal and spanwise
velocity fluctuations, $u^*$, $v^*$, $w^*$ respectively, 
in viscous wall units of
the unstable side using a logscale for $y^*$.
Velocity fluctuations are thus scaled by $u_{\tau u}$ and
$y^* = (y+1) u_{\tau u}/\nu$ since $u_{\tau u}$ appears
to be the most relevant quantity very close to the wall. Note that $u_{\tau u}$
can deviate quite significantly from $u_\tau$, see table \ref{sim_par}.
The peak of $u^*$ on the unstable side
declines and moves towards smaller $y^*$ with $Ro$
whereas the peaks of $v^*$ and $w^*$ grow with $Ro$ until $Ro=1.5$
and then decline. 
This reduction and growth respectively are
caused by an energy redistribution from streamwise to wall-normal
fluctuations by the Coriolis term and pressure-strain correlations
in the Reynolds stress equations, as will be shown later.
The peak of $w^*$ moves towards the wall
with $Ro$ whereas that of $v^*$ is found far away from the wall
at $Ro=0.45$ and comes closer to the wall with increasing $Ro$.
The profile of $w^*$ has two peaks for $Ro \geq 1.2$ which is accompanied
by a double peak in the spectra as shown later.

\begin{figure}
\begin{center}
\setlength{\unitlength}{1cm}
\includegraphics[width=62mm]{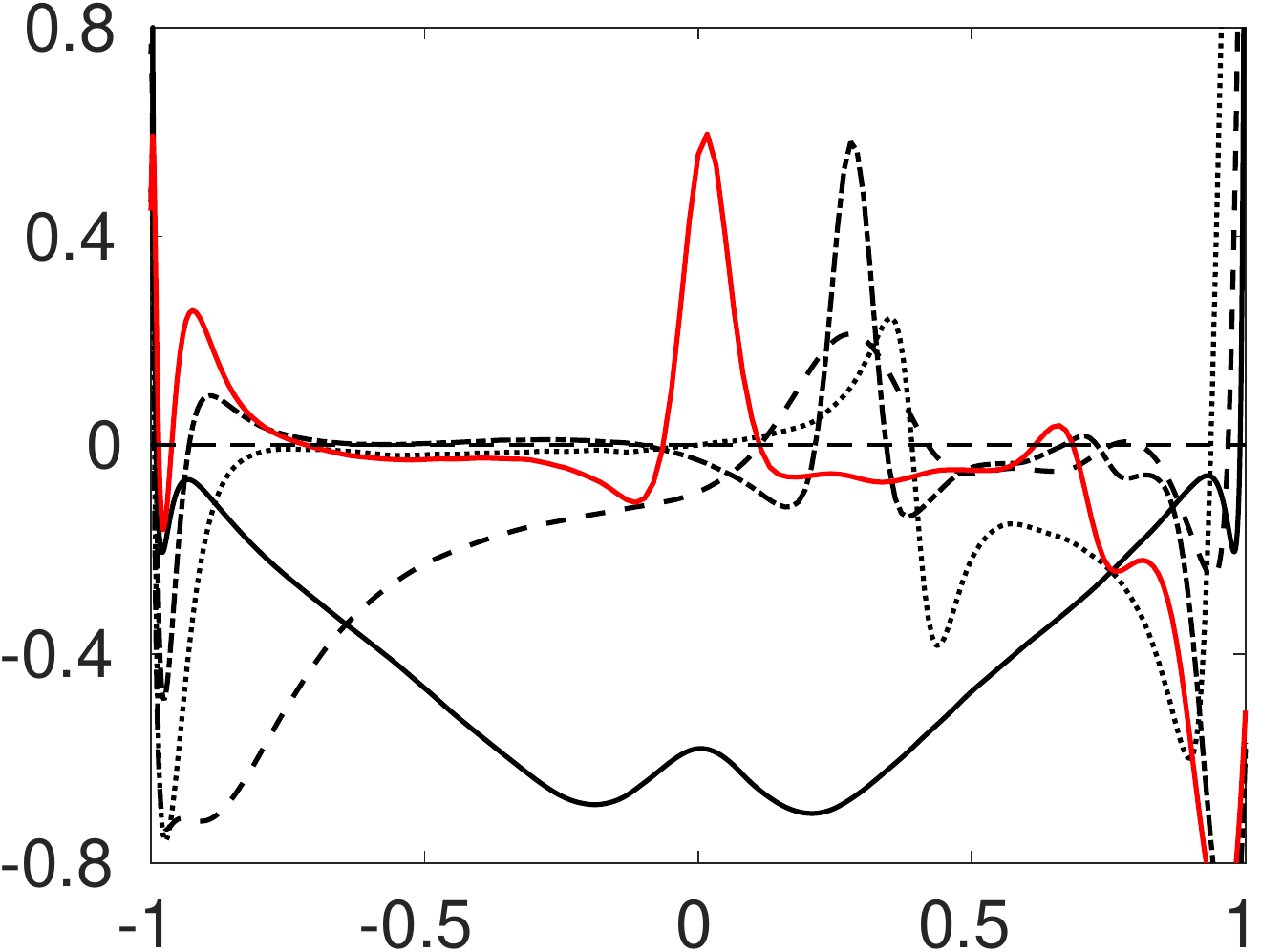}
\put(-0.1,4.3){$\displaystyle (a)$}
\put(-3.1,-0.3){$y$}
\put(-6.9,2.3){$S(u)$}
\hskip9mm
\includegraphics[width=62mm]{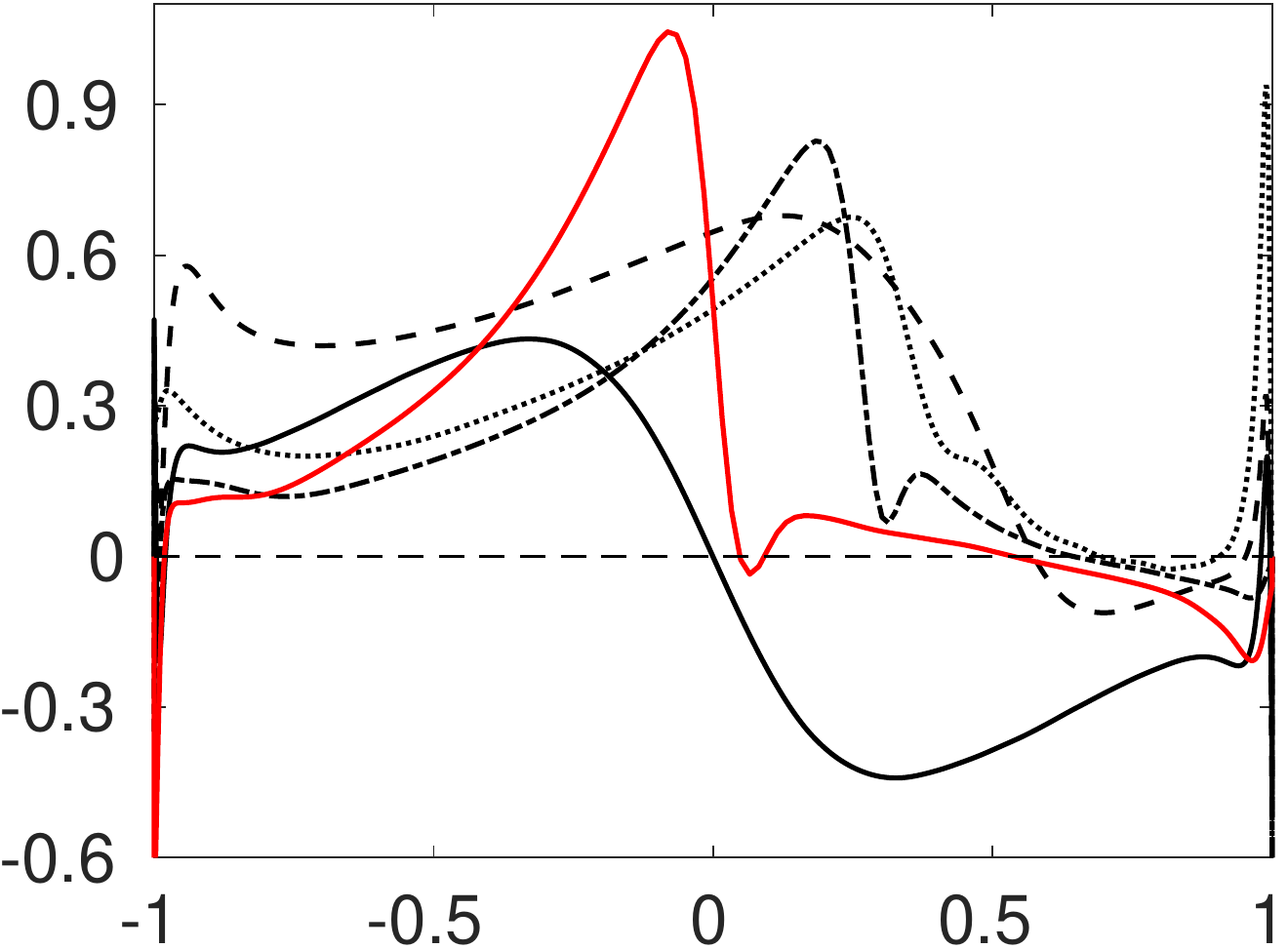}
\put(-0.2,4.2){$\displaystyle (b)$}
\put(-3.1,-0.3){$y$}
\put(-7.0,2.3){$S(v)$}
\end{center}
\caption{(Colour online) Skewness of
({\it a}) $u$ and 
({\it b}) $v$.
$Re=31\,600$ and
($~^{\line(1,0){20}}$) $Ro=0$,
($---$) $Ro=0.45$,
($\cdot\cdot\cdot$) $Ro=0.9$,
($-\cdot-\cdot-$) $Ro=1.2$.
$Re=30\,000$ and
($\textcolor{red}{~^{\line(1,0){20}}}$) $Ro=1.5$.}
\label{skewflat}
\end{figure}

The skewness of the streamwise velocity $S(u)$ and wall-normal velocity
fluctuations $S(v)$ are presented in
figure \ref{skewflat}. 
Profiles of $S(u)$ and $S(v)$ at lower $Re$ are presented by Hsieh \& Biringen (2016).
$S(u)$ is typically negative away from the wall
in non-rotating channel flow owing to ejections of low speed fluid (Kim \etal 1987).
In rotating channel flow $S(u)$ becomes considerably more negative near the wall
at $Ro=0.45$ and 0.9 on the unstable side, except very near the wall.
This could be caused by a reduction of sweeping events of high-speed flow towards
the wall as a consequence of rotation (Kristoffersen \& Andersson 1993).
On the other hand, the analysis by Dai \etal (2016)
indicates that under the influence of rotation the streaks become stronger
on the unstable side, at least up to moderate $Ro$, which implies more or
intenser ejections. The roll cells, observed before, could also play a role.
If $Ro >0.9$, $S(u)$ becomes less negative near the wall, which can be related
to the weaker streaks and related ejections at high $Ro$, as suggested by
Lamballais \etal (1998).
Further away from the wall $S(u)$
attains small values in the rotating cases compared to the
non-rotating case indicating that sweeping and
ejection events are significantly altered by rotation. 
$S(v)$ has a large positive value near the wall on the unstable side
at $Ro=0.45$ compared to
$Ro=0$ which is likely caused by roll cells that induce high-speed
wall-normal velocity away from the wall. 
Similar behaviour of $S(v)$ was observed in the experiments by Nakabayashi \& Kitoh (2015).
Large positive values of $S(v)$, 
indicating events with large positive wall-normal velocities,
are also found in all rotating cases on the unstable side
quite close to the 
position where the slope of $U$ begins to deviate from $2\Omega$.
At $Ro=1.2$ and 1.5, $S(u)$ has large values 
at slightly larger $y$ near the position where $U$ has its maximum value.
The reason for the large values of $S(u)$ and $S(v)$ around these positions
is not fully clear, but visualizations suggest that large values of $v$
are found near the streamwise vortices visualized in figure \ref{vis_lam}.
The large $S(u)$ could be related to the hairpin vortices
seen in figure \ref{vis_lam} since positive values of $u$ 
are found above the head of such hairpin vortices (Christensen \& Adrian 2001, Adrian 2007).

The flatness of the wall-normal velocity $F(v)$ 
attains extreme values in non-rotating channel flow very near the wall 
as a result of intense near-wall vortices (Lenaers \etal 2012). 
The present DNSs show that with faster rotation
$F(v)$ on the unstable channel side
monotonically decays with $Ro$
suggesting that intense near-wall vortices are suppressed.

\section{Influence of the Reynolds number}
\label{sec_rey}

In this section, the influence of the Reynolds number on rotating channel flow
at a fixed $Ro$ is examined and shown to be significant.
In a later section, this influence on the flow structures found on the unstable channel
side is investigated.

\begin{figure}
\begin{center}
\setlength{\unitlength}{1cm}
\includegraphics[width=62mm]{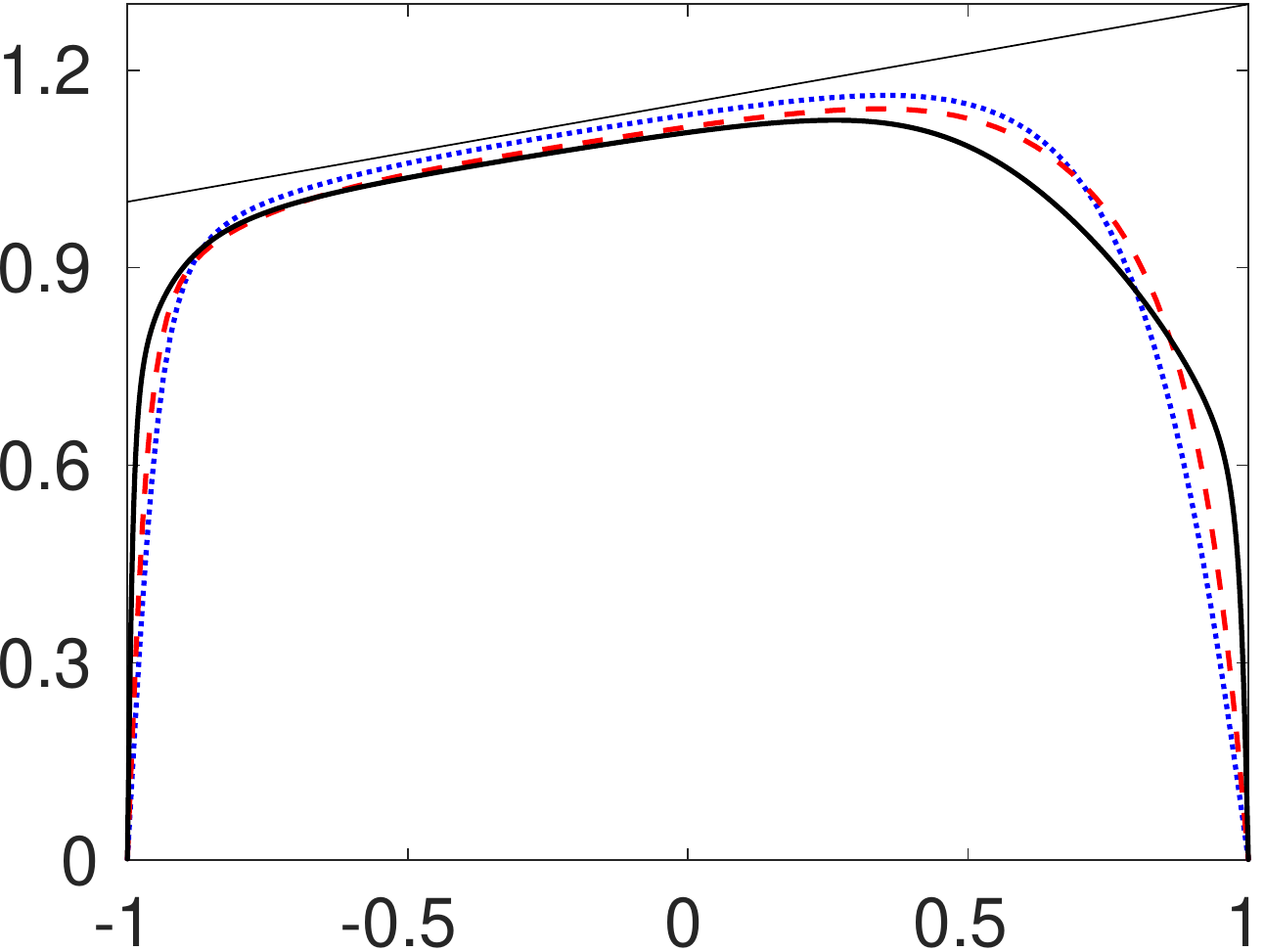}
\put(-0.1,4.3){$\displaystyle (a)$}
\put(-3.1,-0.3){$y$}
\put(-6.9,2.3){$\frac{U}{U_b}$}
\hskip9mm
\includegraphics[width=62mm]{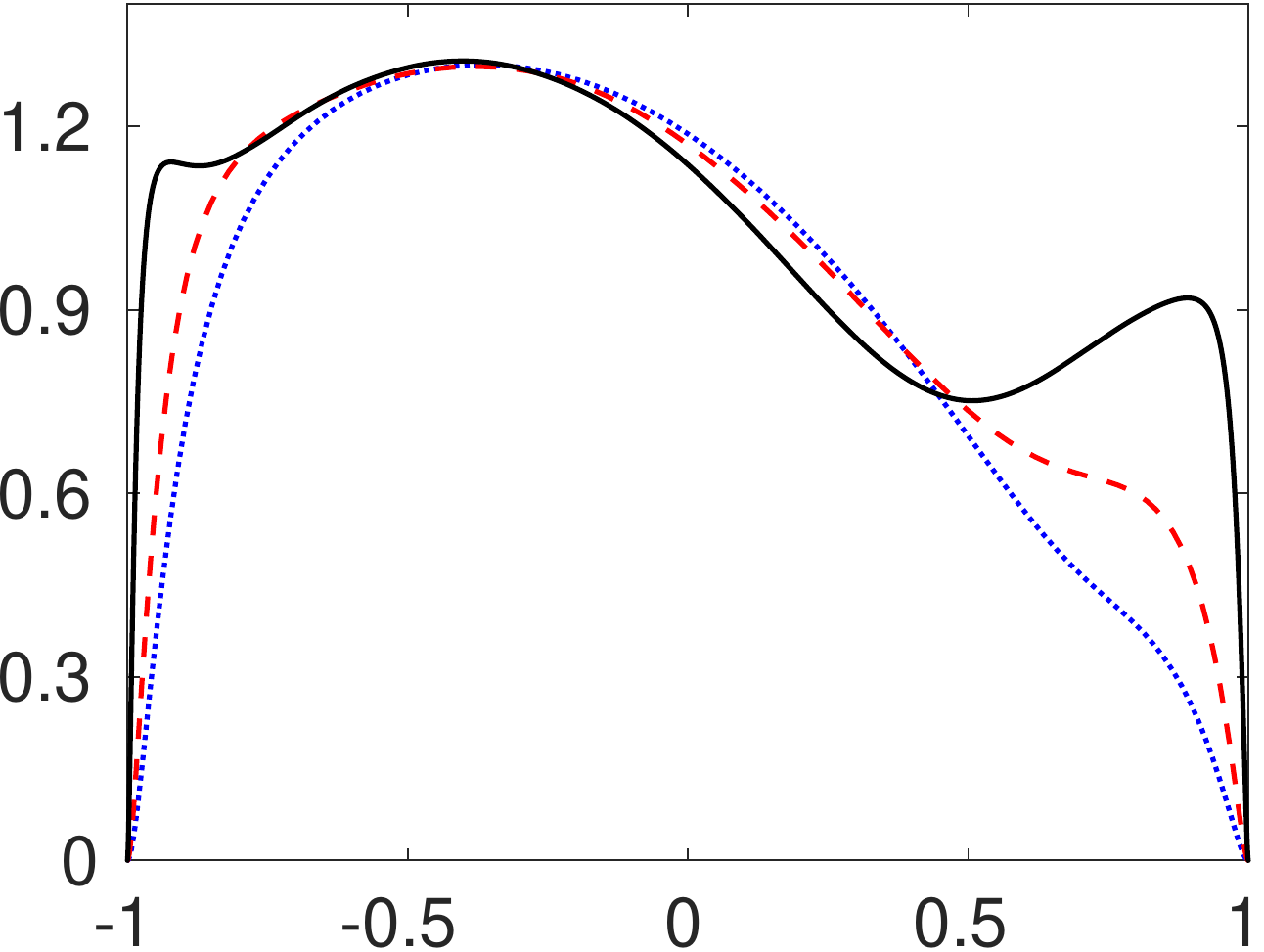}
\put(-0.1,4.3){$\displaystyle (b)$}
\put(-3.1,-0.3){$y$}
\put(-6.9,2.3){$v^+$}

\includegraphics[width=62mm]{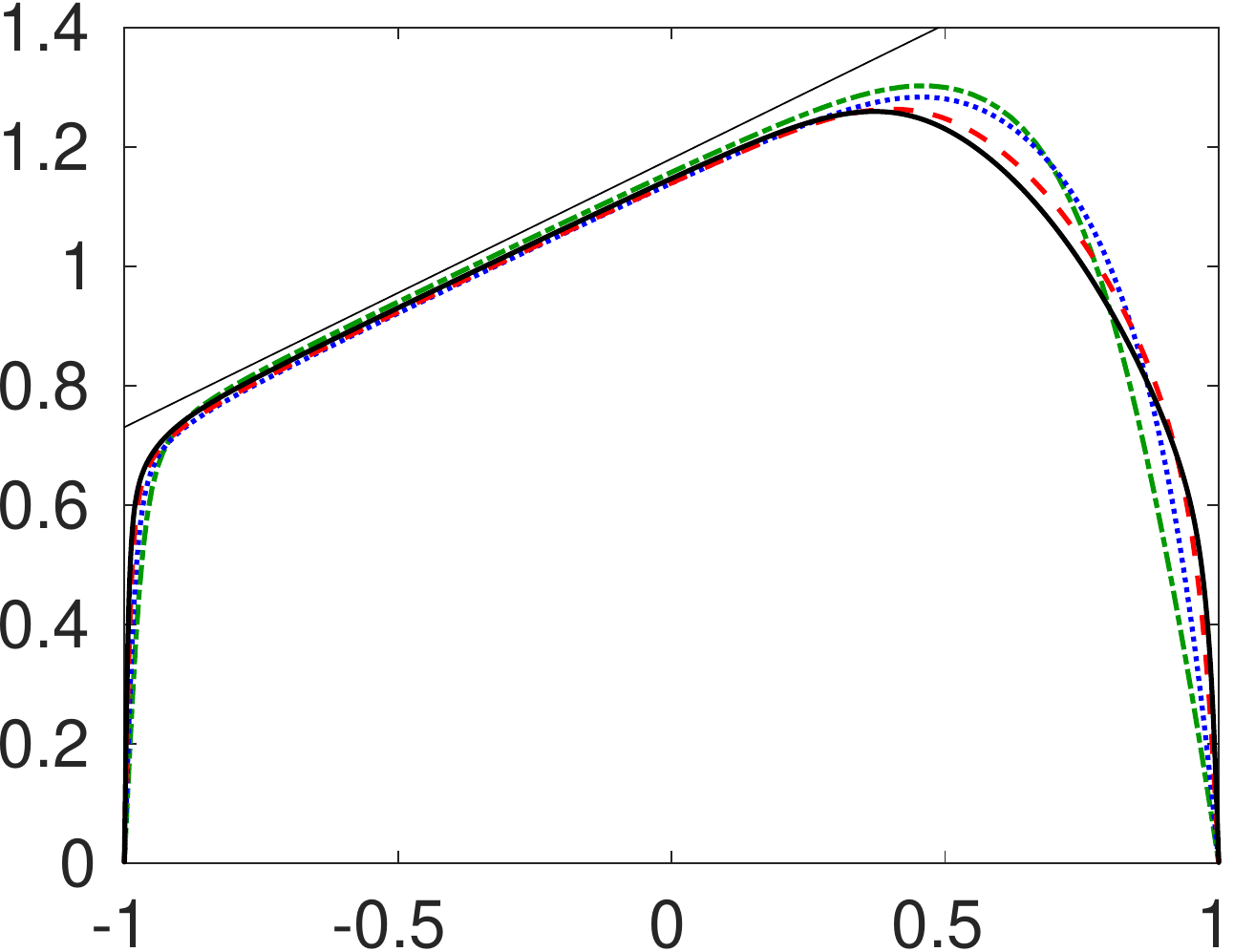}
\put(-0.1,4.3){$\displaystyle (c)$}
\put(-3.1,-0.3){$y$}
\put(-6.9,2.3){$\frac{U}{U_b}$}
\hskip9mm
\includegraphics[width=62mm]{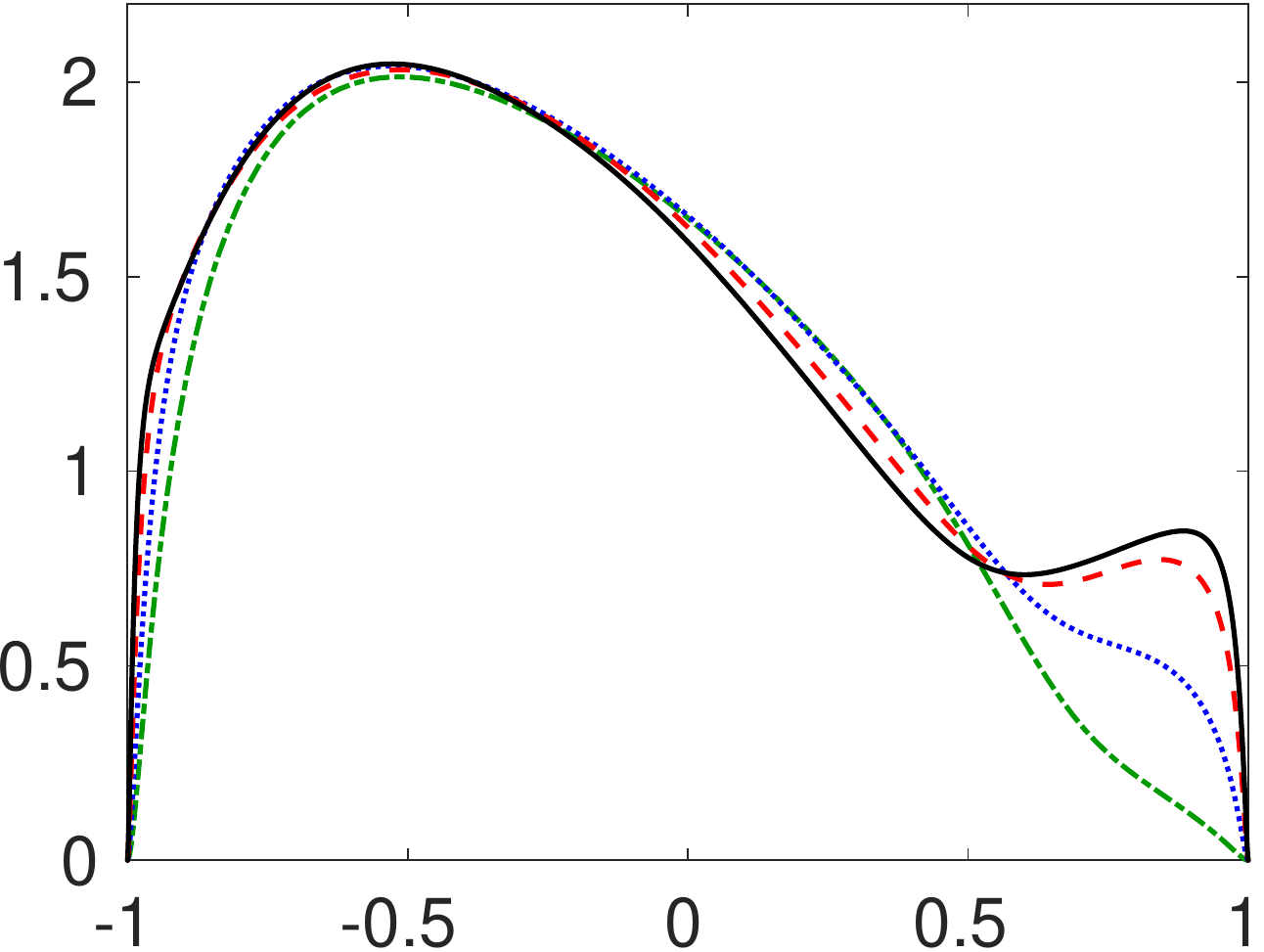}
\put(-0.1,4.3){$\displaystyle (d)$}
\put(-3.1,-0.3){$y$}
\put(-6.9,2.3){$v^+$}

\includegraphics[width=62mm]{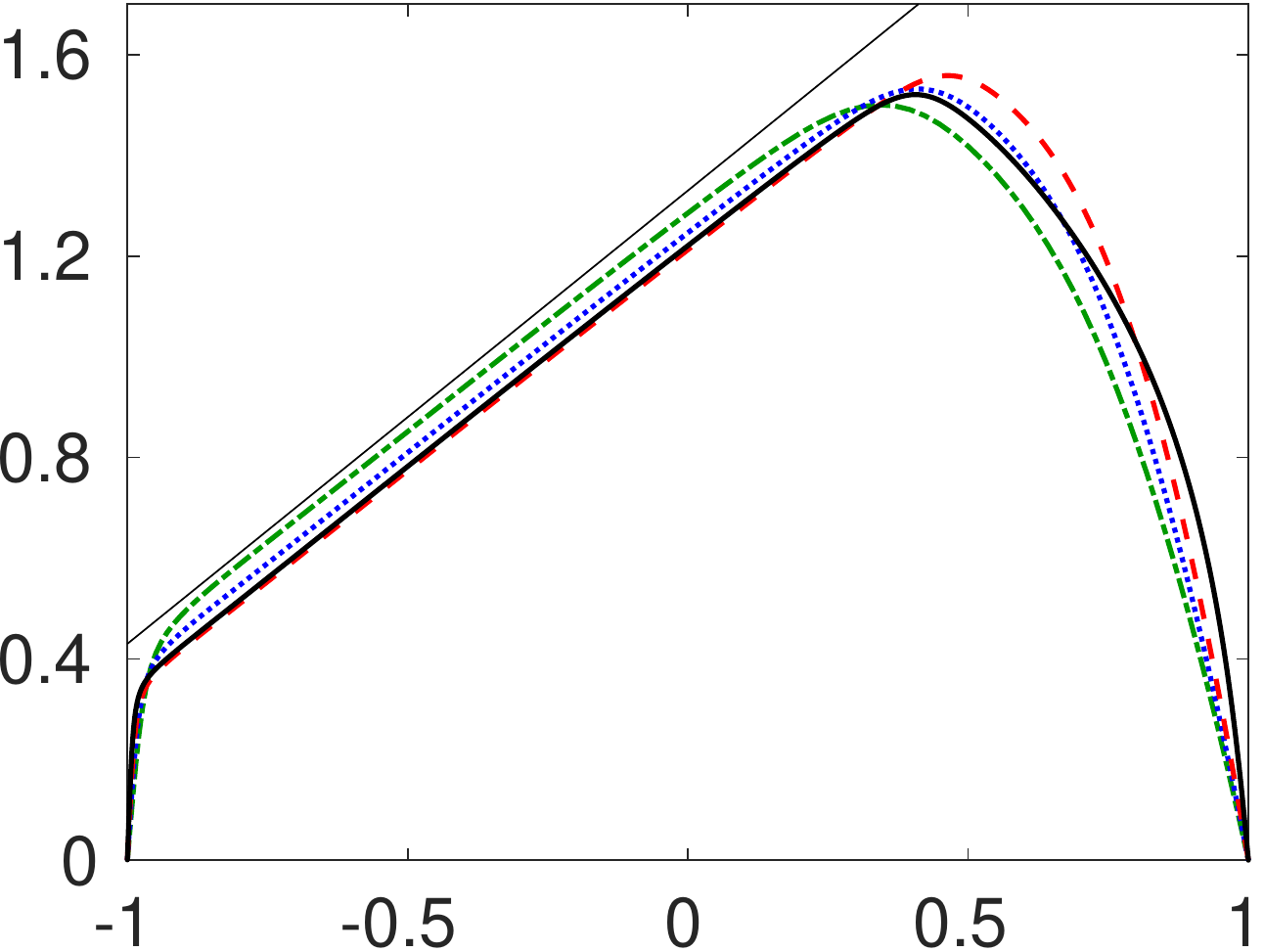}
\put(-0.1,4.3){$\displaystyle (e)$}
\put(-3.1,-0.3){$y$}
\put(-6.9,2.3){$\frac{U}{U_b}$}
\hskip9mm
\includegraphics[width=62mm]{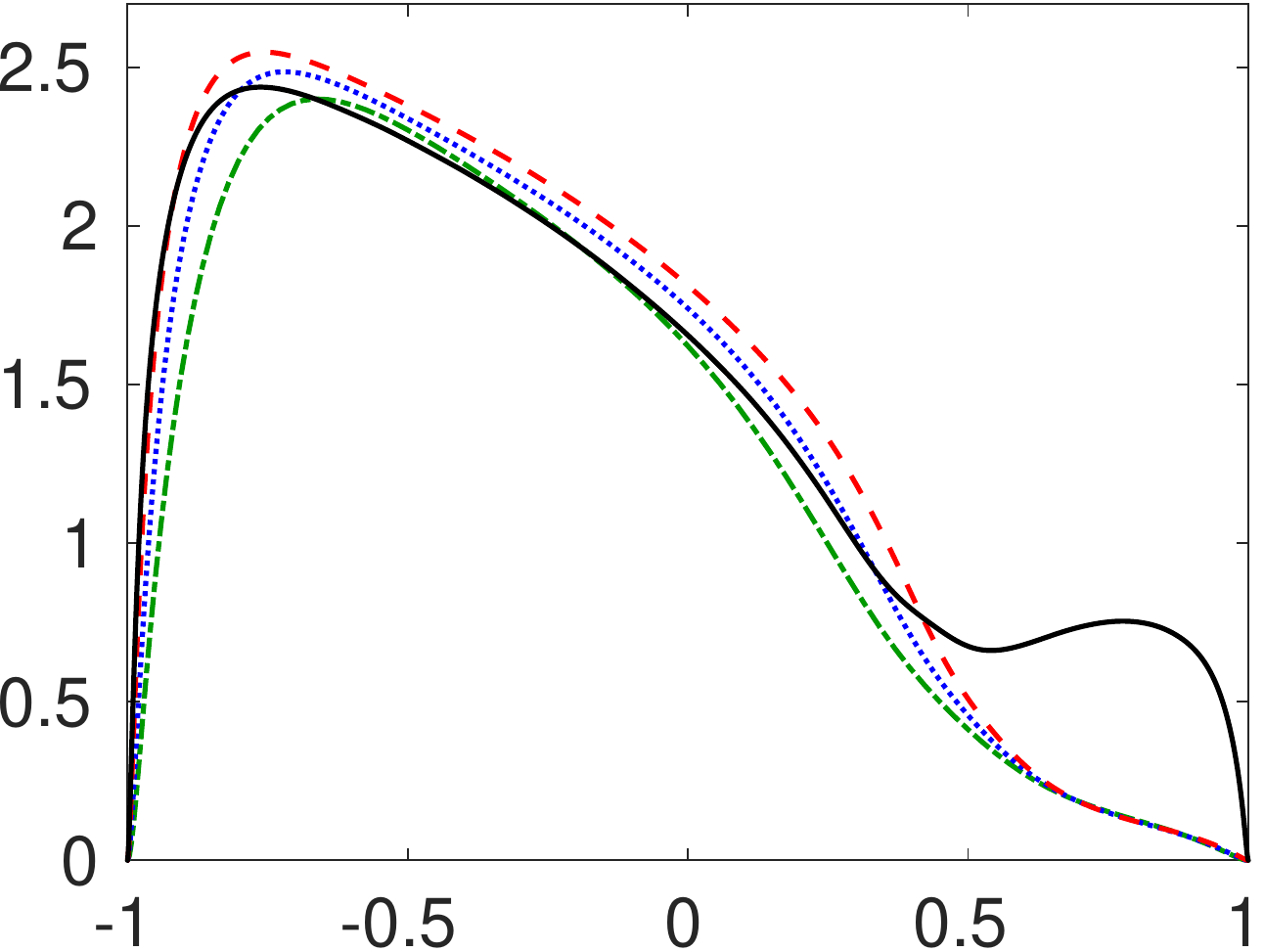}
\put(-0.1,4.3){$\displaystyle (f)$}
\put(-3.1,-0.3){$y$}
\put(-6.9,2.3){$v^+$}
\end{center}
\caption{(Colour online) ({\it a}) Mean velocity profiles scaled by $U_b$ and
({\it b}) rms of $v$ in wall units at $Ro=0.15$ and
$~^{\line(1,0){20}}$, $Re=20\,000$;
$\textcolor{red}{---}$, $Re=5000$;
$\textcolor{blue}{\cdot\cdot\cdot}$, $Re=3000$.
({\it c}) Mean velocity profiles scaled by $U_b$ and
({\it d}) rms of $v$ in wall units at $Ro=0.45$ and
$~^{\line(1,0){20}}$, $Re=31\,600$;
$\textcolor{red}{---}$, $Re=20\,000$;
$\textcolor{blue}{\cdot\cdot\cdot}$, $Re=10\,000$;
$\textcolor{green}{-\cdot-\cdot-}$, $Re=5000$.
({\it e}) and ({\it f}) 
are the same as 
({\it c}) and ({\it d}), respectively, but for $Ro=0.9$.
The straight black line has a slope
$S=-1$.}
\label{re_effect}
\end{figure}

Figure \ref{re_effect} shows profiles of the mean velocity $U/U_b$ and 
fluctuations $v^+$ at $Ro=0.15$, 0.45 and 0.9 for three to four $Re$.
In all cases a region with a linear mean velocity profile where $S \simeq -1$
can be readily recognised demonstrating once more that the appearance of a region
with a zero absolute mean vorticity is a fundamental feature of rotating channel flow.
Its extent and the closeness of 
$\diff U / \diff y$ to $2 \Omega$
do not show much variation 
with $Re$ and 
the same applies to the maximum value of $v^+$ on the unstable
channel side. On the other hand,
$Re$ has an obvious influence on the mean
velocity profile as well as $v^+$ on the stable channel side where at higher $Re$ $v^+$
is considerably larger. 

This Reynolds number effect is also observed in visualizations of the instantaneous
streamwise velocity in a plane parallel and close to the wall on the stable channel side.
\begin{figure}
\begin{center}
\setlength{\unitlength}{1cm}
\includegraphics[width=62mm]{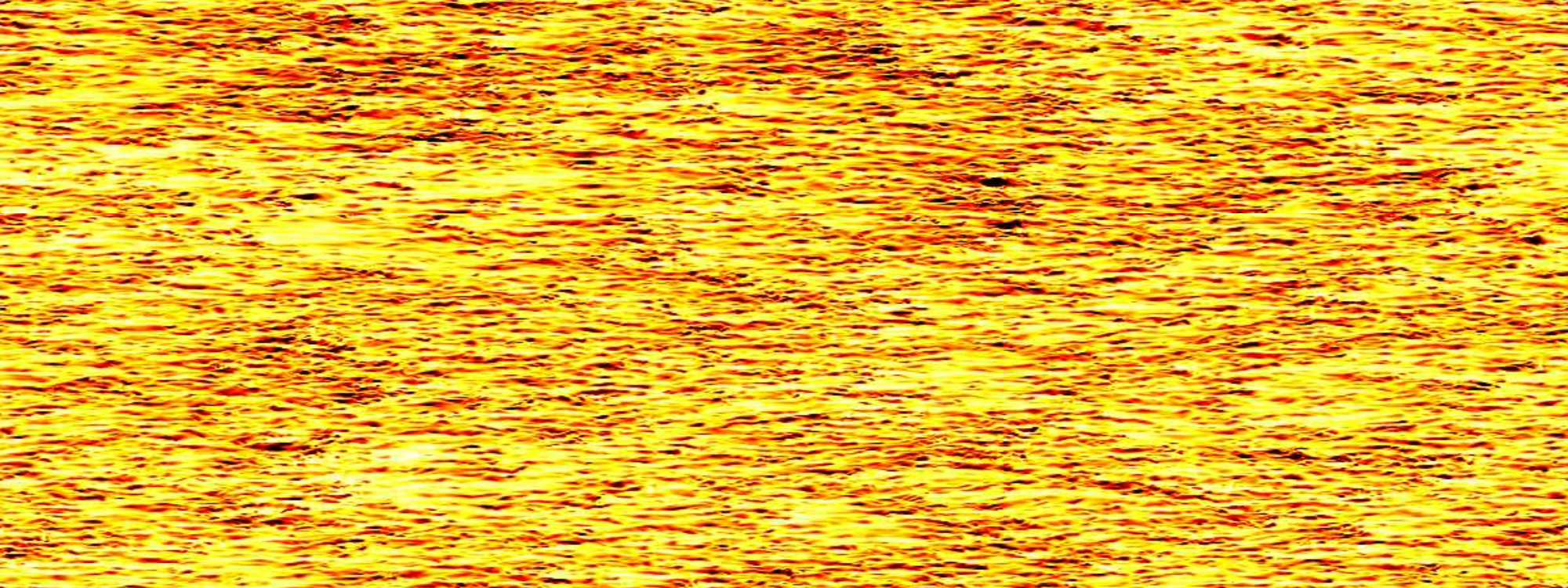}
\put(0.1,2.3){$\displaystyle (a)$}
\vskip2mm
\includegraphics[width=62mm]{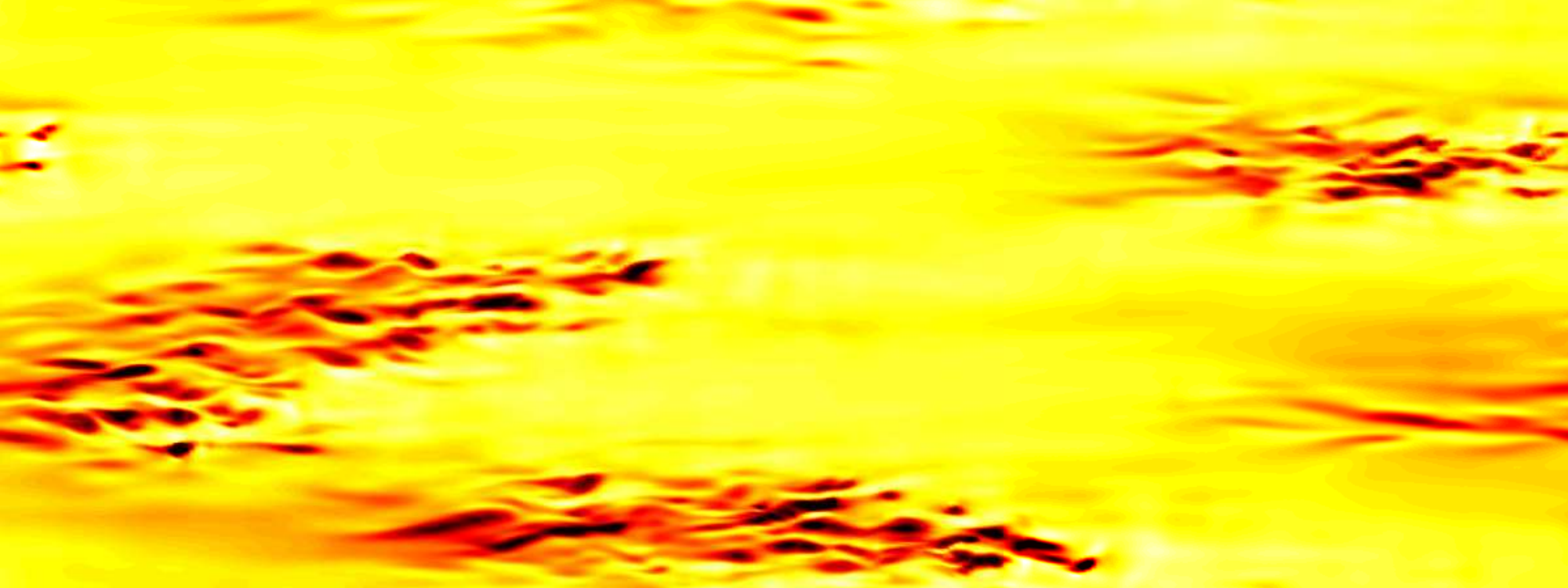}
\put(0.1,2.3){$\displaystyle (b)$}
\vskip2mm
\includegraphics[width=62mm]{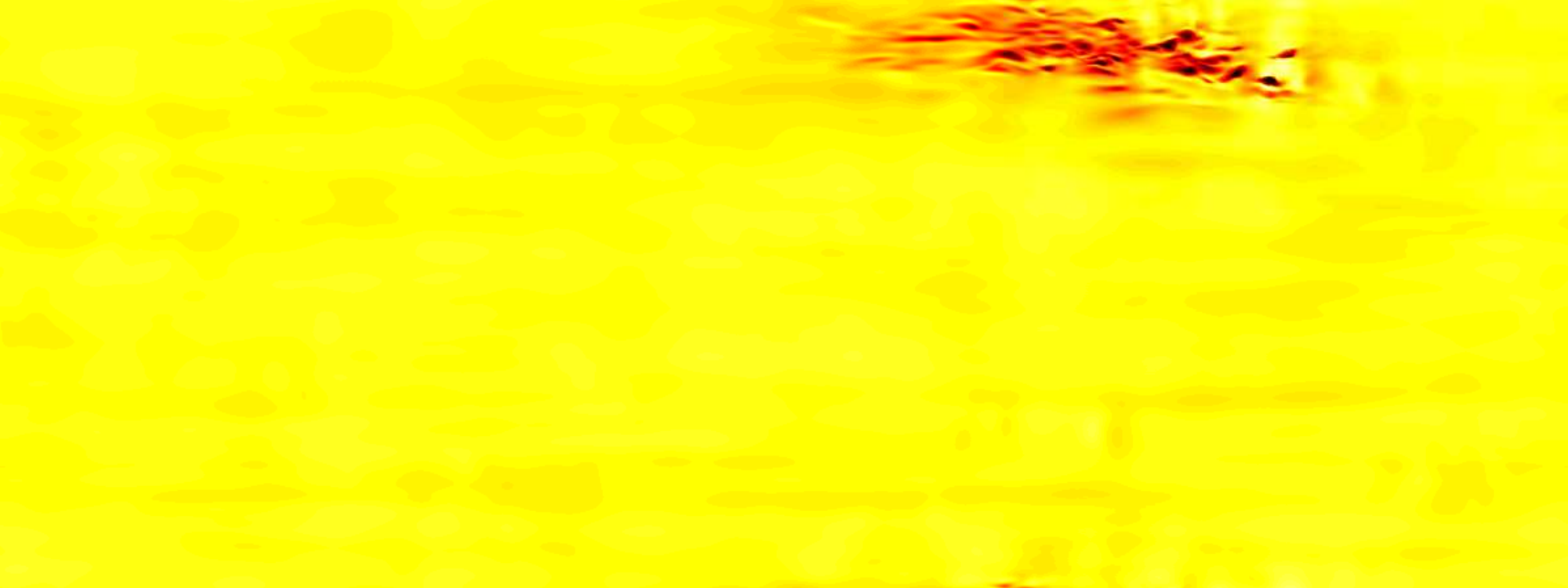}
\put(0.1,2.3){$\displaystyle (c)$}
\end{center}
\caption{(Colour online) Visualizations of the instantaneous streamwise velocity in an $x$-$z$ plane 
at $y^+ \approx 5$ on the stable channel side. 
Dark colours signify high velocities.
({\it a}) $Ro=0.15$ and $Re=20\,000$, 
({\it b}) $Ro=0.15$ and $Re=3000$ and
({\it c}) $Ro=0.45$ and $Re=5000$. 
}
\label{vis_re}
\end{figure}
At $Ro=0.15$ and $Re=20\,000$ the flow is fully turbulent on the stable channel side 
(figure \ref{vis_re}.{\it a})
whereas at a lower $Re=3000$ the flow is largely laminar and contains only some turbulent spots 
(figure \ref{vis_re}.{\it b}) resulting in much lower turbulence levels on the stable
channel side
(figure \ref{re_effect}.{\it b}). 
At $Ro=0.45$ and $Re=31\,600$ a large fraction of the flow on the stable side
is turbulent and turbulent-laminar patterns develop, as shown before in
figure \ref{vis_xz}.({\it b}), 
whereas at the same $Ro$ but lower $Re=5000$ the flow is predominantly laminar and
only one turbulent spot can be observed
(figure \ref{re_effect}.{\it c}). 
At a lower $Re=3000$ the spot is even smaller.
Also at $Ro=0.9$ and $Re=31\,600$ regions with small-scale turbulence 
can be observed
(figure \ref{vis_xz}.{\it c}), but these regions with small-scale turbulence
are completely absent at lower $Re$ (not shown here). 
$Re$ has thus a marked influence on the flow, especially
on the stable channel side, with 
a growing turbulent fraction and stronger turbulence
at fixed $Ro$ when $Re$ gets higher. 
In the present study,
oblique turbulent-laminar patterns have only been observed at higher $Re$.
A speculation is that they also
exist at low $Re$ although at a lower $Ro$ when the stabilizing Coriolis
force is less strong. But at a lower $Re$ they can have a longer wave length, 
as indicated by Brethouwer \etal (2012), implying that very large 
computational domains are required to resolve them.

Profiles of the rms of the velocity fluctuations in terms of wall units of
the unstable side, i.e., velocity fluctuations scaled by $u_{\tau u}$
and $y^*=y u_{\tau u}/\nu$, for different $Ro$ and $Re$ are 
not presented here. However, the peak of the streamwise and spanwise
fluctuations in general increases with $Re$ at fixed $Ro$
whereas the maximum of the wall-normal fluctuations is quite independent
of $Re$ but moves towards larger $y^*$ with $Re$.

\begin{figure}
\begin{center}
\setlength{\unitlength}{1cm}
\includegraphics[width=62mm]{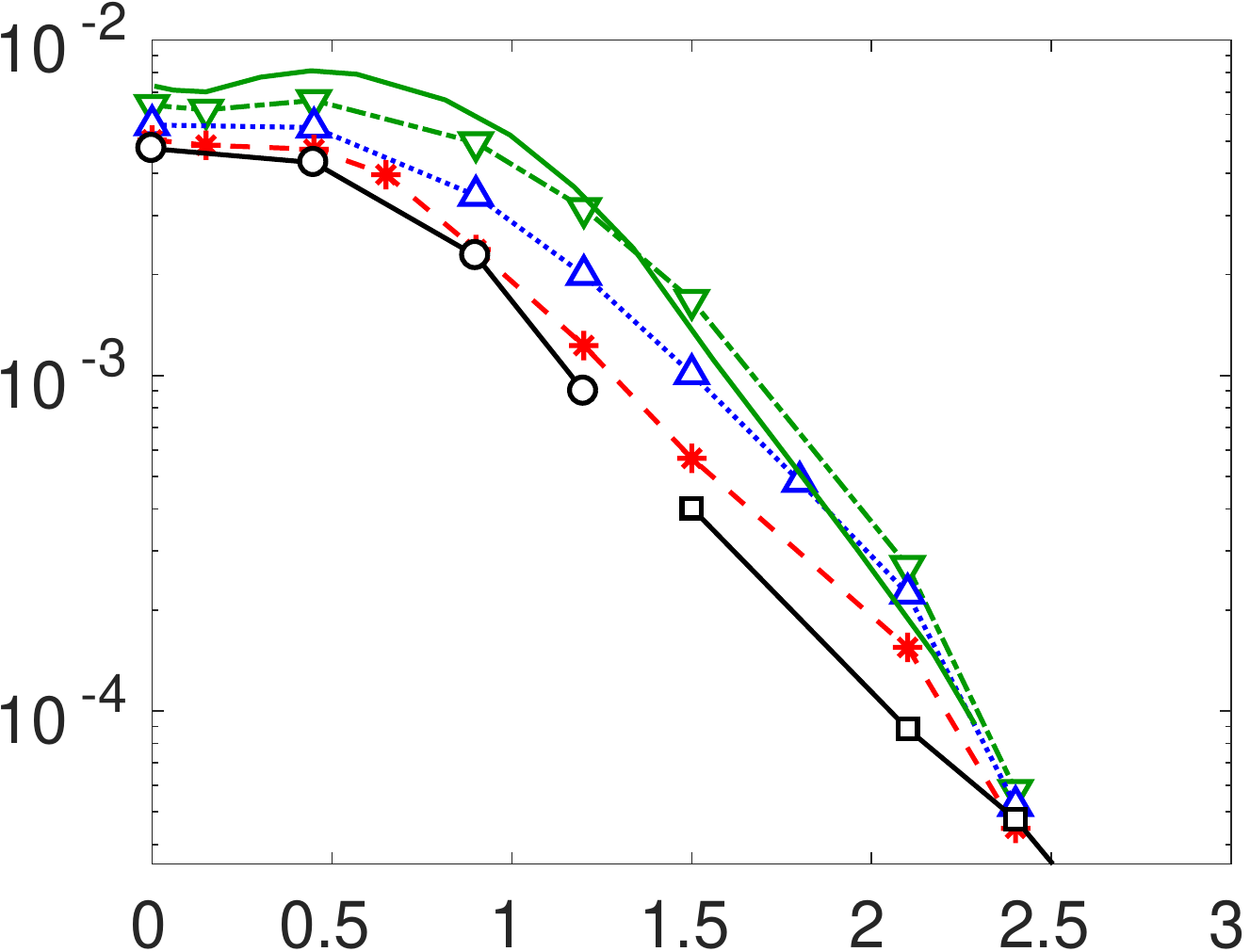}
\put(-0.1,4.3){$\displaystyle (a)$}
\put(-3.1,-0.3){$Ro$}
\put(-6.9,2.3){$\frac{K}{U^2_b}$}
\hskip9mm
\includegraphics[width=62mm]{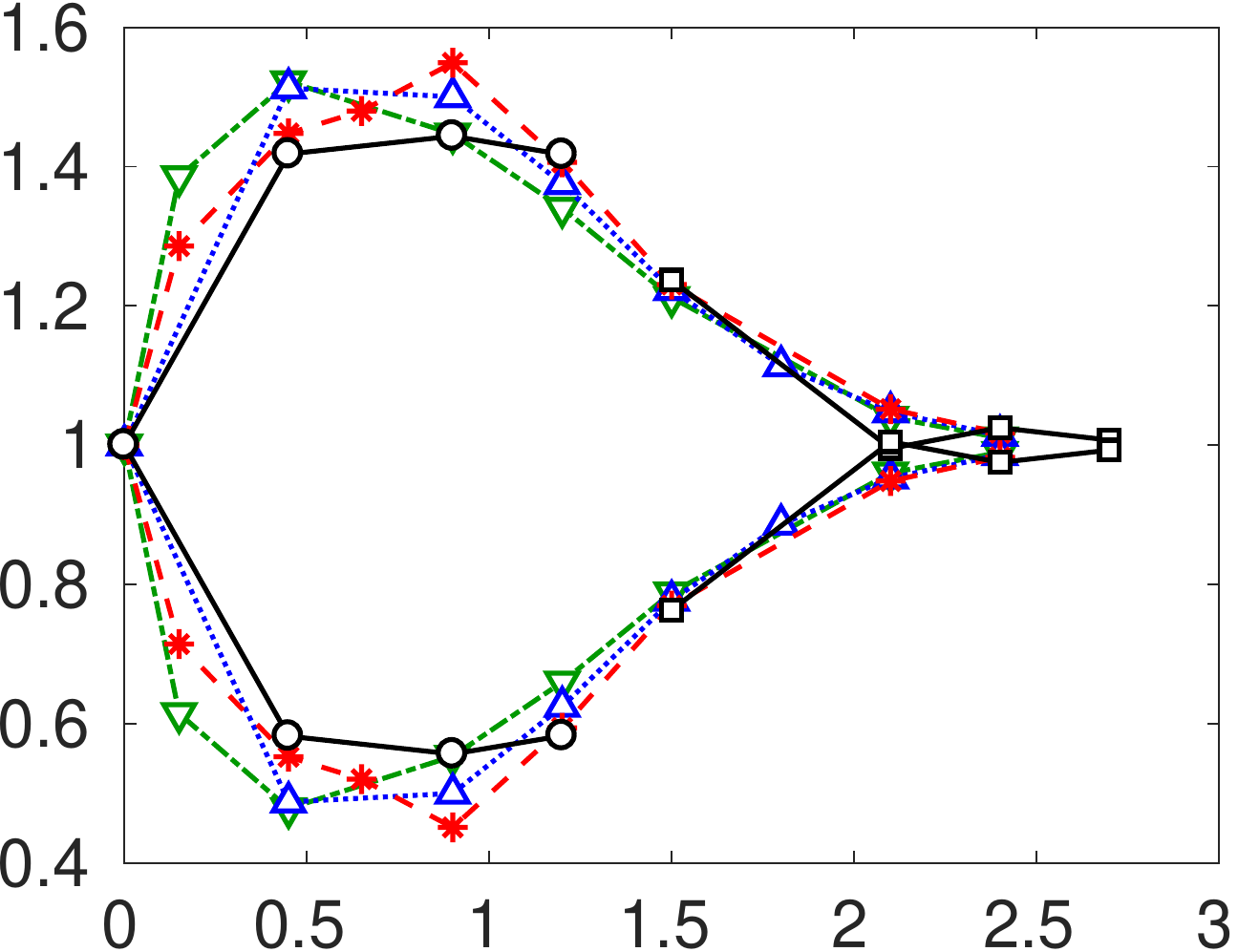}
\put(-0.1,4.3){$\displaystyle (b)$}
\put(-3.1,-0.3){$Ro$}
\put(-6.9,2.3){$\frac{\tau^i_w}{\tau_w}$}

\includegraphics[width=62mm]{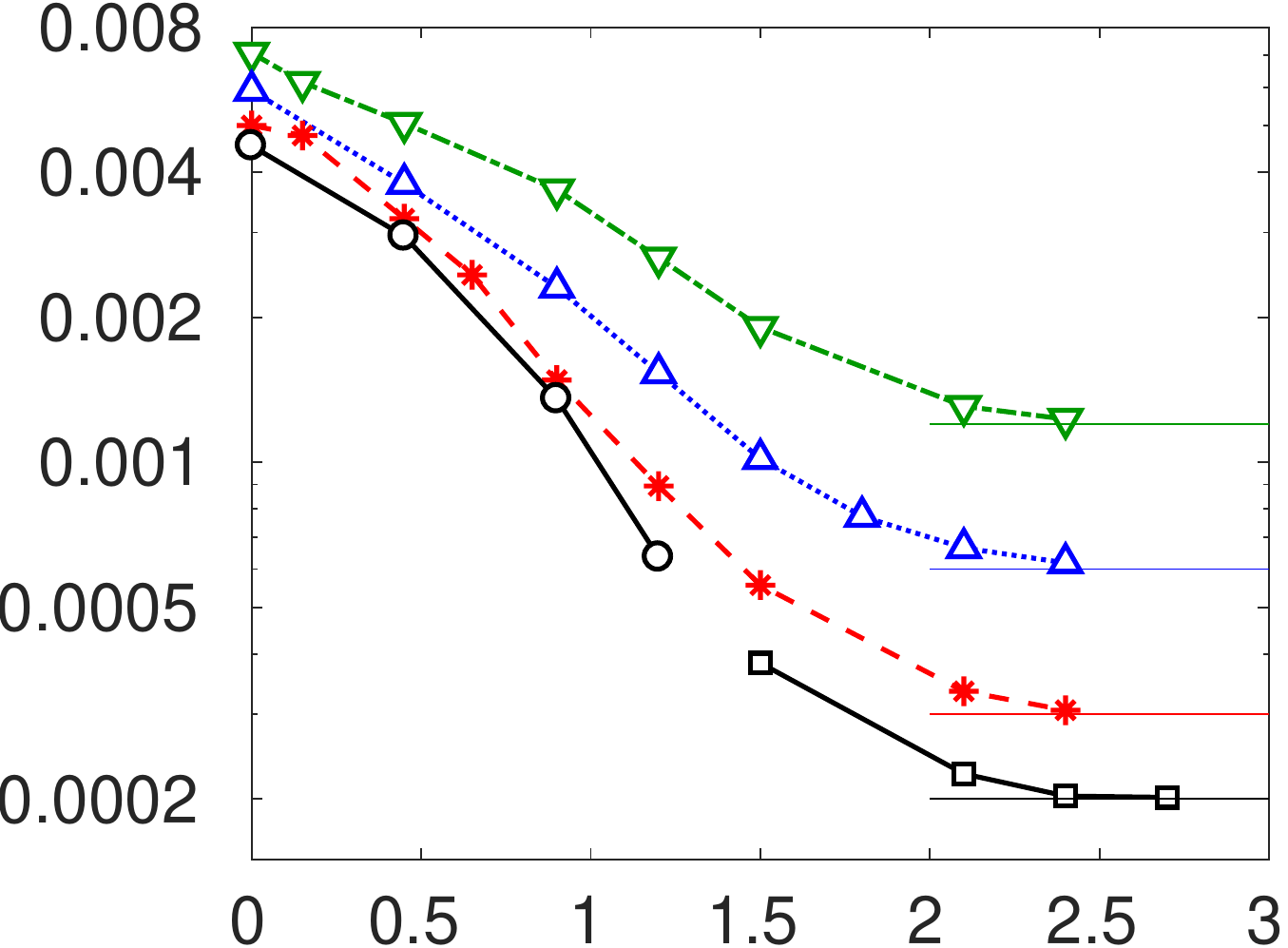}
\put(-0.1,4.3){$\displaystyle (c)$}
\put(-3.1,-0.3){$Ro$}
\put(-6.9,2.3){$C_f$}
\end{center}
\caption{(Colour online) ({\it a}) $K_m/(U^2_b)$,
({\it b}) $\tau^u_w/\tau_w$ (upper curves) and $\tau^s_w/\tau_w$,
({\it c}) $C_f$.
$\circ^{\line(1,0){20}}\circ$, $Re=31\,600$;
$\boxempty^{\line(1,0){20}}\boxempty$, $Re=30\,000$;
$\textcolor{red}{\ast---\ast}$, $Re=20\,000$;
$\textcolor{blue}{\Delta\cdot\cdot\cdot\Delta}$, $Re=10\,000$;
$\textcolor{green}{\nabla -\cdot-\cdot- \nabla}$, $Re=5000$;
$\textcolor{green}{~^{\line(1,0){20}}}$, $Re_\tau=180$ (Xia \etal 2016).
The horizontal line in ({\it c}) represent $C_f$ for a laminar Poiseuille flow.}
\label{skin}
\end{figure}

The volume-averaged turbulent kinetic energy $K_m$ 
scaled by $U^2_b$, mean wall shear stresses 
$\tau^u_w$ and $\tau^s_w$
on the unstable and stable channel sides respectively,
and skin friction coefficient $C_f = \tau_w
/ (\frac{1}{2} \rho U^2_b)$ with $\tau_w = \frac{1}{2}(\tau^u_w + \tau^s_w)$ are shown in figure \ref{skin}
at different $Ro$ and $Re$.
Figure \ref{skin}.({\it a}) shows that overall the turbulence intensity 
decays with $Ro$ at fixed $Re$ 
and becomes very weak at high $Ro$ with similar trends for all $Re$.
The data of Xia \etal (2016) are also included in the figure and
show first a 15\% growth in $K_m/U^2_b$ until $Ro=0.44$
and then a quite similar decay as in the other cases, but 
note that in their DNS series $Re_\tau = 180$ and constant and therefore $Re$ varies.
The difference in $\tau_w$ on the stable and unstable side first
grows substantially with $Ro$ until its maximum found around $Ro=0.9$,
with the highest $\tau_w$ naturally occurring on the unstable side, 
but then diminishes and disappears at high $Ro$
when the flow becomes laminar (figure \ref{skin}.{\it b}).
Again, the trends are similar at different $Re$ but the maximum difference appears at lower
$Ro$ when $Re$ is lower and is smaller at the highest $Re$, possibly because the flow
relaminarizes less fast on the stable side at higher $Re$, as discussed before.
The skin friction $C_f$ decays monotonically with $Ro$, like $K_m$, and
nearly equals the value for laminar Poiseuille flow for $Ro \geq 2.4$ at all $Re$
(figure \ref{skin}.{\it c}).

%
%
%
%

\section{Balances}
\label{sec_bal}

In this section, the balance terms in the transport equations of the
turbulent kinetic energy and Reynolds stresses are studied.
Grundestam \etal (2008) have also presented budgets for rotating channel flow,
although only for one high $Ro=1.5$ when the turbulence is very weak,
while Xia \etal (2016) have only presented production terms.
The present study contributes with a study of several budgets terms at a significantly
higher $Re$ covering a wide $Ro$ range.

From equation (\ref{stress_approx}) 
follows that the production of turbulent kinetic energy in
the part of the channel where $\diff U / \diff y \simeq 2 \Omega$
is approximately (see also Xia \etal 2016)
\begin{equation}
P_K =
-\overline{uv} \frac{\diff U}{\diff y} \simeq
-2 \overline{uv} \Omega = 
-\overline{uv} \frac{U_b}{h} Ro  \simeq
\left [ u^2_{\tau u}- u^2_\tau ( y +1)
 - U^2_b \frac{Ro}{Re}\right ] \frac{U_b}{h} Ro 
\label{turb_prod}
\end{equation}
Consequently,
\begin{equation}
\frac{ h P_K }{u^2_\tau U_b Ro} \simeq
-(y+1)+
\left( \frac{u_{\tau u}}{u_\tau} \right)^2 -
\left( \frac{U_b}{u_\tau} \right)^2 \frac{Ro}{Re}.
\label{pk_norm}
\end{equation}
Thus, the profile of $P_K$ scaled with $u^2_\tau U_b Ro / h
= (u^4_\tau/\nu) (Ro_\tau /Re_\tau)$, 
where $u^4_\tau/\nu$ is the usual
viscous wall unit scaling, is expected to be
approximately linear in $y$ with a slope $-1$ in the part of the
channel where $\diff U / \diff y \simeq 2 \Omega$.

\begin{figure}
\begin{center}
\setlength{\unitlength}{1cm}
\includegraphics[width=62mm]{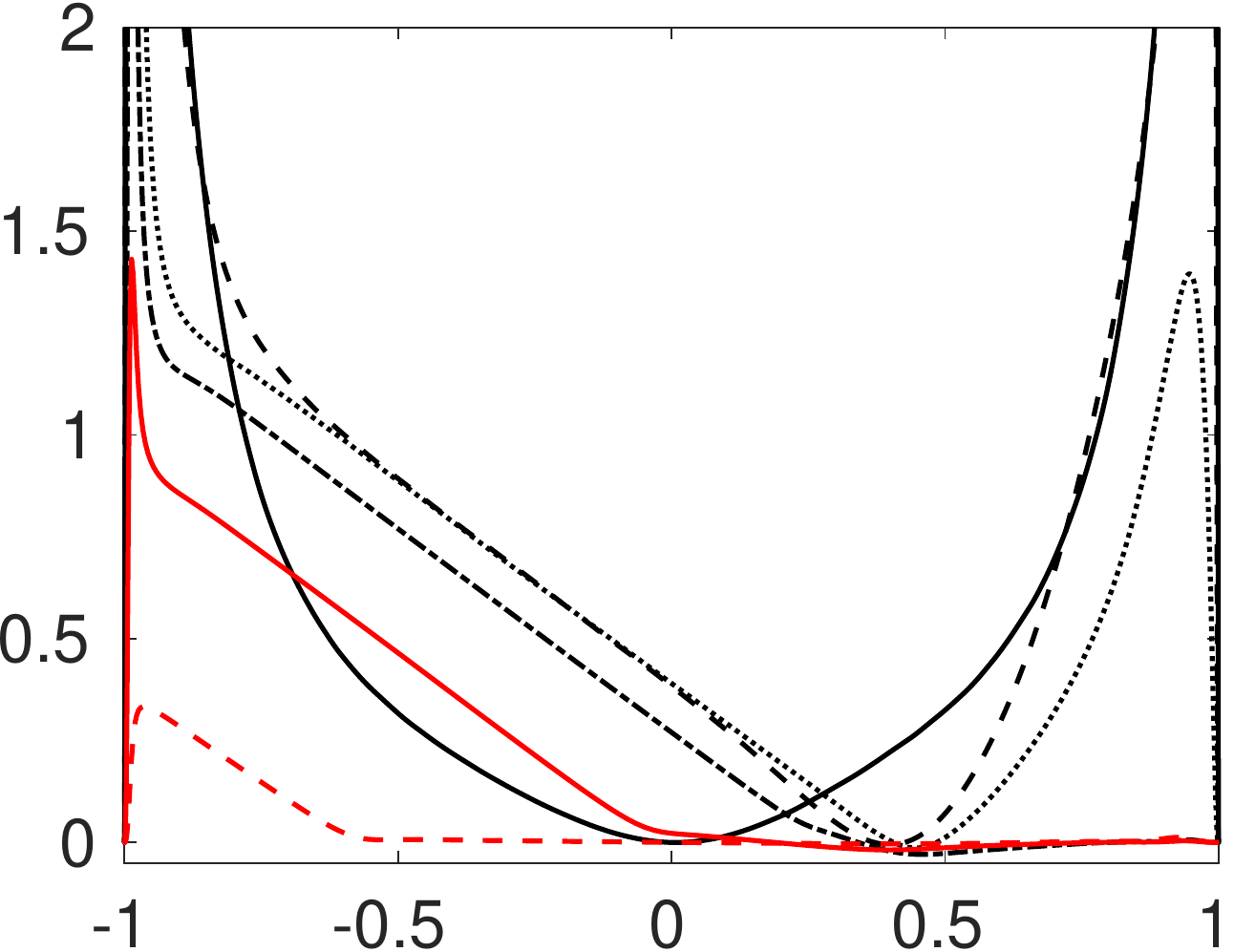}
\put(-0.1,4.3){$\displaystyle (a)$}
\put(-3.1,-0.3){$y$}
\put(-6.9,2.3){$P_K$}
\hskip9mm
\includegraphics[width=62mm]{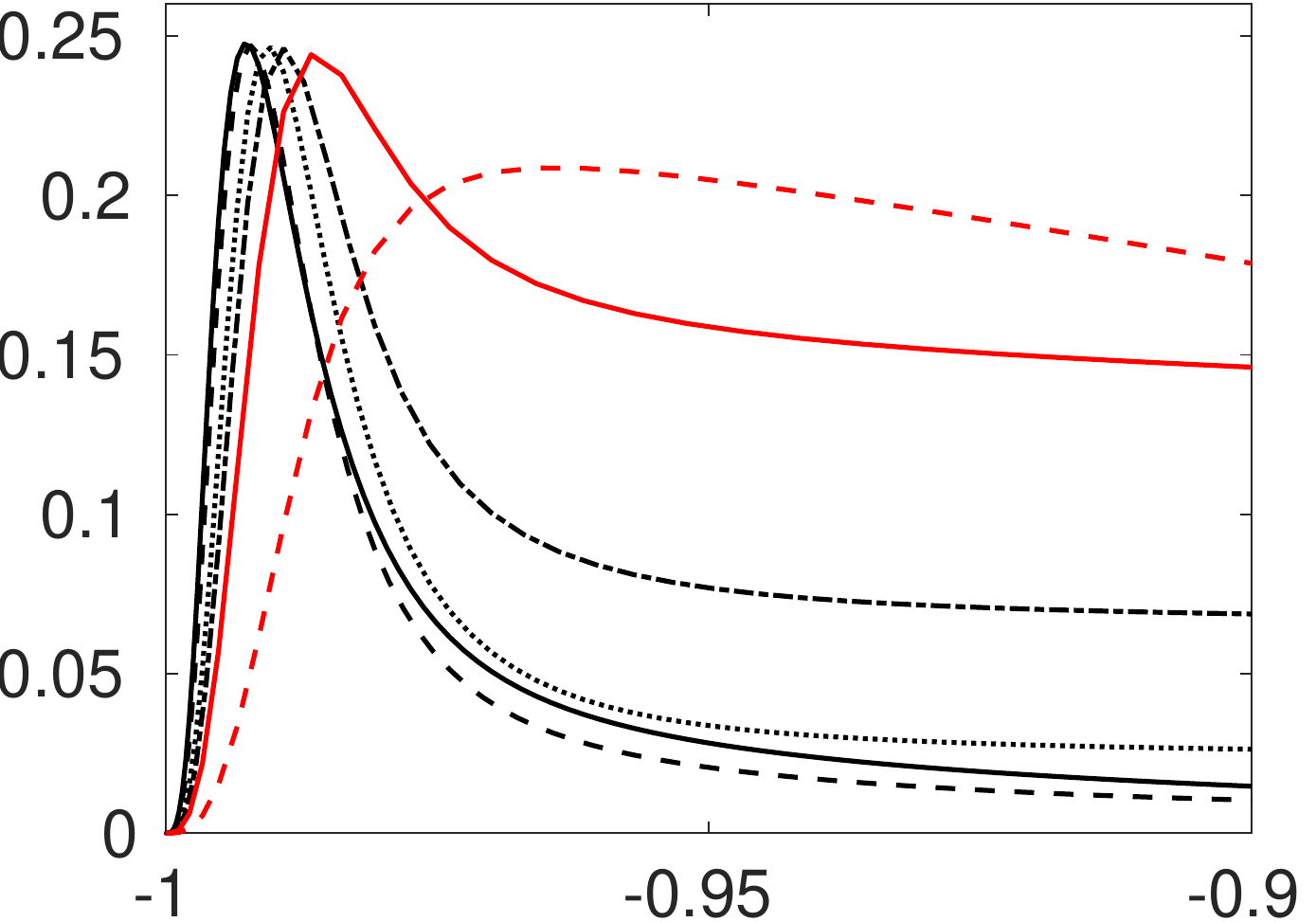}
\put(-0.1,4.3){$\displaystyle (b)$}
\put(-3.1,-0.3){$y$}
\put(-6.9,2.3){$P_K$}

\includegraphics[width=62mm]{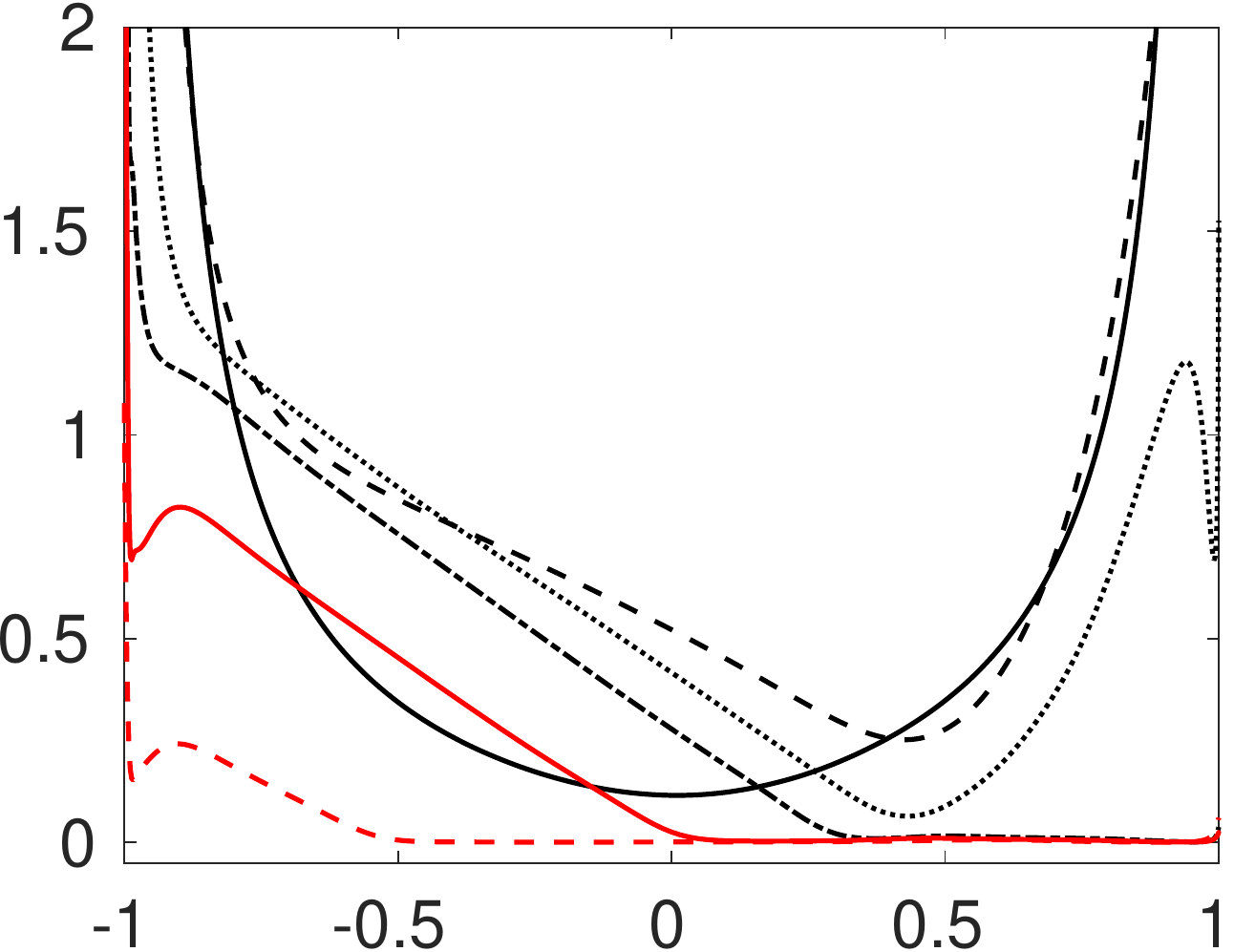}
\put(-0.1,4.3){$\displaystyle (c)$}
\put(-3.1,-0.3){$y$}
\put(-6.9,2.3){$\varepsilon$}
\hskip9mm
\includegraphics[width=62mm]{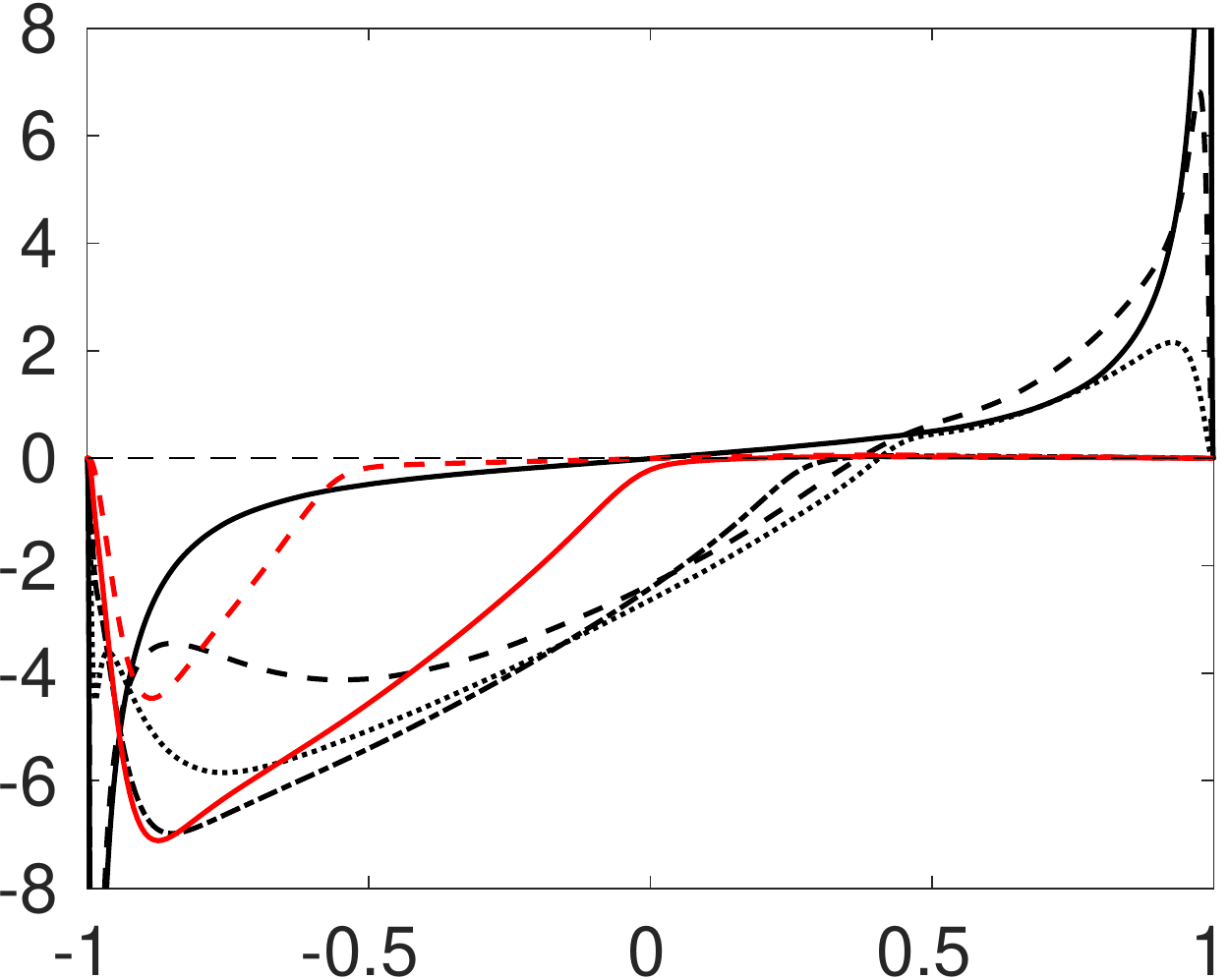}
\put(-0.1,4.3){$\displaystyle (d)$}
\put(-3.1,-0.3){$y$}
\put(-6.9,2.3){$P_{uv}$}

\includegraphics[width=62mm]{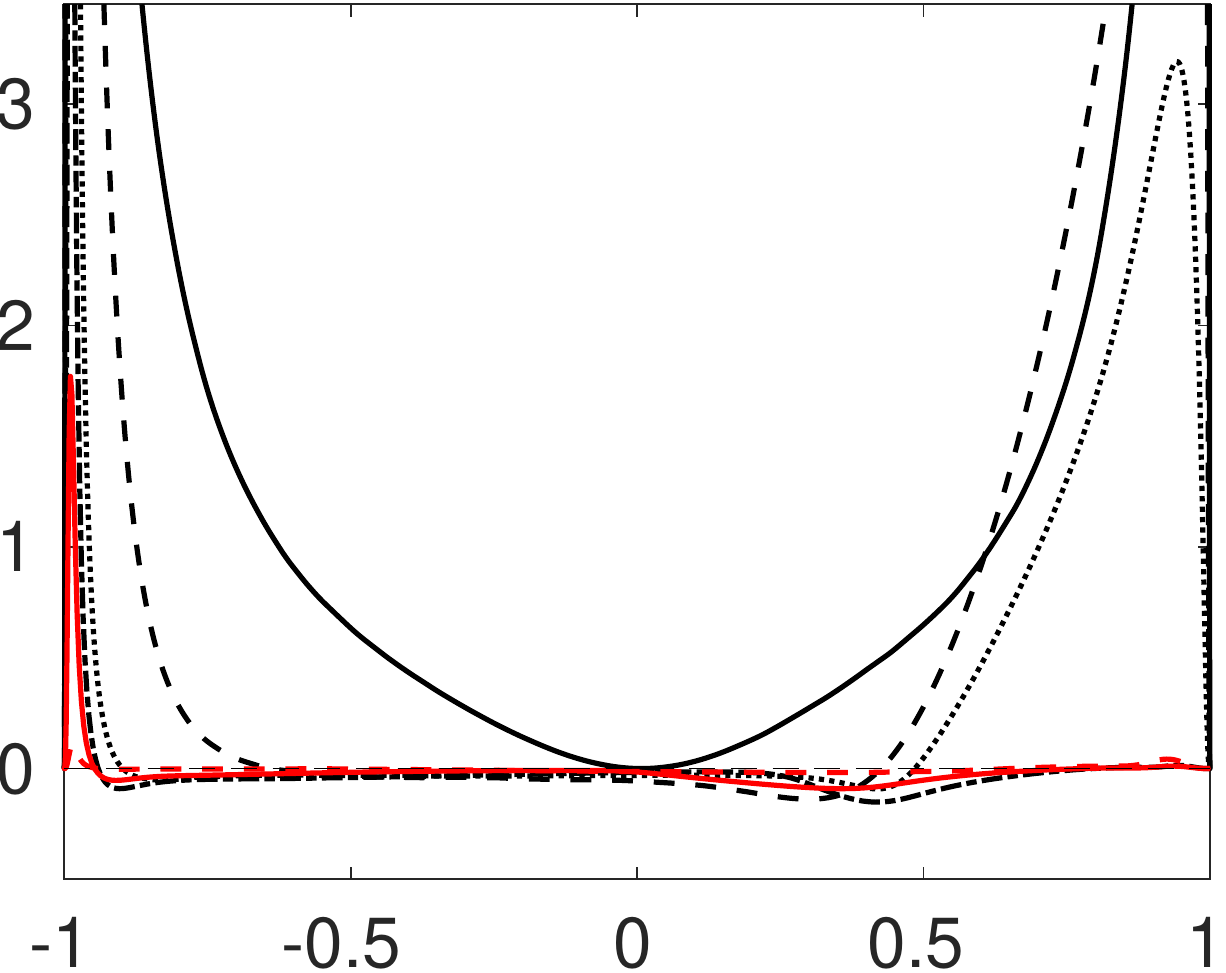}
\put(-0.1,4.3){$\displaystyle (e)$}
\put(-3.1,-0.3){$y$}
\put(-6.9,2.3){$P^{tot}_{uu}$}
\hskip9mm
\includegraphics[width=62mm]{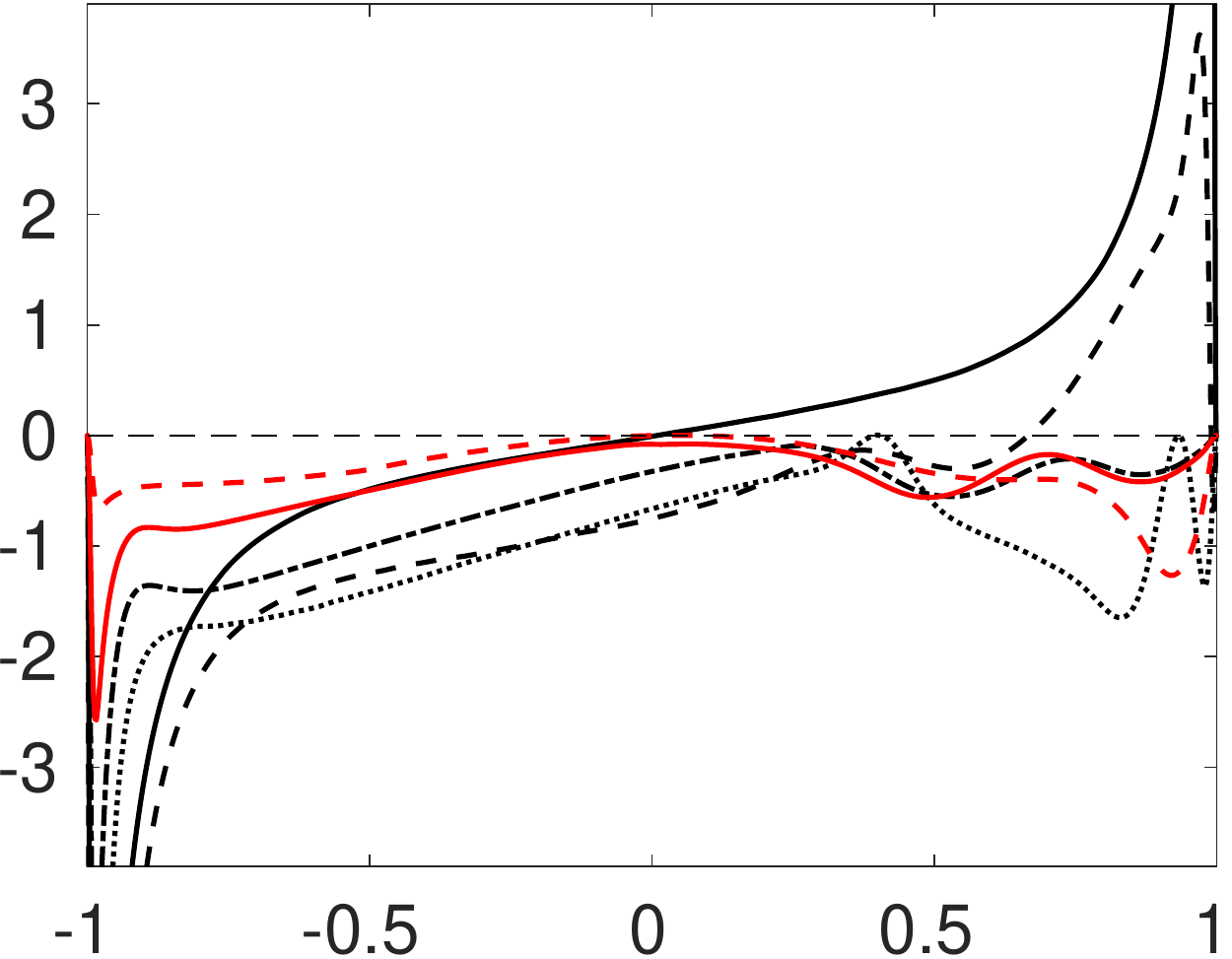}
\put(-0.1,4.3){$\displaystyle (f)$}
\put(-3.1,-0.3){$y$}
\put(-6.9,2.3){$P^{tot}_{uv}$}
\end{center}
\caption{(Colour online)
({\it a}) $P_K$ scaled by $u^2_\tau U_b Ro/h$ and
({\it b}) $P_K$ scaled by $u^4_{\tau u}/\nu$.
({\it c}) $\varepsilon$ and
({\it d}) $P_{uv}$ and
({\it e}) $P^{tot}_{uu} = P_{uu}+C_{uu}$ and
({\it f}) $P^{tot}_{uv} = P_{uv}+C_{uv}$ 
scaled by $u^2_\tau U_b Ro/h$.
$Re=31\,600$ and
($~^{\line(1,0){20}}$) $Ro=0$,
($---$) $Ro=0.45$,
($\cdot\cdot\cdot$) $Ro=0.9$,
($-\cdot-\cdot-$) $Ro=1.2$.
$Re=30\,000$ and
($\textcolor{red}{~^{\line(1,0){20}}}$) $Ro=1.5$,
($\textcolor{red}{---}$) $Ro=2.1$.}
\label{eps_rot}
\end{figure}

Figure \ref{eps_rot}.({\it a}) 
shows $P_K$ scaled by $u^2_\tau U_b Ro/h$ for $Ro$ up to 2.1.
The DNS results for $Ro=0$ are included and scaled by 
$ 0.45 u^2_\tau U_b/h$, i.e., the same scaling as used for the $Ro=0.45$ case.
The profiles of $P_K$ in the rotating cases show as expected a significant 
linear part with a $-1$ slope. In this part, the scaled $P_K$ follows 
closely the prediction given by the right-hand-side of equation (\ref{pk_norm}).
Similar linear profiles of $P_K$ were observed in DNSs of
rotating channel flow at $Re_\tau = 180$ by Xia \etal (2016).
Near the maximum of $U$
slight negative values of $P_K$ are observed 
at some $Ro$, as in a DNS by Grundestam \etal (2008),
which implies that some energy is transferred from turbulence to 
the mean flow in this region.
The maximum value of $P_K$ near the wall on the unstable side
scales instead with viscous walls units since
the peak of $P_K$ lies between 0.244 and 0.248 for $Ro \leq 1.5$ when scaled by
$u^4_{\tau u}/\nu$ (figure \ref{eps_rot}.{\it b}).

In shear flows, the dissipation rate of turbulent kinetic energy, $\varepsilon$,
is often approximately equal to the production of turbulent kinetic energy, $P_K$,
which suggests that $\varepsilon$, like $P_K$, scales with
$u^2_\tau U_b Ro/h$.
Figure \ref{eps_rot}.({\it c}) affirms that in rotating channel flow profiles of
$\varepsilon$, scaled by $u^2_\tau U_b Ro/h$, 
display a linear part.
The DNS results for $Ro=0$ are again included and scaled by 
$ 0.45 u^2_\tau U_b/h$.
The slopes at $Ro=0.45$ and $Ro=2.1$ deviate from $-1$, but
in the other three rotating cases it follows closely 
the prediction given by the right-hand-side of equation (\ref{pk_norm}).
For $Ro\geq 0.9$, $P_K$ and $\varepsilon$ are in fact strikingly similar in the outer layer
on the unstable channel side, much more so than at $Ro=0$.  
The close balance between $P_K$ and $\varepsilon$ implies that the sum
of turbulent, pressure and viscous diffusion
in the equation for turbulent kinetic energy is small compared
to $P_K$ and $\varepsilon$.
In terms of wall units, $P_K$ and $\varepsilon$ are large on the unstable
side away from the wall in the rotating cases compared to the $Ro=0$ case
whereas on the stable side both are very small if $Ro \geq 1.2$.

The observed scaling of $P_K$ and $\varepsilon$ with 
$u^2_\tau U_b Ro / h$ 
suggests that this scaling is meaningful as well for the budget
terms in the balance equation for the Reynolds stresses. This equation reads
\begin{equation}
\frac{\p \overline{u_i u_j}}{\p t} + U_k \frac{\p \overline {u_i u_j}}{\p x_k} =
P_{ij} +\varepsilon_{ij} + C_{ij}+ \Pi_{ij} + D_{ij}, 
\label{rey_eq}
\end{equation}
where the terms on the right-hand-side are the production, dissipation,
Coriolis, pressure-strain and diffusion term respectively (Grundestam \etal 2008).
The Coriolis terms in the equation for $\overline{uu}$, 
$\overline{vv}$, 
$\overline{ww}$ 
and $\overline{uv}$ stresses are
$C_{uu} = 4 \overline{uv}\Omega$,
$C_{vv} = -4 \overline{uv}\Omega$,
$C_{ww} = 0$
and $C_{uv} = 2 (\overline{vv}-\overline{uu})\Omega$, respectively.
These terms do not perform work but transfer energy between the Reynolds stress
components (Kawata \& Alfredsson 2016).
Note that $P_{uu}\,= \,2 P_K$.

\begin{figure}
\begin{center}
\setlength{\unitlength}{1cm}
\includegraphics[width=62mm]{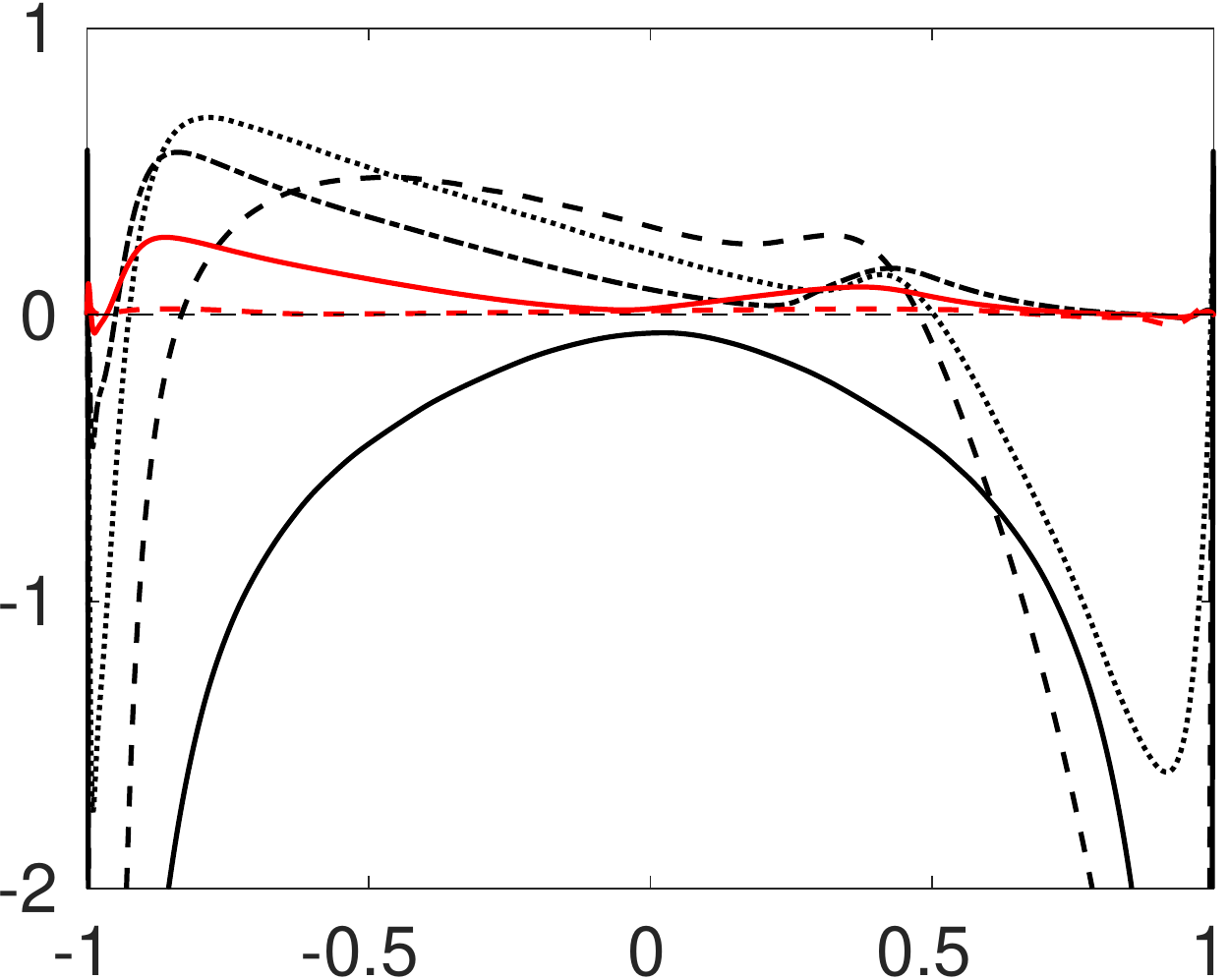}
\put(-0.1,4.3){$\displaystyle (a)$}
\put(-3.1,-0.3){$y$}
\put(-6.9,2.3){$\Pi_{uu}$}
\hskip9mm
\includegraphics[width=62mm]{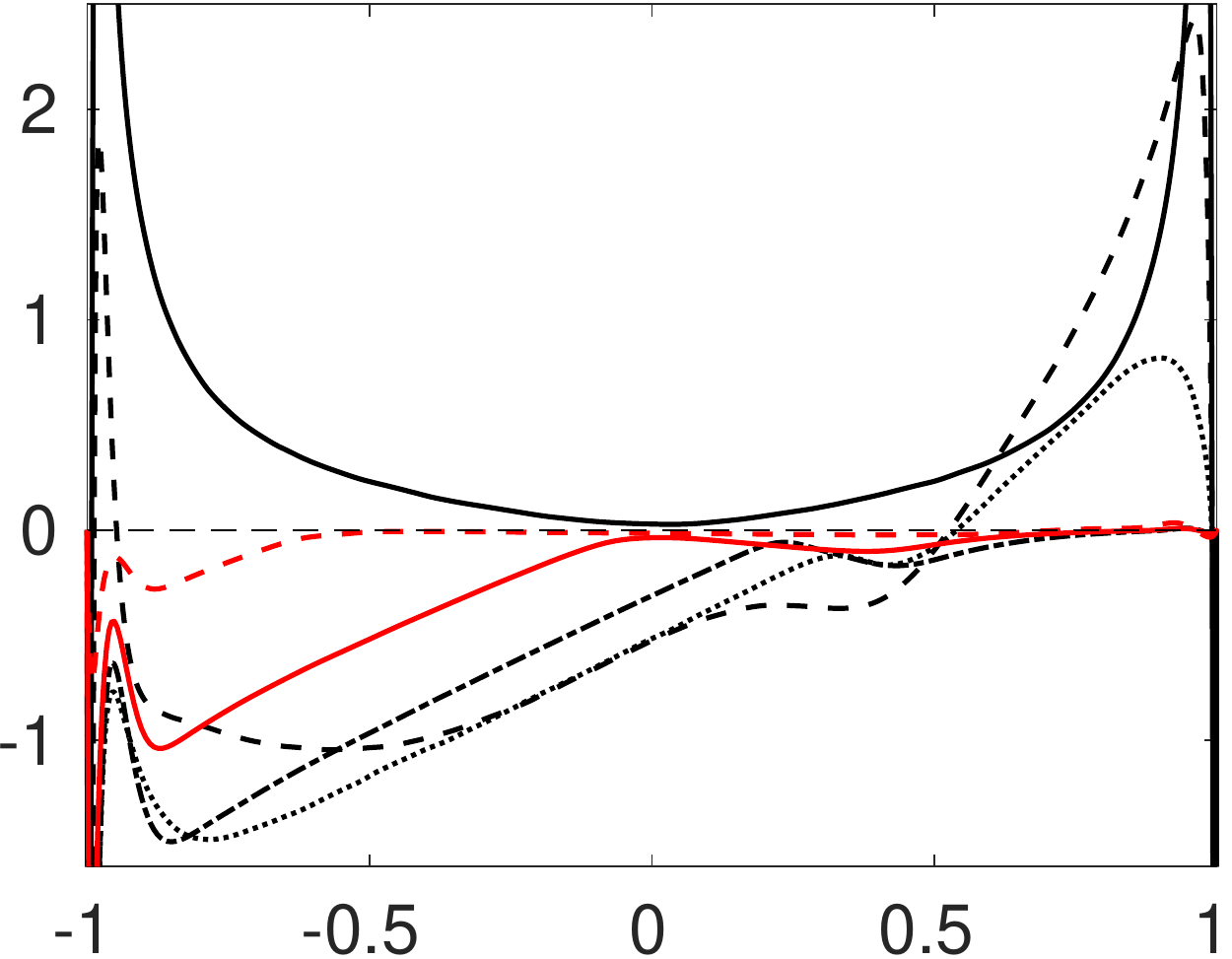}
\put(-0.1,4.3){$\displaystyle (b)$}
\put(-3.1,-0.3){$y$}
\put(-6.9,2.3){$\Pi_{vv}$}

\includegraphics[width=62mm]{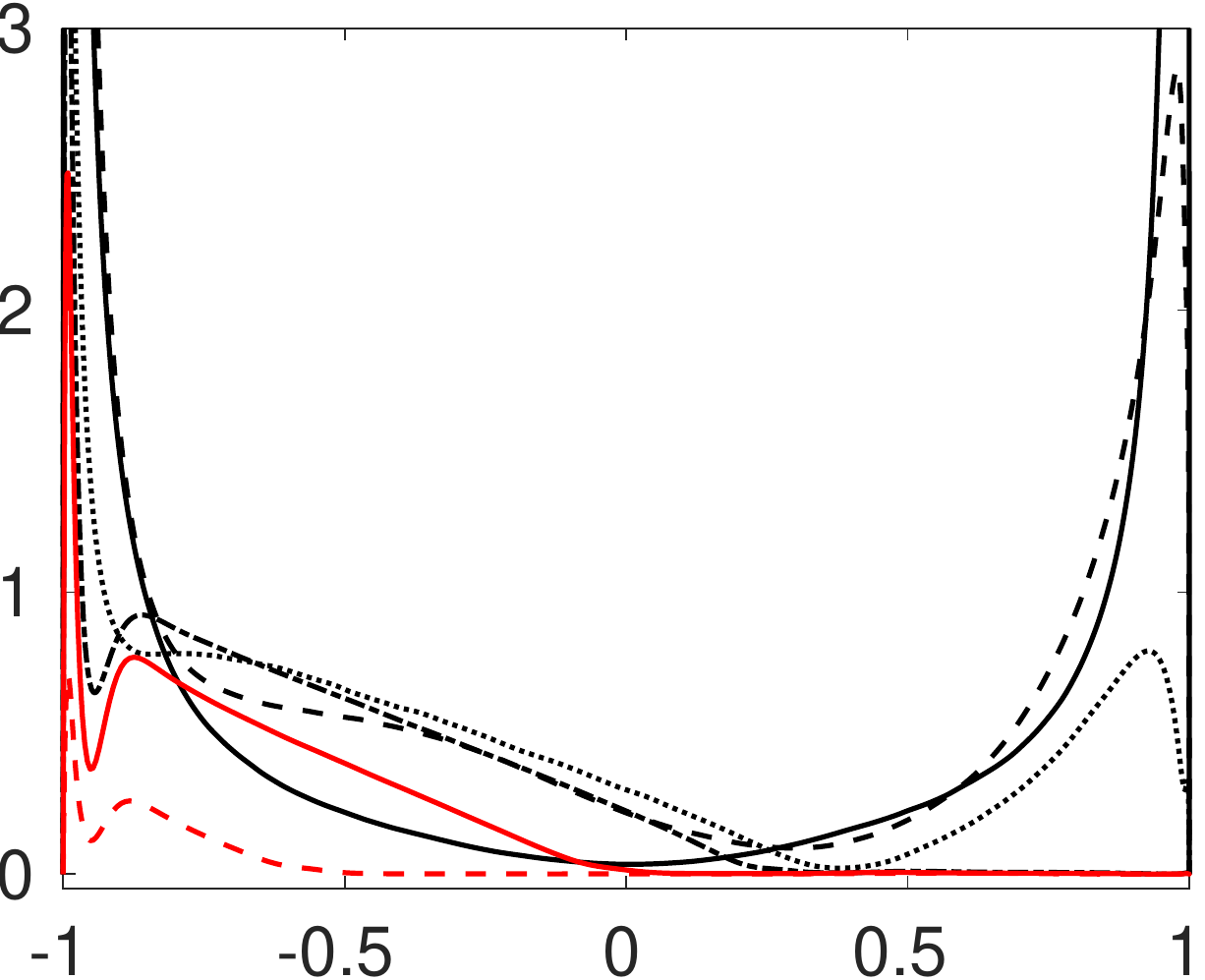}
\put(-0.1,4.3){$\displaystyle (c)$}
\put(-3.1,-0.3){$y$}
\put(-6.9,2.3){$\Pi_{ww}$}
\hskip9mm
\includegraphics[width=62mm]{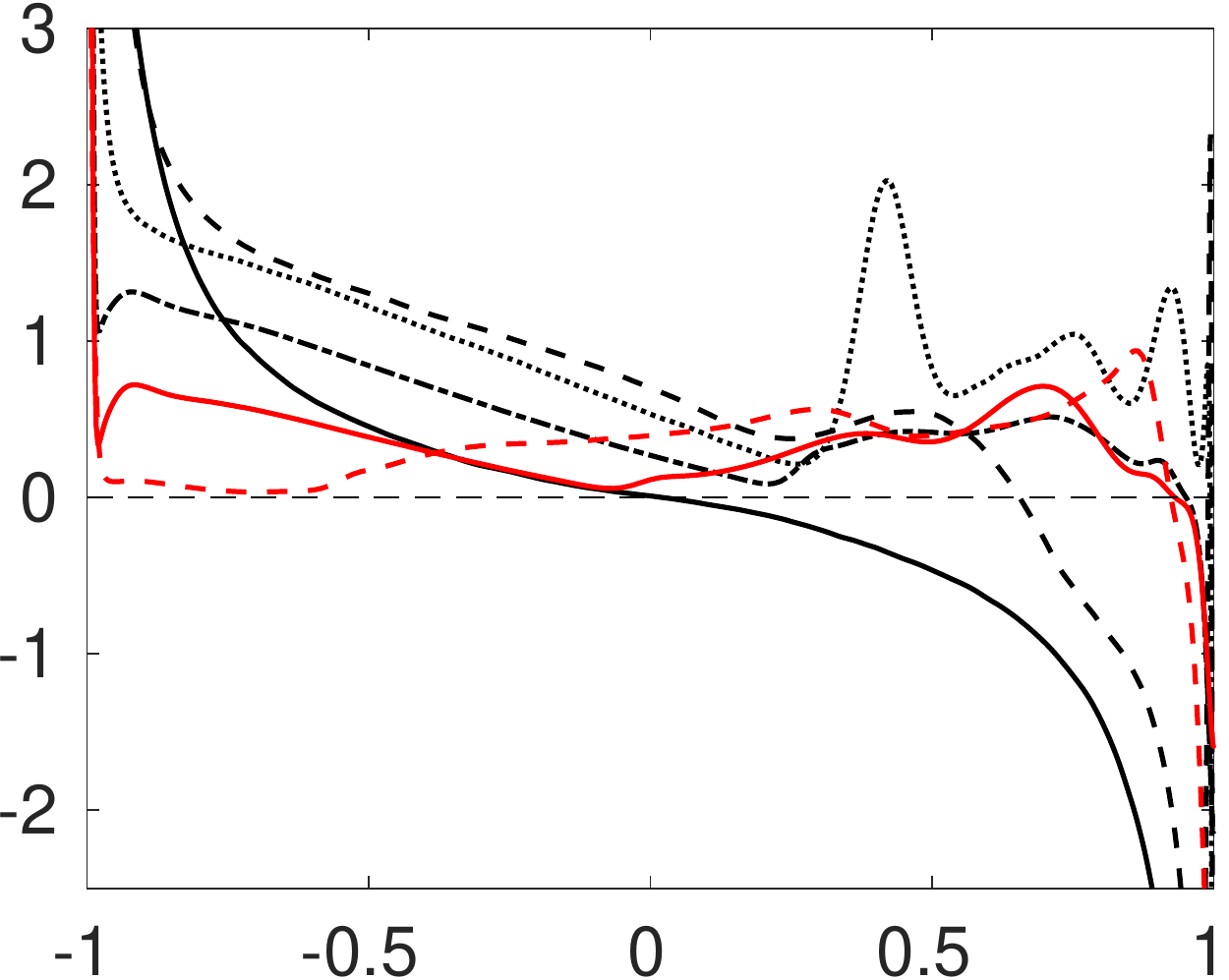}
\put(-0.1,4.3){$\displaystyle (d)$}
\put(-3.1,-0.3){$y$}
\put(-6.9,2.3){$\Pi_{uv}$}
\end{center}
\caption{(Colour online)
({\it a}) $\Pi_{uu}$, 
({\it b}) $\Pi_{vv}$,
({\it c}) $\Pi_{ww}$
and ({\it d}) $\Pi_{uv}$ scaled by
$u^2_\tau U_b Ro/h$.
Lines used as in figure \ref{eps_rot}.}
\label{pstrain_rot}
\end{figure}

Figure \ref{eps_rot}.({\it d}) shows the production of 
$\overline{uv}$ Reynolds stresses, $P_{uv}\,=\, -\overline{vv} \diff U/\diff y$, scaled by
$u^2_\tau U_b Ro / h$.
Also in this figure and figure \ref{pstrain_rot}, data for $Ro=0$ are included using the scaling
$ 0.45 u^2_\tau U_b/h$ to make a comparison with data of rotating channel
flow possible.
The profiles of $P_{uv}$ show less spreading than when using the ordinary wall unit scaling
and also reveal an approximately linear slope on the unstable side if $Ro \geq 0.9$.
On the stable side $P_{uv}$ is significant at $Ro=0.45$ but negligible for $Ro \geq 1.2$.

The Coriolis terms in the balance equations of
the $\overline{uu}$ and $\overline{vv}$ Reynolds stresses
are $C_{uu}=2 \overline{uv}/u^2_\tau$ and $C_{vv}=-2 \overline{uv}/u^2_\tau = -C_{uu}$ 
respectively, when scaled by $u^2_\tau U_b Ro / h$. 
Profiles of $\overline{uv}/u^2_\tau$ were already presented in figure \ref{urms}.({\it d}).
From these profiles it follows that on the unstable side the 
Coriolis term transfers energy from $\overline{uu}$ to the $\overline{vv}$ component
in rotating channel flow.
On the stable side it is the other way around for $Ro=0.45$ and 0.9
whereas it is small in the other cases.

If $\diff U / \diff y \simeq 2 \Omega$,
$P_{uu}  \simeq -4 \overline{uv} \Omega= - C_{uu}$, which means that the Coriolis term
closely balances the production term and, consequently,  the dissipation term approximately
balances the pressure-strain term in the equation for the $\overline{uu}$ component
since the transport terms are small.
This is confirmed by the DNSs but not shown here.
In fact, the sum $P_{ij} + C_{ij} \equiv P^{tot}_{ij}$ may be considered as a total production term.
When $\diff U / \diff y \simeq 2 \Omega$, 
$P^{tot}_{uu} \simeq 0$,
$P^{tot}_{vv} \simeq P_{uu}$ and
$P^{tot}_{ww} \simeq 0$
since
$P_{uu} \simeq -C_{uu} = C_{vv}$ as explained above and $P_{vv}=P_{ww}=C_{ww}=0$.
Figure \ref{eps_rot}.({\it e}) shows $P^{tot}_{uu}$ scaled
by $u^2_\tau U_b Ro / h$ and confirms that in a large part of a rotating channel flow
it is nearly zero.
Thus, in a non-rotating flow, production feeds energy in the $\overline{uu}$ component
and then energy is redistributed to the other components. By contrast,
in rotating channel flow on the unstable side $P^{tot}_{ij}$ feeds
energy mainly in the $\overline{vv}$ component and pressure-strain correlations
redistribute this energy to the $\overline{uu}$ and $\overline{ww}$ components.
This may explain the strong wall-normal velocity fluctuations observed
in rotating channel flow on the unstable side.
Very near the wall on the unstable side energy is still fed into the 
$\overline{uu}$ component. The same applies to the stable side for $Ro=0.45$ and 0.9
but if $Ro \geq 1.2$ the sum is nearly zero. Note that $P^{tot}_{uu}$
is slightly negative in rotating channel flow in a region around the maximum of $U$.

When scaled by $u^2_\tau U_b Ro / h$, 
$C_{uv}=(\overline{vv}-\overline{uu})/u^2_\tau$. 
The total production $P^{tot}_{uv}$ scaled by
$u^2_\tau U_b Ro / h$, 
shown in figure \ref{eps_rot}.({\it f}),
is balanced by $\Pi_{uv}$ since the dissipation and diffusion terms
are relatively small in a large part of the channel, in particular on the unstable side.
On the unstable side, $P^{tot}_{uv}$ scaled is negative, but it becomes less
negative for $Ro \geq 0.9$. On the stable side it is positive at $Ro=0.45$
and negative at larger $Ro$
meaning that it destroys $\overline{uv}$ correlations, leading
to the low turbulent shear stresses on the stable side observed before.

Pressure-strain correlations $\Pi_{uu}$, $\Pi_{vv}$, $\Pi_{ww}$ and $\Pi_{uv}$ 
scaled by
$u^2_\tau U_b Ro / h$ 
are presented in figure \ref{pstrain_rot}.
When $Ro=0$ energy is transferred from 
$\overline{uu}$ to $\overline{vv}$ and $\overline{ww}$,
whereas in the rotating cases energy is transferred from
$\overline{vv}$ to $\overline{uu}$ and $\overline{ww}$
by pressure-strain correlations
on the unstable side, except very close to the wall where $\Pi_{uu}$
is still negative. 
Especially at high $Ro$, $\Pi_{ww}$ is large which explains the
strong spanwise fluctuations in rapidly rotating channel flows.
On the stable side there is a significant
energy transfer from $\overline{uu}$ to the other two components
by pressure-strain correlations for $Ro \leq 0.9$.
At larger $Ro$, $\Pi_{uu}$ and $\Pi_{vv}$
are small but not negligible on the stable side while
$\Pi_{uv}$ is positive 
(figure \ref{pstrain_rot}.{\it d})
and approximately balanced by the Coriolis term $C_{uv}$ as discussed before.
The slope of the pressure-strain profiles on the unstable side
is also approximately linear for $Ro\geq 0.9$ and scales with
$u^2_\tau U_b Ro/h$. 

Turbulent velocity fluctuations are not insignificant
on the stable side (figure \ref{urms}), 
although $P_K$ is very small there for $Ro \geq 1.2$.
Grundestam \etal (2008) have suggested that the turbulence
on the stable side is forced by the turbulence on the unstable side
through the pressure diffusion term of the $\overline{vv}$ component
which then redistributes the energy through pressure-strain correlations.
The pressure diffusion of $\overline{vv}$ has indeed a small positive value
(not shown here) and
figure \ref{pstrain_rot}.({\it a}) and ({\it b}) show
that, albeit small, $\Pi_{vv}$ is negative and $\Pi_{uu}$ positive at $Ro=1.2$ and 1.5
on the stable side. The latter is approximately balanced by a negative 
$P^{tot}_{uu}$.

%
\section{Spectra and two-point correlations}
\label{sec_spec}

In order to study the effect of rotation on turbulence structures in
a quantitative way 
Kristoffersen \& Andersson (1993), Alvelius (1999) and Grundestam \etal (2008)
computed two-point correlations at $Re_\tau = 180$ to 194.
Kristoffersen \& Andersson observed that the spanwise near-wall streak spacing
in wall units narrows with $Ro$ whereas Alvelius observed the opposite trend.
Generally, relatively short streamwise two-point correlations of 
streamwise fluctuations on the unstable side
and relatively long ones on the other laminar-like side 
were observed when the channel was rotating.
At low $Ro$ Kristoffersen \& Andersson and Alvelius could also notice an impact
of Taylor-G{\"o}rtler vortices on correlations on the unstable side.

Here, I present spanwise two-point correlations of the wall-normal
velocity fluctuations as well as spectra since they reveal better the impact of rotation on the
different flow scales, especially the large scales which are known to become
important at higher $Re$ in wall flows (Smits \etal 2011).
Spectra of the streamwise velocity fluctuations are presented since these most
clearly reveal the presence of large-scale motions in non-rotating wall flows
as well as spectra of the spanwise and wall-normal velocity fluctuations since
these may uncover the presence of roll cells. 
The presented co-spectra give information on the scales that contribute to
the momentum transfer and production of turbulent kinetic energy.
The focus is on the strongly turbulent unstable side of the channel.
Spectra and two-point correlations
are less instructive for the stable side when relaminarization occurs.
First, I consider the trends with $Ro$ and next with $Re$.

\subsection{Spectra at $Re=20\,000$}

\begin{figure}
\begin{center}
\setlength{\unitlength}{1cm}
\includegraphics[width=55mm]{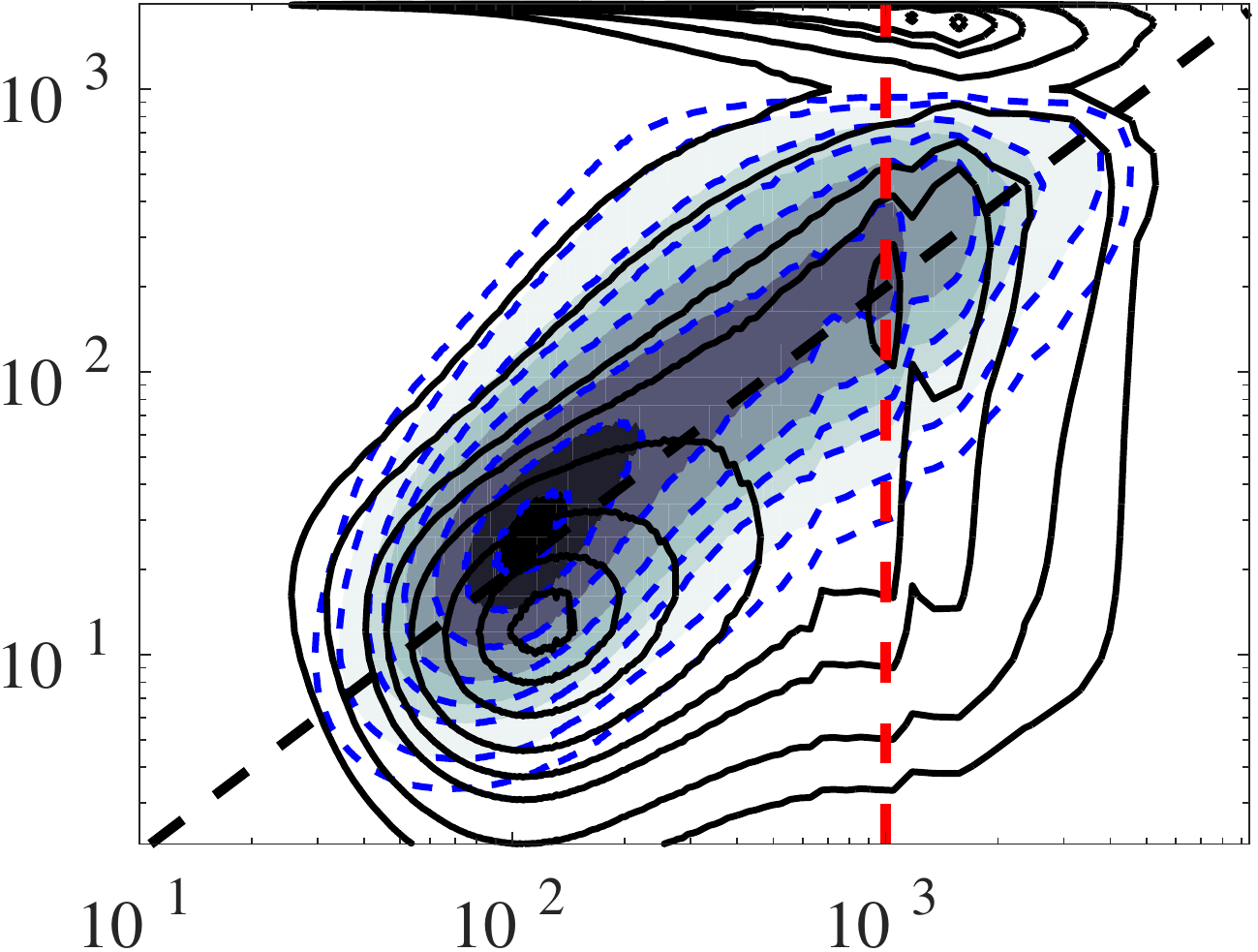}
\put(-0.1,3.9){$\displaystyle (a)$}
\put(-2.9,-0.3){$\lambda^*_z$}
\put(-6.0,2.1){$y^*$}
\hskip9mm
\includegraphics[width=55mm]{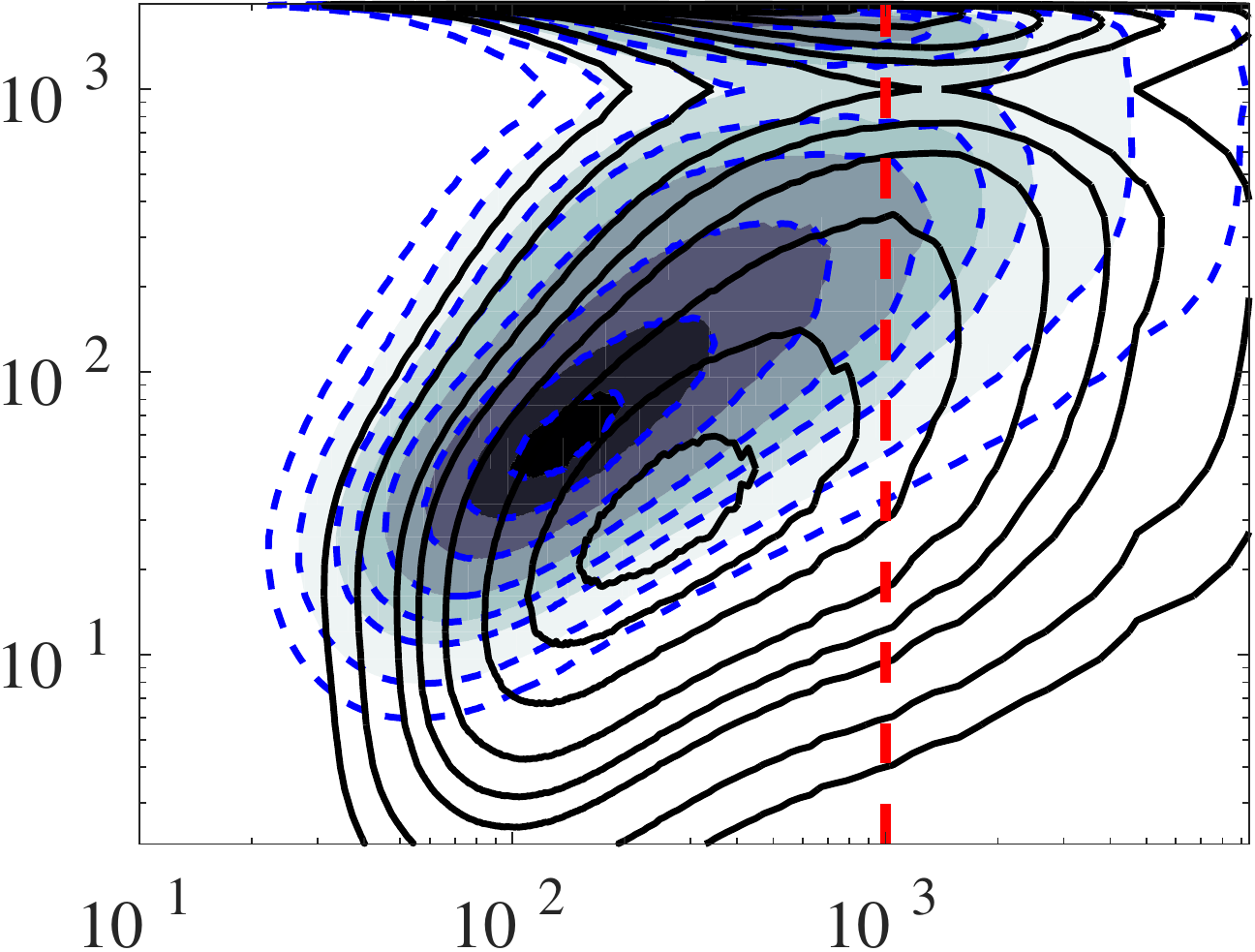}
\put(-0.1,3.9){$\displaystyle (b)$}
\put(-2.9,-0.3){$\lambda^*_z$}
\put(-6.0,2.1){$y^*$}

\includegraphics[width=55mm]{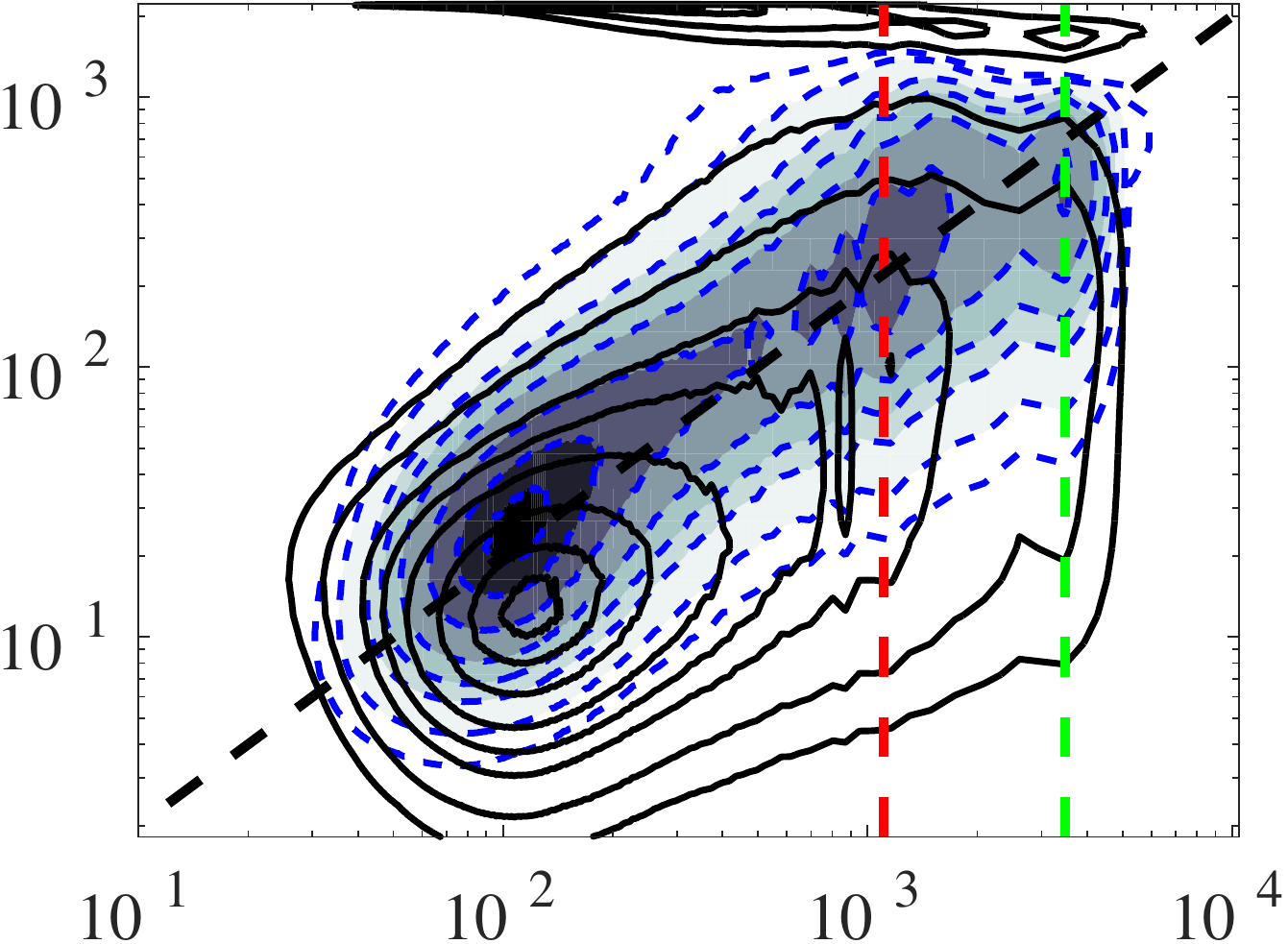}
\put(-0.1,3.9){$\displaystyle (c)$}
\put(-2.9,-0.3){$\lambda^*_z$}
\put(-6.0,2.1){$y^*$}
\hskip9mm
\includegraphics[width=55mm]{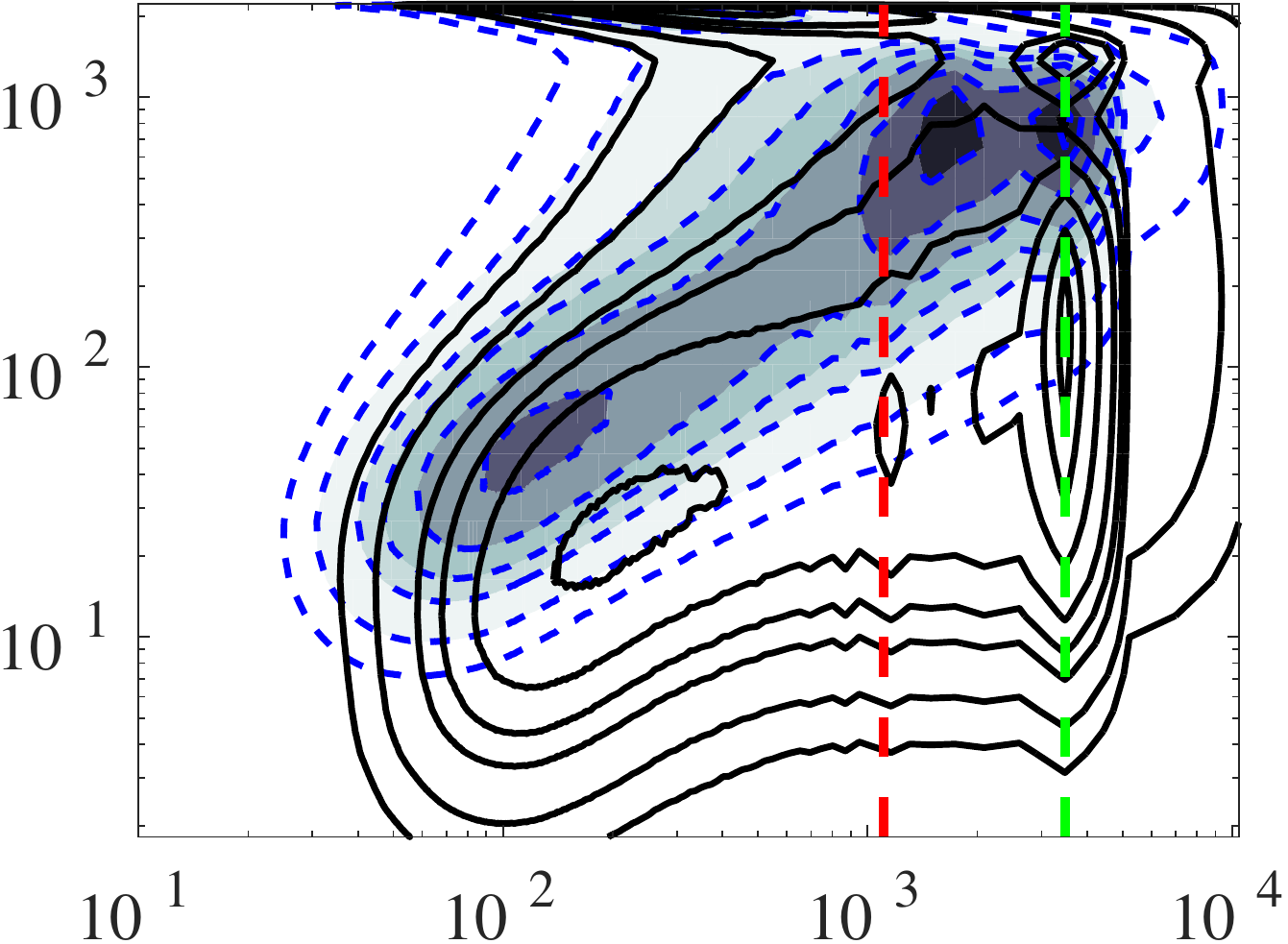}
\put(-0.1,3.9){$\displaystyle (d)$}
\put(-2.9,-0.3){$\lambda^*_z$}
\put(-6.0,2.1){$y^*$}

\includegraphics[width=55mm]{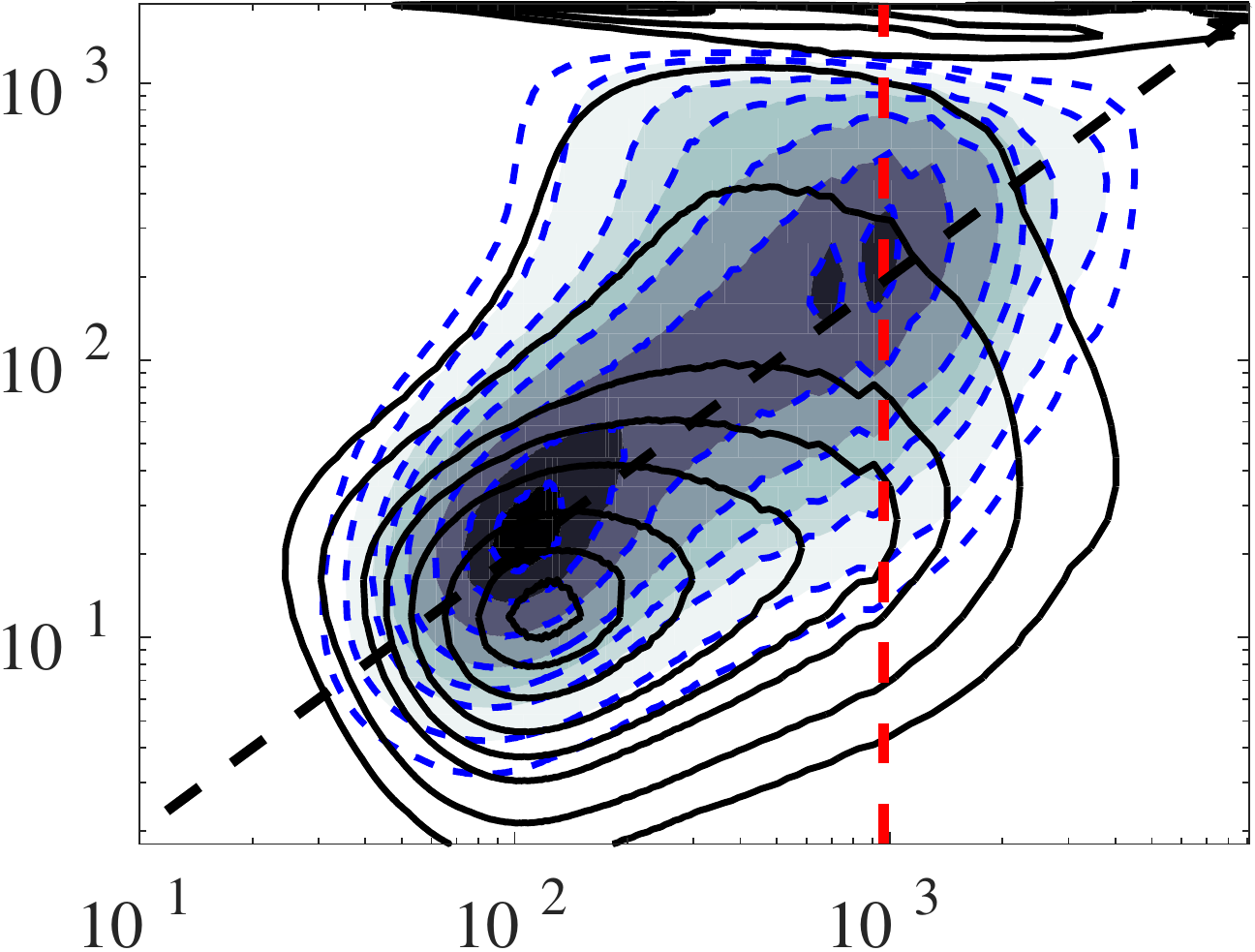}
\put(-0.1,3.9){$\displaystyle (e)$}
\put(-2.9,-0.3){$\lambda^*_z$}
\put(-6.0,2.1){$y^*$}
\hskip9mm
\includegraphics[width=55mm]{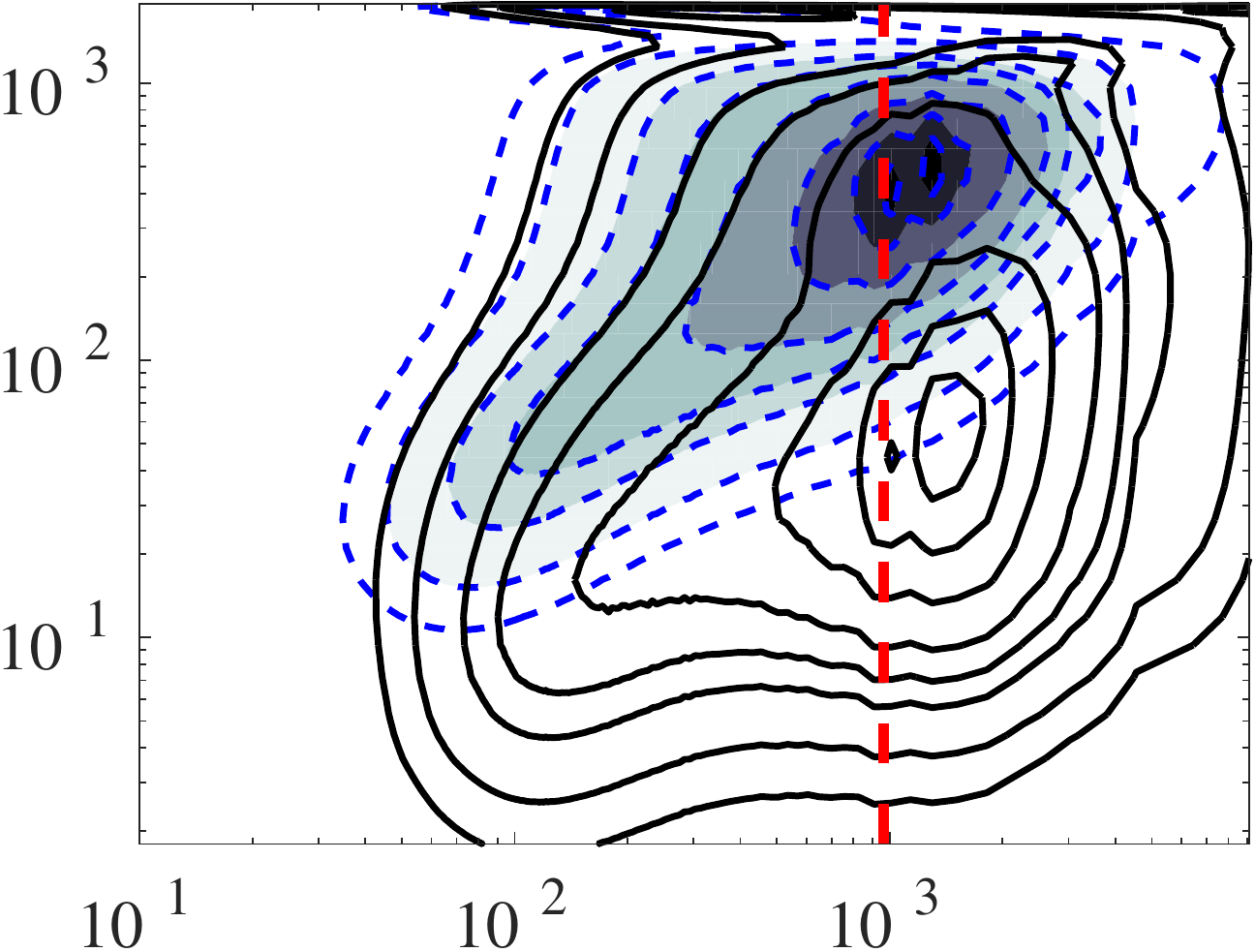}
\put(-0.1,3.9){$\displaystyle (f)$}
\put(-2.9,-0.3){$\lambda^*_z$}
\put(-6.0,2.1){$y^*$}

\includegraphics[width=55mm]{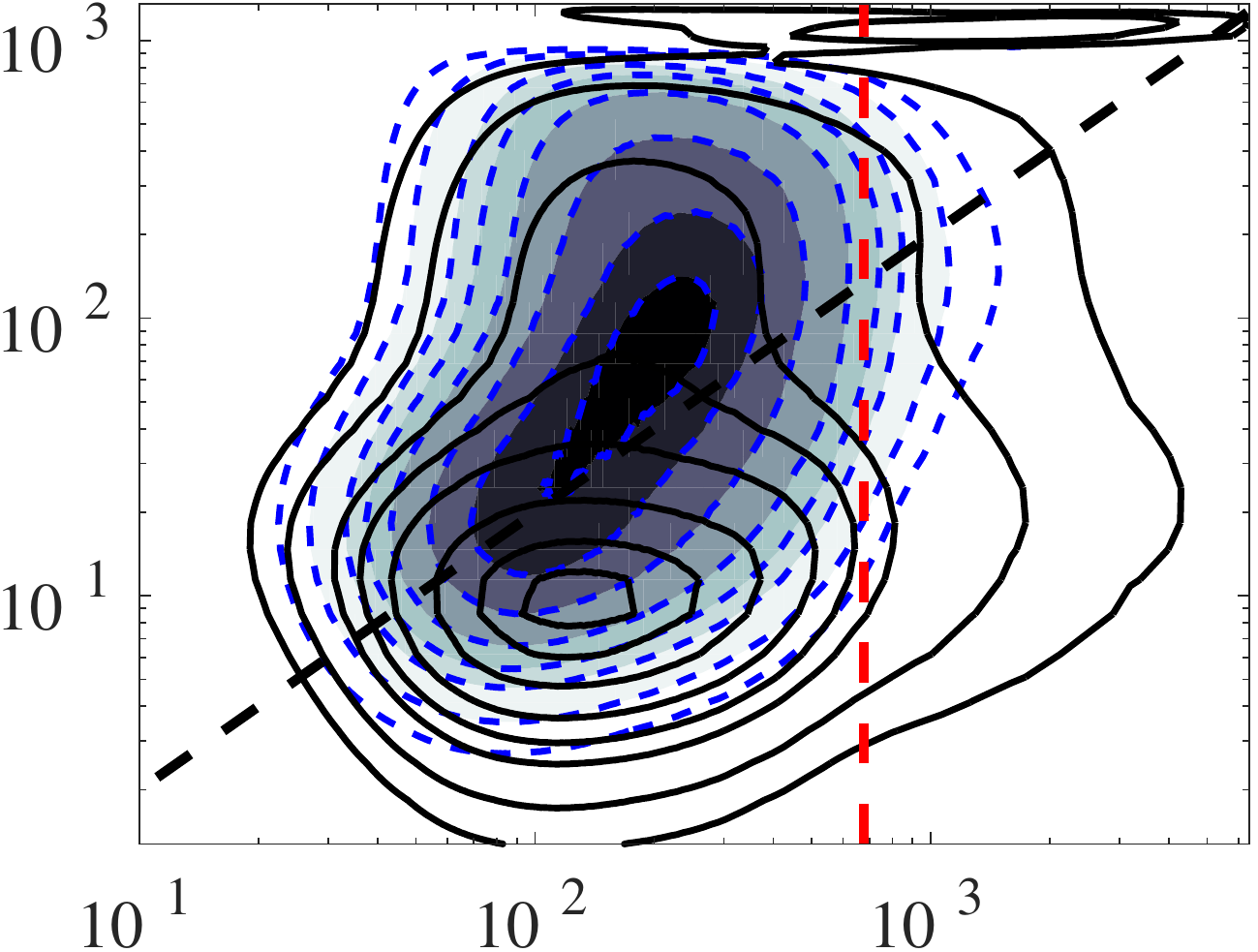}
\put(-0.1,3.9){$\displaystyle (g)$}
\put(-2.9,-0.3){$\lambda^*_z$}
\put(-6.0,2.1){$y^*$}
\hskip9mm
\includegraphics[width=55mm]{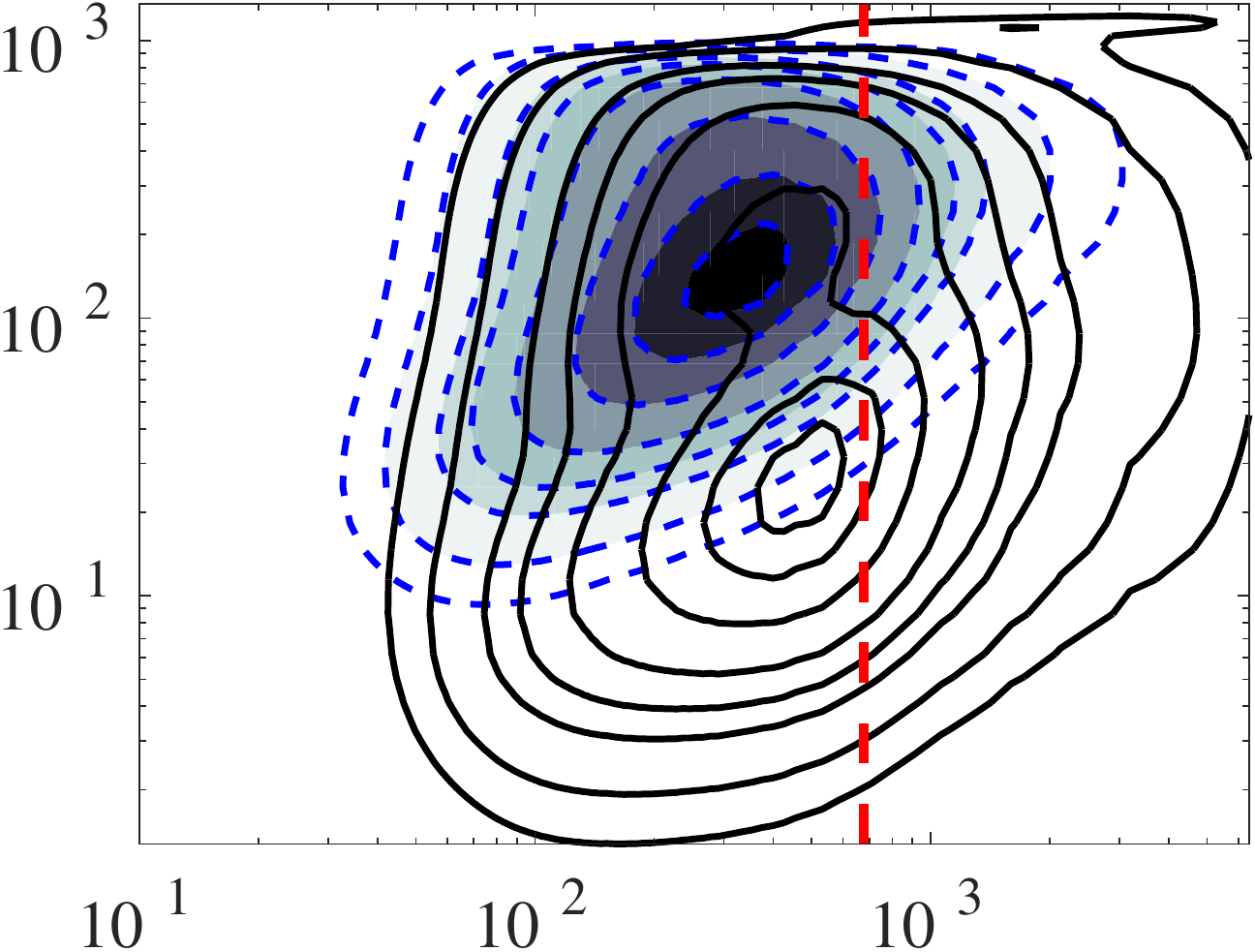}
\put(-0.1,3.9){$\displaystyle (h)$}
\put(-2.9,-0.3){$\lambda^*_z$}
\put(-6.0,2.1){$y^*$}
\end{center}
\caption{(Colour online) Maps of premultiplied one-dimensional energy spectra as function of spanwise
wavelength $\lambda^*_z$
and distance from the wall $y^*$ at $Re=20\,000$ for $Ro=0$ ({\it a, b}), $Ro=0.15$ ({\it c, d}), $Ro=0.45$ ({\it e, f})
and $Ro=0.9 $ ({\it g, h}).
Left column shows $k_z \Phi_{uu}$ (black lines) and cospectra $k_z \Phi_{uv}$ (blue dashed lines and colours).
Right column shows $k_z \Phi_{vv}$ (blue dashed lines and colours) and $k_z \Phi_{ww}$ (black lines).
Contour levels from innermost to outermost are 0.94, 0.8, 0.6, 0.4, 0.28, 0.2, 0.1 and 0.05.
Spectra are scaled with their maximum value.
The straight black dashed line follows the relation $y=0.1 \lambda_z$.
}
\label{specz}
\end{figure}

\begin{figure}
\begin{center}
\setlength{\unitlength}{1cm}
\includegraphics[width=55mm]{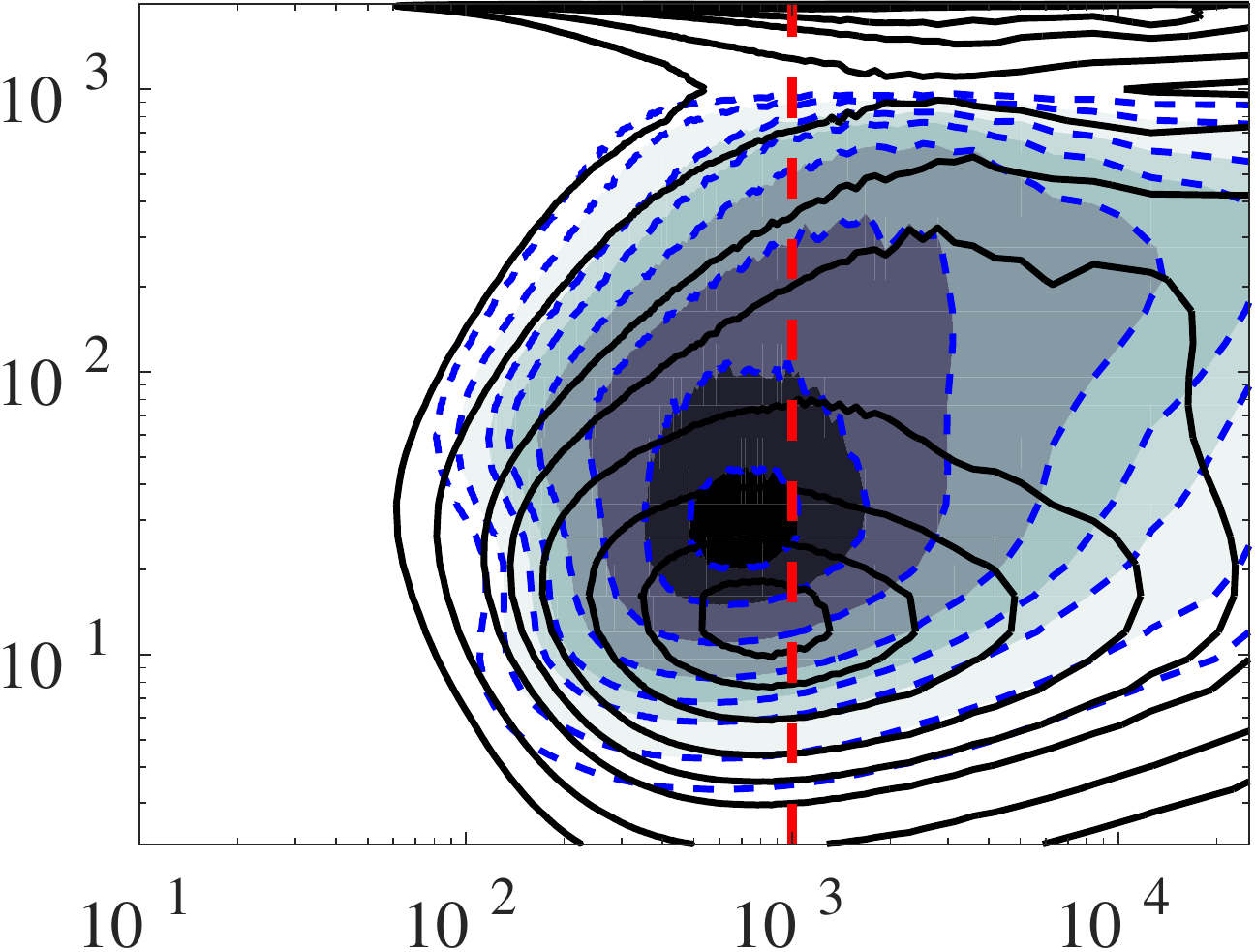}
\put(-0.1,3.9){$\displaystyle (a)$}
\put(-2.9,-0.3){$\lambda^*_x$}
\put(-6.0,2.1){$y^*$}
\hskip9mm
\includegraphics[width=55mm]{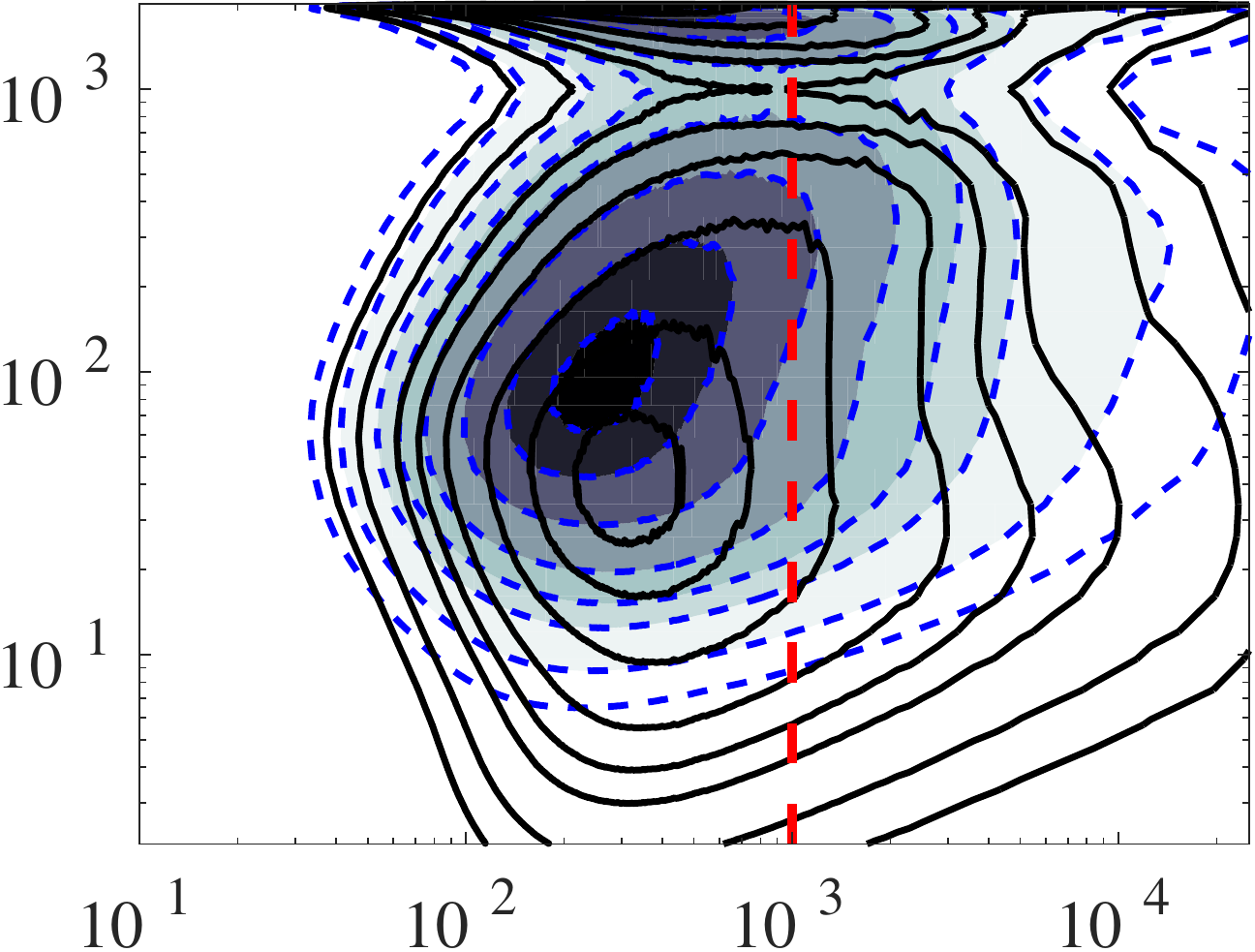}
\put(-0.1,3.9){$\displaystyle (b)$}
\put(-2.9,-0.3){$\lambda^*_x$}
\put(-6.0,2.1){$y^*$}

\includegraphics[width=55mm]{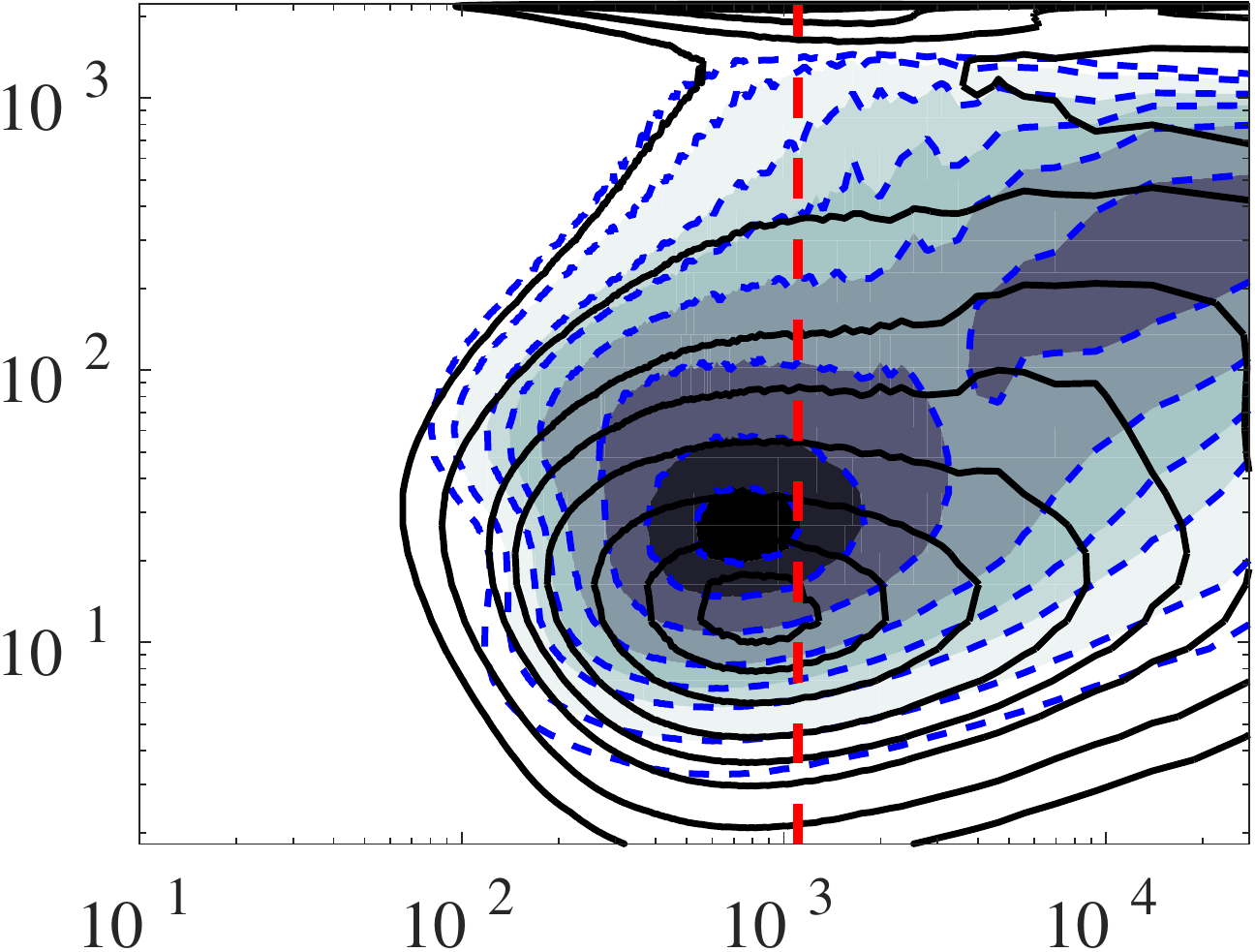}
\put(-0.1,3.9){$\displaystyle (c)$}
\put(-2.9,-0.3){$\lambda^*_x$}
\put(-6.0,2.1){$y^*$}
\hskip9mm
\includegraphics[width=55mm]{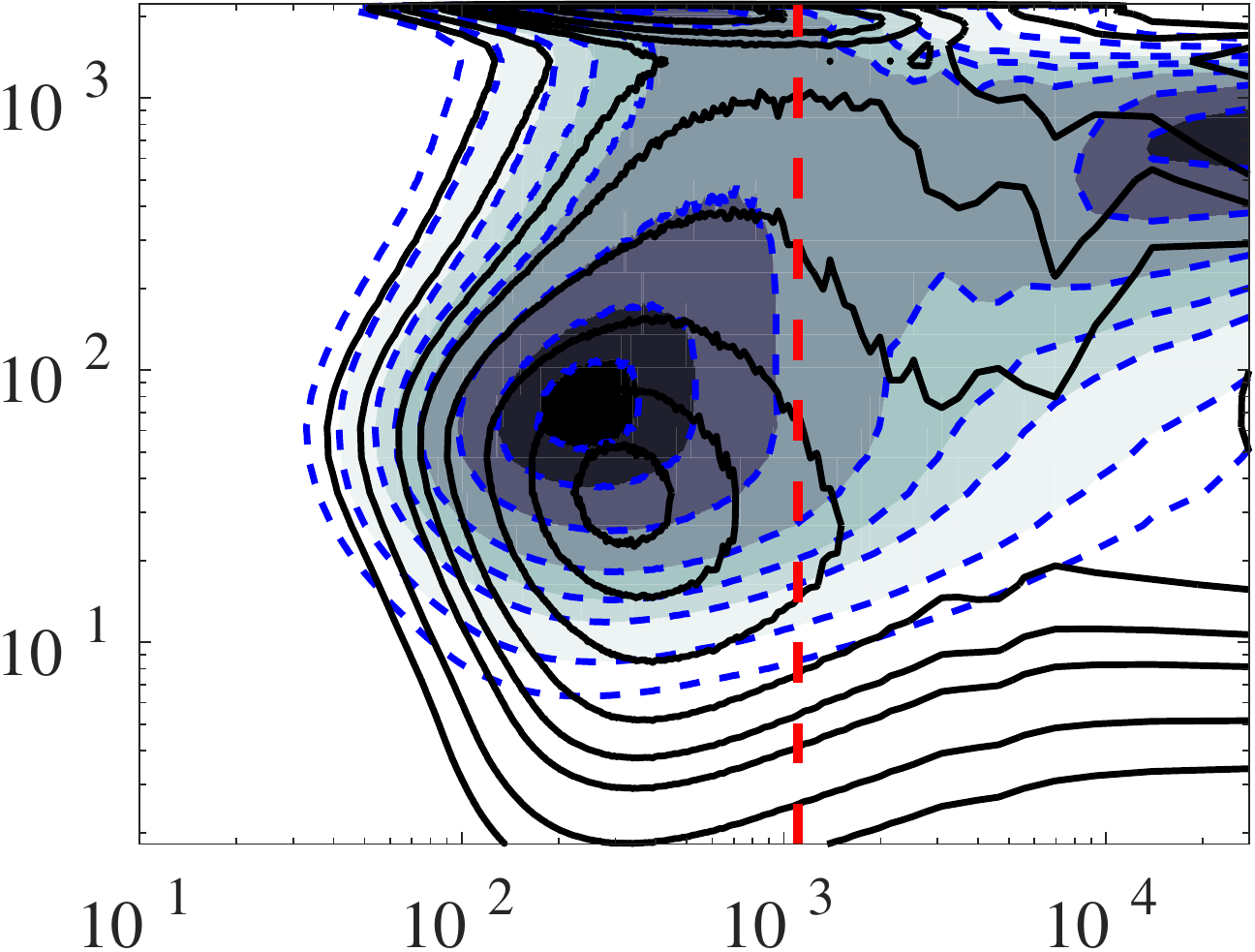}
\put(-0.1,3.9){$\displaystyle (d)$}
\put(-2.9,-0.3){$\lambda^*_x$}
\put(-6.0,2.1){$y^*$}

\includegraphics[width=55mm]{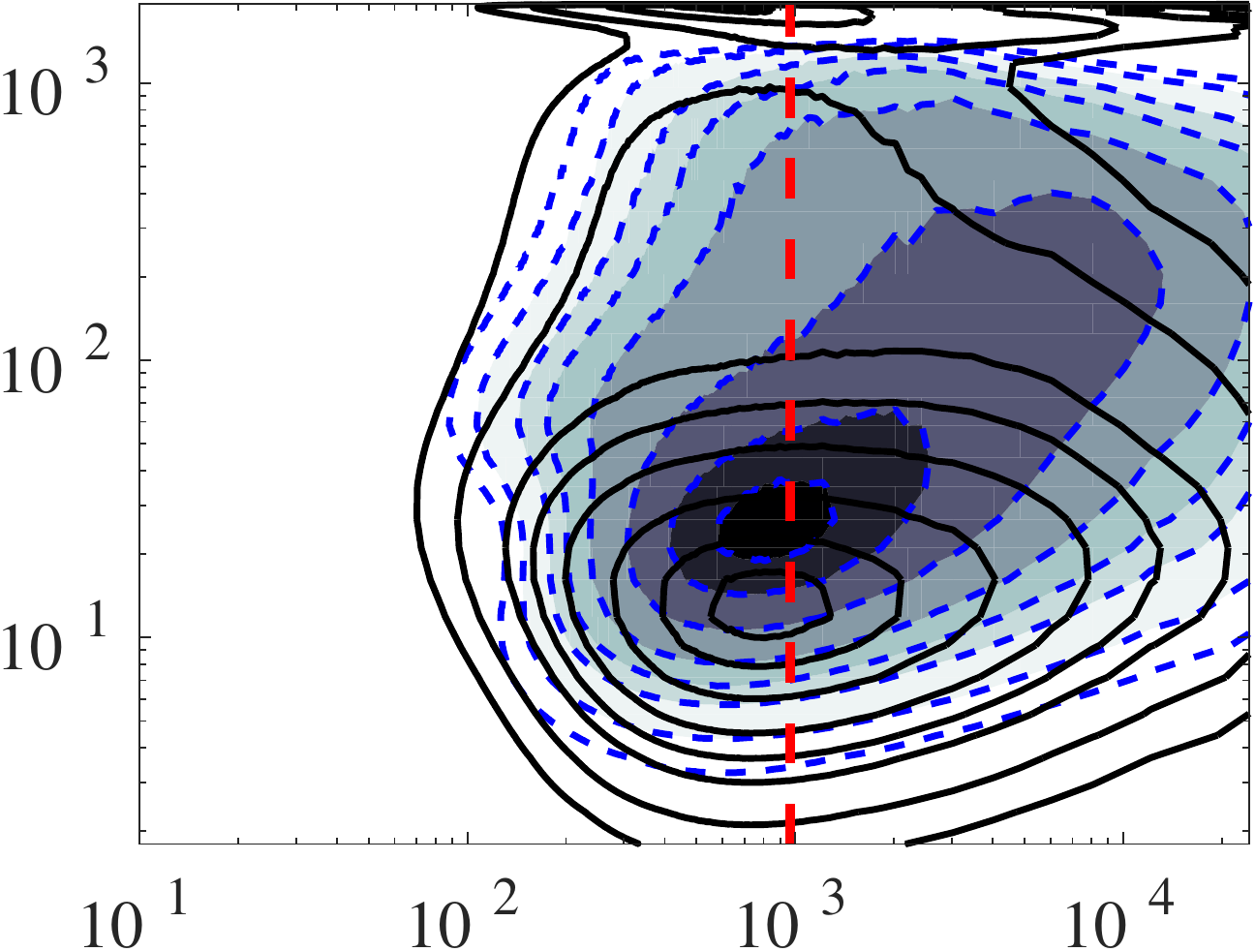}
\put(-0.1,3.9){$\displaystyle (e)$}
\put(-2.9,-0.3){$\lambda^*_x$}
\put(-6.0,2.1){$y^*$}
\hskip9mm
\includegraphics[width=55mm]{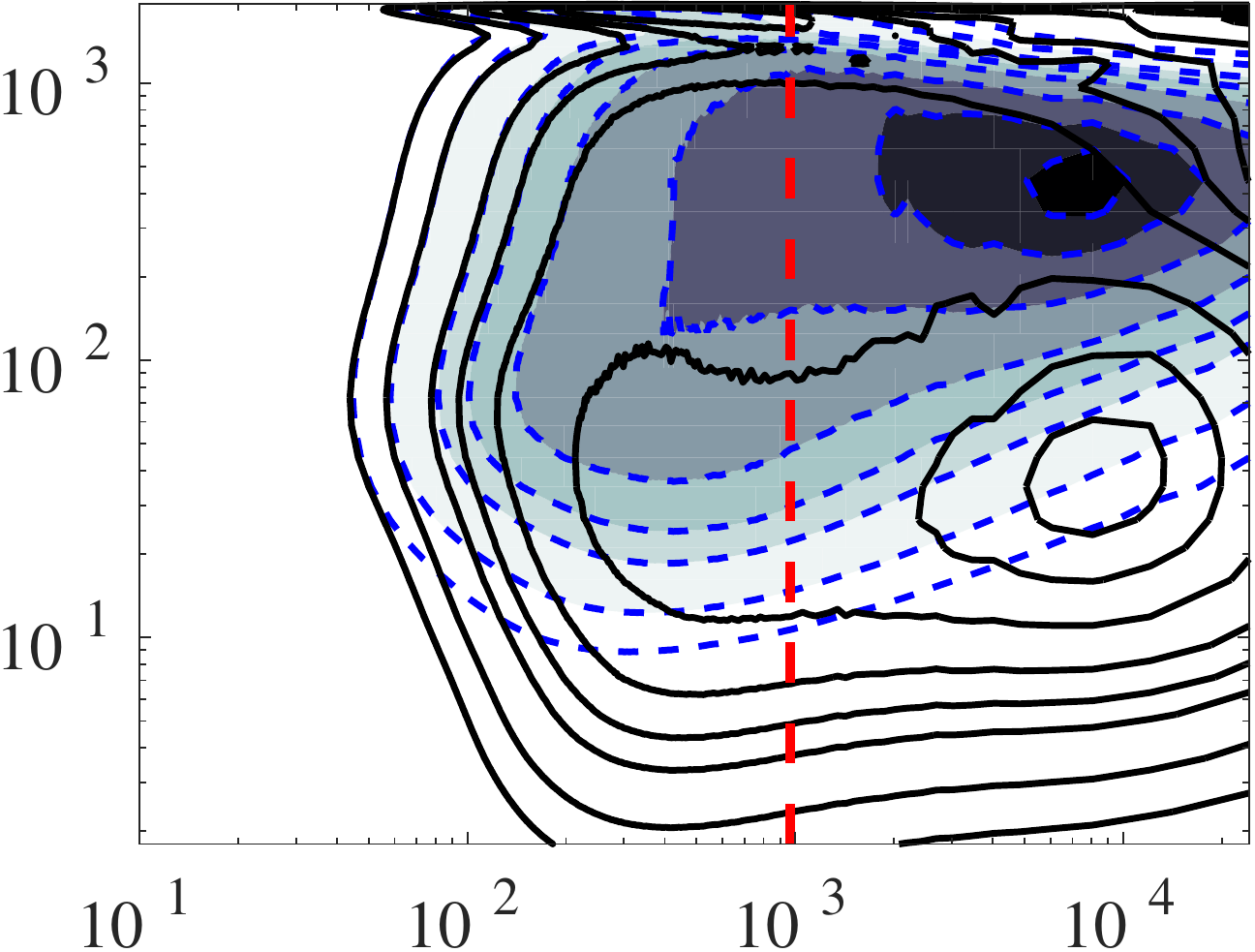}
\put(-0.1,3.9){$\displaystyle (f)$}
\put(-2.9,-0.3){$\lambda^*_x$}
\put(-6.0,2.1){$y^*$}

\includegraphics[width=55mm]{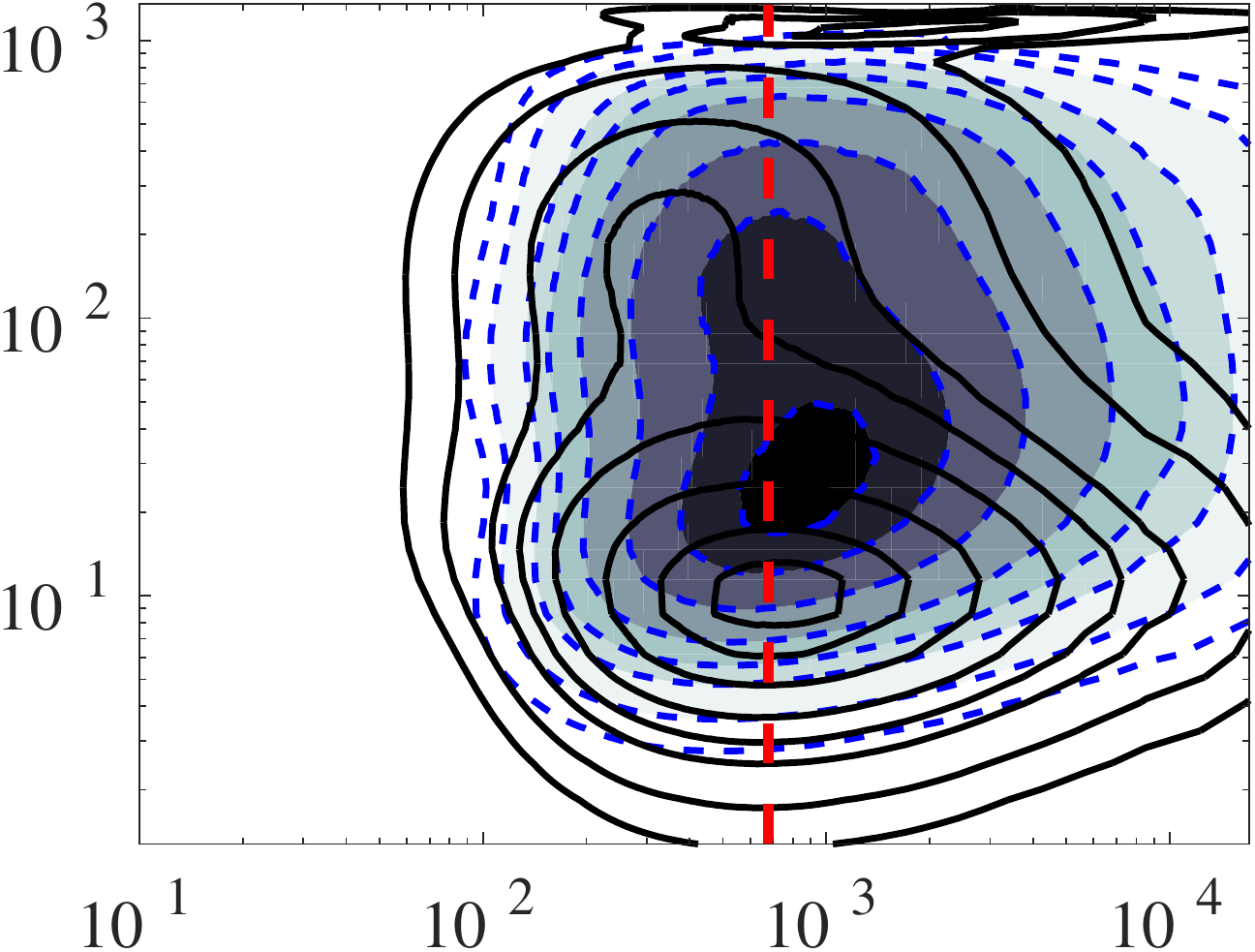}
\put(-0.1,3.9){$\displaystyle (g)$}
\put(-2.9,-0.3){$\lambda^*_x$}
\put(-6.0,2.1){$y^*$}
\hskip9mm
\includegraphics[width=55mm]{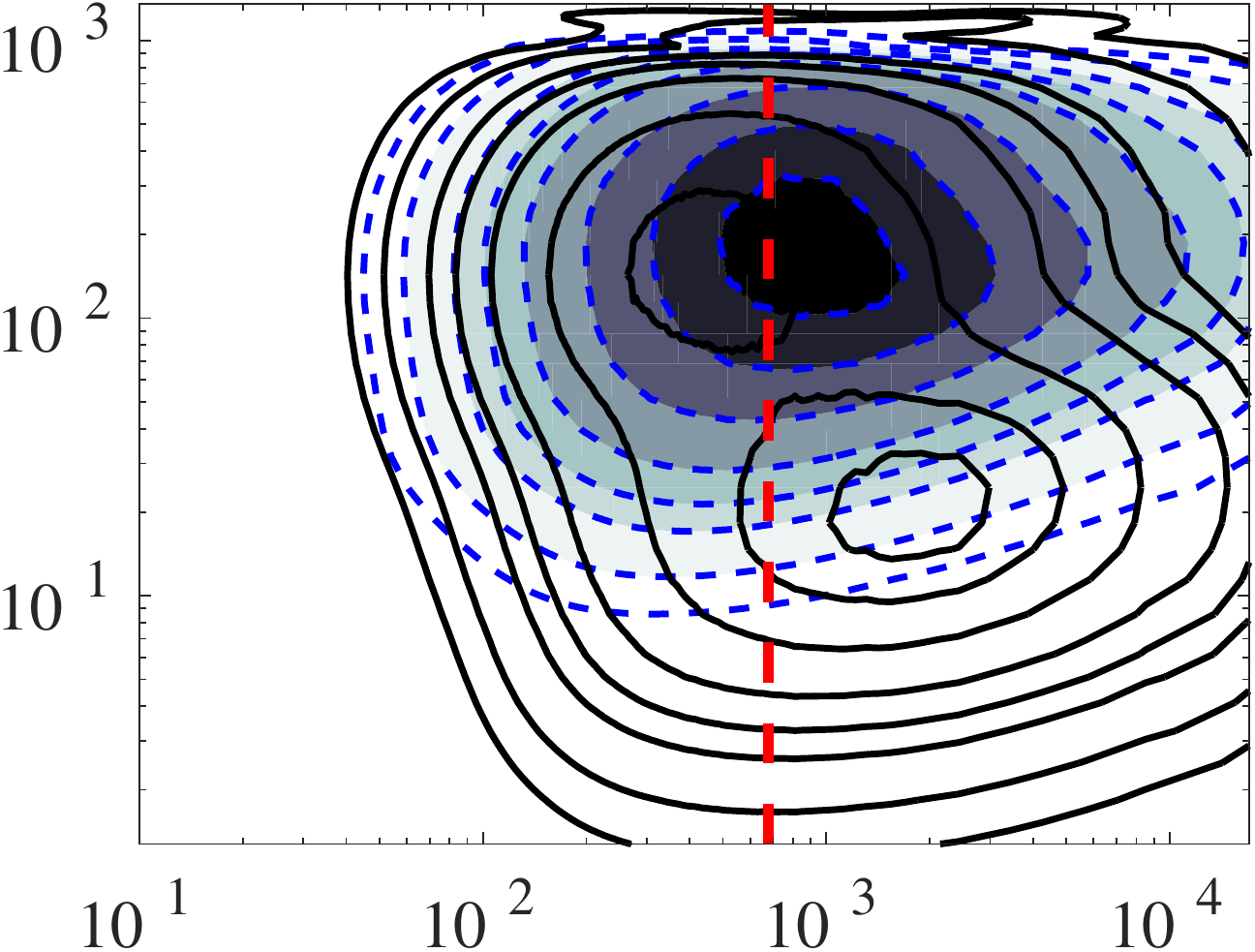}
\put(-0.1,3.9){$\displaystyle (h)$}
\put(-2.9,-0.3){$\lambda^*_x$}
\put(-6.0,2.1){$y^*$}
\end{center}
\caption{(Colour online) Maps of premultiplied one-dimensional energy spectra as function of streamwise
wavelength $\lambda^*_x$
and distance from the wall $y^*$ at $Re=20\,000$ for $Ro=0$ ({\it a, b}), $Ro=0.15$ ({\it c, d}), $Ro=0.45$ ({\it e, f})
and $Ro=0.9 $ ({\it g, h}).
Left column shows $k_x \Phi_{uu}$ (black lines) and cospectra $k_x \Phi_{uv}$ (blue dashed lines and colours).
Right column shows $k_x \Phi_{vv}$ (blue dashed lines and colours) and $k_x \Phi_{ww}$ (black lines).
Contour levels from innermost to outermost are 0.94, 0.8, 0.6, 0.4, 0.28, 0.2, 0.1 and 0.05.
Spectra are scaled with their maximum value.
}
\label{specx}
\end{figure}

Spanwise premultiplied energy spectra, 
$k_z \Phi_{uu}$, $k_z \Phi_{vv}$ and $k_z \Phi_{ww}$ of the streamwise, wall-normal and spanwise
velocity fluctuations, respectively, and cospectra, $k_z \Phi_{uv}$, at $Re=20\,000$ are shown
in figure \ref{specz}. 
Streamwise premultiplied energy spectra,
$k_x \Phi_{uu}$, $k_x \Phi_{vv}$ and $k_x \Phi_{ww}$ of the streamwise, wall-normal and spanwise
velocity fluctuations, respectively, and cospectra, $k_x \Phi_{uv}$, at $Re=20\,000$ are shown
in figure \ref{specx}. Spectra are shown for $Ro=0$, 0.15, 0.45 and 0.9,
and are presented
as function of the wall distance $y^* = (y+1) u_{\tau u}/\nu$ 
and the spanwise and streamwise wave length
$\lambda^*_z$ and $\lambda^*_x$, respectively, both scaled in term of the viscous
length scale of the unstable side, $l^* = \nu/u_{\tau u}$.
The red dashed line in each plot indicates scales with a wavelength $h$.
Spectra are scaled with their maximum value, 
which are given in terms of $u_{\tau u}$ in table \ref{maxval}.
\begin{table}
\begin{center}
\def~{\hphantom{0}}
\begin{tabular}{llrrrrrrrr}
$Re$ & $Ro$ & $k_z\frac{\Phi_{uu}}{u^2_{\tau u}}^\star$ &
$k_z\frac{\Phi_{vv}}{u^2_{\tau u}}^\star$ &
$k_z\frac{\Phi_{ww}}{u^2_{\tau u}}^\star$ &
$k_z\frac{\Phi_{uv}}{u^2_{\tau u}}^\star$ &
$k_x\frac{\Phi_{uu}}{u^2_{\tau u}}^\star$ &
$k_x\frac{\Phi_{vv}}{u^2_{\tau u}}^\star$ &
$k_x\frac{\Phi_{ww}}{u^2_{\tau u}}^\star$ &
$k_x\frac{\Phi_{uv}}{u^2_{\tau u}}^\star$ \\[3pt]
$20\,000$ & 0   &	3.83	&0.56	&0.73	&0.58	&2.13	&0.40	&0.66	&0.28\\
$20\,000$ & 0.15&	3.85	&0.75	&1.00	&0.57	&2.15	&0.32	&0.56	&0.28\\
$20\,000$ & 0.45&	3.01	&1.60	&1.39	&0.49	&1.89	&0.59	&0.64	&0.27\\
$20\,000$ & 0.9 &	1.67	&2.13	&1.86	&0.46	&1.24	&1.04	&0.95	&0.29\\[3pt]
$5000$ & 0   & ~ &0.50	&0.64	&~	&~	&0.33	&0.57	&~\\
$5000$ & 0.45   & ~ &1.61 &1.32	&~	&~	&0.61	&0.61	&~\\
$5000$ & 0.9   & ~ &2.50 &1.85	&~	&~	&1.16	&0.96	&~\\
\end{tabular}
\caption{Maximum values of the premultiplied one-dimensional energy spectra.
Superscript $\star$ means the maximum value of the quantity.
The spectra are scaled with the friction velocity of the unstable side, $u_{\tau u}$.
}
\label{maxval}
\end{center}
\end{table}

Roll cells produce significant wall-normal and spanwise motions
in rotating channel flow, as shown by e.g. Dai \etal (2016),
and that can be expected to lead to energetic large-scale 
modes in the spectra of the spanwise and especially the
wall-normal velocity away from the wall.
I therefore interpret distinctly energetic large-scale modes
in the spanwise $k_z \Phi_{vv}$ and $k_z \Phi_{ww}$ spectra
as well as in the streamwise $k_x \Phi_{vv}$ spectra as signs of roll cells.
If there are $n$ pairs of counter-rotating roll cells in the computational domain, which
has a spanwise size of $L_z = 3 \pi h$ in these DNSs, the spanwise spectrum 
$k_z \Phi_{vv}$ can be expected to have a peak at $\lambda_z = 3 \pi h/n$.

In non-rotating channel flow the spanwise spectra reveal, besides the 
streaks at $\lambda^*_z \approx 100$ indicated by the near-wall peak
in $k_z \Phi_{uu}$, the signature of large wide structures
of wavelength $\lambda_z \approx h$ in $k_z \Phi_{uu}$ and $k_z \Phi_{uv}$ further away from the wall,
(figure \ref{specz}.{\it a}),
like in the DNS by Lee \& Moser (2015). 
In the streamwise spectra
$k_x \Phi_{uu}$ and $k_x \Phi_{uv}$ 
energetic peaks are seen at
$\lambda^*_x \approx 1000$ near the wall related to the near-wall cycle
(Monty \etal 2009). In the outer layer weakly energetic large-scale structures with wavelengths
$\lambda_x \approx \pi h$ to $4 \pi h$ are observed in
$k_x \Phi_{uu}$ and $k_x \Phi_{uv}$ 
(figure \ref{specx}.{\it a}), which become more energetic at higher $Re$
(Lee \& Moser 2015). 

The near-wall peak in the spanwise spectrum $k_z \Phi_{uu}$ and streamwise spectrum
$k_x \Phi_{uu}$ 
changes little in the rotating cases if $Ro \leq 0.45$ so that a change of the streak spacing 
induced by rotation, as
suggested by Kristoffersen \& Andersson (1993) and Alvelius (1999), cannot be confirmed.
However, at $Ro=0.15$, 
$k_z \Phi_{vv}$ 
has three peaks caused by near-wall structures of wavelength $\lambda^*_z \approx 150$ 
around $y^* = 50$ and
large structures with wavelengths
$\lambda_z \approx \pi h/2$ and
$\lambda_z \approx \pi h$ (the latter indicated by the green dashed line) in the outer layer
of the unstable side far away from the wall at $y \approx -0.3$
(figure \ref{specz}.{\it d}).
A strong large-scale peak is also observed in
$k_z \Phi_{ww}$ 
at wavelength $\lambda_z \approx \pi h$ in the outer layer. 
The energetic large-scale modes in the outer layer at $\lambda_z \approx \pi h$ 
are almost certainly a consequence of 3 pairs of counter-rotating
roll cells of spanwise size $\pi h/2$ seen previously in figure
\ref{vis_roll}.({\it a}). The other peak in 
$k_z \Phi_{vv}$ at $\lambda_z = \pi h/2$ shows that there also smaller large-scale
structures, possibly roll cells that are half the size as the largest ones, indicating
that there might be roll cells of different sizes.
Scales with the maximum possible streamwise wavelength, i.e. $\lambda_x = L_x$,
are obviously energetic according to the streamwise spectra in
figure \ref{specx}.({\it c}) and ({\it d}), implying that at least some of
the roll cells span the whole domain in the streamwise direction.

At a higher $Ro=0.45$
the peaks in the spanwise spectra
$k_z \Phi_{vv}$, $k_z \Phi_{ww}$
have, compared to the spectra at $Ro=0$,
clearly shifted towards larger scales 
(figure \ref{specz}.{\it f})
and these of the streamwise spectra $k_x \Phi_{vv}$, $k_x \Phi_{ww}$ 
to a streamwise wavelength 
$\lambda_x \approx 8 \pi h/3$ 
(figure \ref{specx}.{\it f}) in the outer layer. 
The energetic large-scale modes observed in the spanwise spectrum
$k_z \Phi_{vv}$ at $\lambda_z \approx 3\pi h/8$ and $y \approx -0.5$
are not present at $Ro=0$, therefore,
I interpret them as the signatures of roll cells. Two-point correlations, presented later,
provide further evidence that they exist.
The roll cells are smaller than at $Ro=0.15$
and do not appear to span the whole streamwise domain according to the streamwise spectra
since the peak in the streamwise $k_x \Phi_{vv}$ is at $\lambda_x < L_x$.
These observations are consistent with the results shown by the visualizations
in figure \ref{vis_roll} before. Dai \etal (2016) noted too that the roll cell diminished
in size with $Ro$ in their DNS at a much lower $Re=2800$.
In the present DNSs,
energetic wide and long structures in terms of outer units
are observed as well in the spanwise and streamwise
spectra $\Phi_{uu}$ 
and especially the co-spectra $\Phi_{uv}$ at $Ro=0.45$ 
and even more so at $Ro=0.15$ in the outer layer 
(figure \ref{specz}.{\it c, e} and \ref{specx}.{\it c, e}), 
showing that the roll cells induce
large-scale momentum transfer.

Previous studies have found strong support for the
hypothesis that the logarithmic region of wall-bounded turbulent flows 
is populated by attached eddies (Perry \& Chong 1982) whose size grows with $y$,
see e.g. Smits \etal (2011) and Hwang (2015). 
Some evidence of attached eddies can indeed be found in the present spanwise
spectrum $k_z \Phi_{uu}$ and cospectrum $k_z \Phi_{uv}$
of the non-rotating channel flow (figure \ref{specz}.{\it a}) since they
show that size of the structures grows with the distance
to the wall in the logarithmic region. The cospectrum
follows roughly the linear relation $y = 0.2 \lambda_z -1$ shown by the straight
black dashed line in the figure, indicating that the size of eddies is
approximately proportional to the distance to the wall, in agreement
with the attached eddy hypothesis (Hwang 2015).
The spanwise spectra $k_z \Phi_{uu}$ and cospectra $k_z \Phi_{uv}$
at $Ro=0.15$ and 0.45 (figure \ref{specz}.{\it c, e}) 
show similar characteristics with growing scales as $y$ increases.
The cospectrum follows also in these cases roughly the linear 
relation $y = 0.2 \lambda_z -1$. This provides support for the idea
that also rotating wall-bounded flows are populated by attached 
eddies, at least up to moderate rotation rates.

At higher rotation rates large-scale structures become progressively less energetic and smaller.
At $Ro=0.9$ no energetic large scale structures are seen in 
the spanwise spectra
$k_z \Phi_{uu}$ and $k_z \Phi_{uv}$ while 
$k_z \Phi_{vv}$ has a peak at spanwise wavelength 
$\lambda_z \approx \pi h/6$ in the outer layer at $y \approx -0.8$
(figure \ref{specz}.{\it g} and {\it h}). 
The streamwise spectrum of the streamwise velocity $k_x \Phi_{uu}$
has still a near-wall peak 
and that of the wall-normal velocity $k_x \Phi_{vv}$
has an outer peak at
streamwise wave length $\lambda_x \approx 2 \pi h/5$, while
the spectrum of the spanwise velocity $k_x \Phi_{ww}$ has two energetic
peaks: an outer peak at
streamwise wave length $\lambda_x \approx \pi h/5$ and a near-wall peak at
$\lambda_x \approx 4 \pi h/5$ 
(figure \ref{specx}.{\it g} and {\it h}). 
Energetic large-scale roll cells are thus not present according to the spectra and
correspondingly, no large structures like roll cells are observed in visualizations
and there are less energetic large-scale motions that contribute to momentum transfer
than in a non-rotating channel flow.

Similar observations (not shown for brevity) are made at yet higher $Ro$, i.e., no energetic wide 
or long structures are present according to the spectra and co-spectra in terms of outer units. 
The peak in the spanwise
spectrum  
$k_z \Phi_{vv}$ shifts towards shorter wavelengths
and the double peak in the streamwise spectrum $k_x \Phi_{ww}$ persists.
The double peak is also observed in the rms profiles (figure \ref{urms}.{\it c}).
The spanwise spectra $k_z \Phi_{uu}$ and 
$k_z \Phi_{uv}$ have still a spectral peak at 
$\lambda^*_z \approx 100$ but the peak moves closer to the wall in terms of 
viscous wall units.
The streamwise spectra $k_x \Phi_{uu}$ and 
$k_x \Phi_{uv}$ have a spectral peak at $\lambda_x \approx h$ when $Ro \geq 0.9$,
but in terms of the viscous wall units the structures become shorter.
This means that the near-wall streaky structures 
move closer to the wall and become shorter,
in agreement with the results of
Lamballais \etal (1998), but they
do not clearly confirm their finding that these structures become
weaker at high $Ro$.

According to the spanwise 
spectrum $k_z \Phi_{uu}$ and cospectrum $k_z \Phi_{uv}$
at $Ro=0.9$ (figure \ref{specz}.{\it g}) the structures do not
become much larger when the distance to the wall increases and
at higher $Ro$ the scales appear to grow even less with $y$. 
The absence of structures whose size grows with the distance to the
walls when $Ro\gtrsim 0.9$ implies that attached eddies 
are much less prominent in very rapidly rotating wall-bounded flows.
This suggests that the interaction between the inner and outer layer
is weak at high $Ro$. Pressure and turbulent diffusion of turbulent
kinetic energy are accordingly very small on the unstable side in that case.

Spectra at $Re=20\,000$ and $Ro=0.15$ 
at the stable side of the channel (not shown here)
show that the roll cells observed on the unstable side
do not deeply penetrate the stable channel side although
the study by Dai \etal (2016) shows that they occasionally can become larger
and may even affect the flow up to wall at the stable channel side.
The spectra also show energetic near-wall modes at 
$\lambda^*_z \approx 100$ 
and $\lambda^*_x \approx 800$, as on the unstable side,
implying that the near-wall cycle is maintained.
On the other hand, 
although the flow is fully turbulent on the stable side at
this low $Ro$, the spectra on the stable channel side
reveal no or much less energetic large-scale modes than at $Ro=0$,
showing that cyclonic rotation weakens large-scale structures
in wall-bounded flows. 
This agrees with the observation that in plane Couette flow
even very weak cyclonic spanwise rotation eliminates large-scale
structures (Komminaho \etal 1996).


\subsection{Reynolds number dependence}

In several other DNS studies of spanwise rotating channel flow 
large streamwise roll cells or Taylor-G{\"o}rtler vortices
have been found and examined, as mentioned before.
Kristoffersen \& Andersson (1993),
Lamballais \etal (1998), 
Hsieh \& Biringen (2016) and Dai \etal (2016) have 
observed them in DNSs at $Re_\tau \leq 200$ and $Ro \leq 0.5$.
Hsieh \& Biringen have observed them also at $Re_\tau = 406$ and $Ro = 0.2$. 
It is important that the roll cells are properly resolved since 
flow statistics deviate significantly if they are suppressed (Hsieh \& Biringen 2016).
Helical spectra computed from DNSs at $Re_\tau = 180$ by Yang \& Wu (2012)
suggest large roll cells if $0.03 \leq Ro \leq 0.57$ with the strongest signal at $Ro=0.15$. 
The size of the roll cells appears to diminish with $Ro$ for $Ro \geq 0.15$
and there is no sign that they exist at $Ro=0.94$ and higher.
On the other hand, visualizations and spanwise 
two-point correlations computed by Grundestam \etal (2008) at
$Re_\tau = 180$ indicate that roll cells are present at $Ro=1.27$,
but the statistics show that the roll cells are non-stationary and do not
have a long correlation length in the streamwise direction and
their size appears to be smaller than at lower $Ro$.

Thus, there is strong evidence that large roll cells exist
in spanwise rotating channel flow at low Reynolds numbers for $Ro \lesssim 0.6$.
At higher $Ro$ roll cells are less certain and more difficult to detect
owing to their unsteadiness, but there are indications of their occurrence (Grundestam \etal 2008).
Spectra presented before reveal large streamwise roll cells in a DNS
at $Re=20\,000$ and $Ro=0.15$ and clearly hint at their occurrence at $Ro=0.45$, but at $Ro=0.9$ and
higher there is no strong evidence of their presence.
This suggests that at lower Reynolds numbers roll cells exist in a wider $Ro$ range.
Note, that in previous DNSs of rotating channel flow the computational domain size
was $4 \pi \times 2 \pi$ in the streamwise and spanwise direction, respectively, or less,
which is smaller than in the present DNSs. This may influence
the size and enhance the coherency of the roll cells, 
especially in combination with periodic boundary conditions. 
Indeed, in the DNS by Kristoffersen \& Andersson (1993)
at $Ro=0.15$ the roll cells were very steady whereas in the experiments by Johnston \etal (1972)
the boundary conditions were different and the roll cells were much more unsteady.

\begin{figure}
\begin{center}
\setlength{\unitlength}{1cm}
\includegraphics[width=55mm]{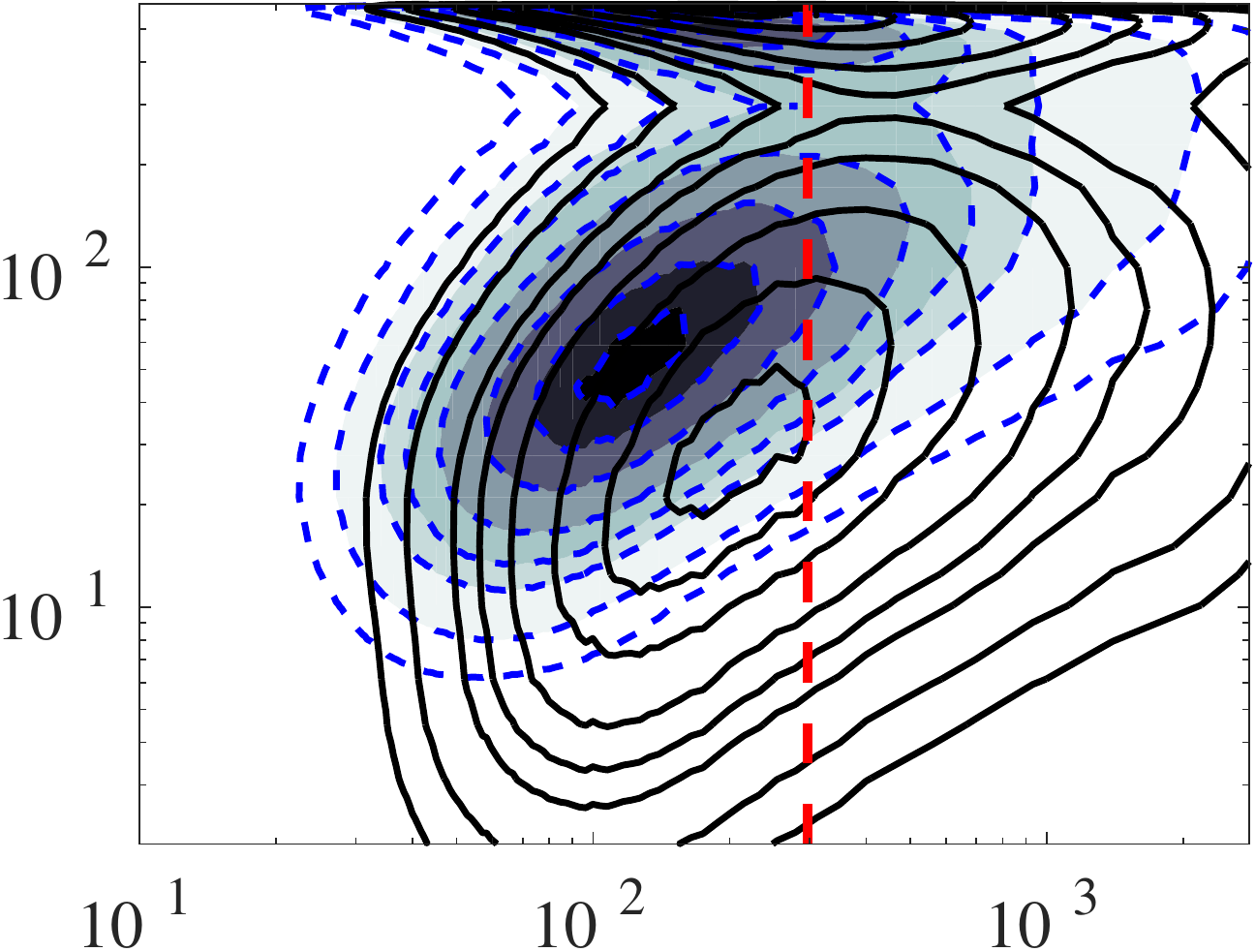}
\put(-0.1,3.9){$\displaystyle (a)$}
\put(-2.9,-0.3){$\lambda^*_z$}
\put(-6.0,2.1){$y^*$}
\hskip9mm
\includegraphics[width=55mm]{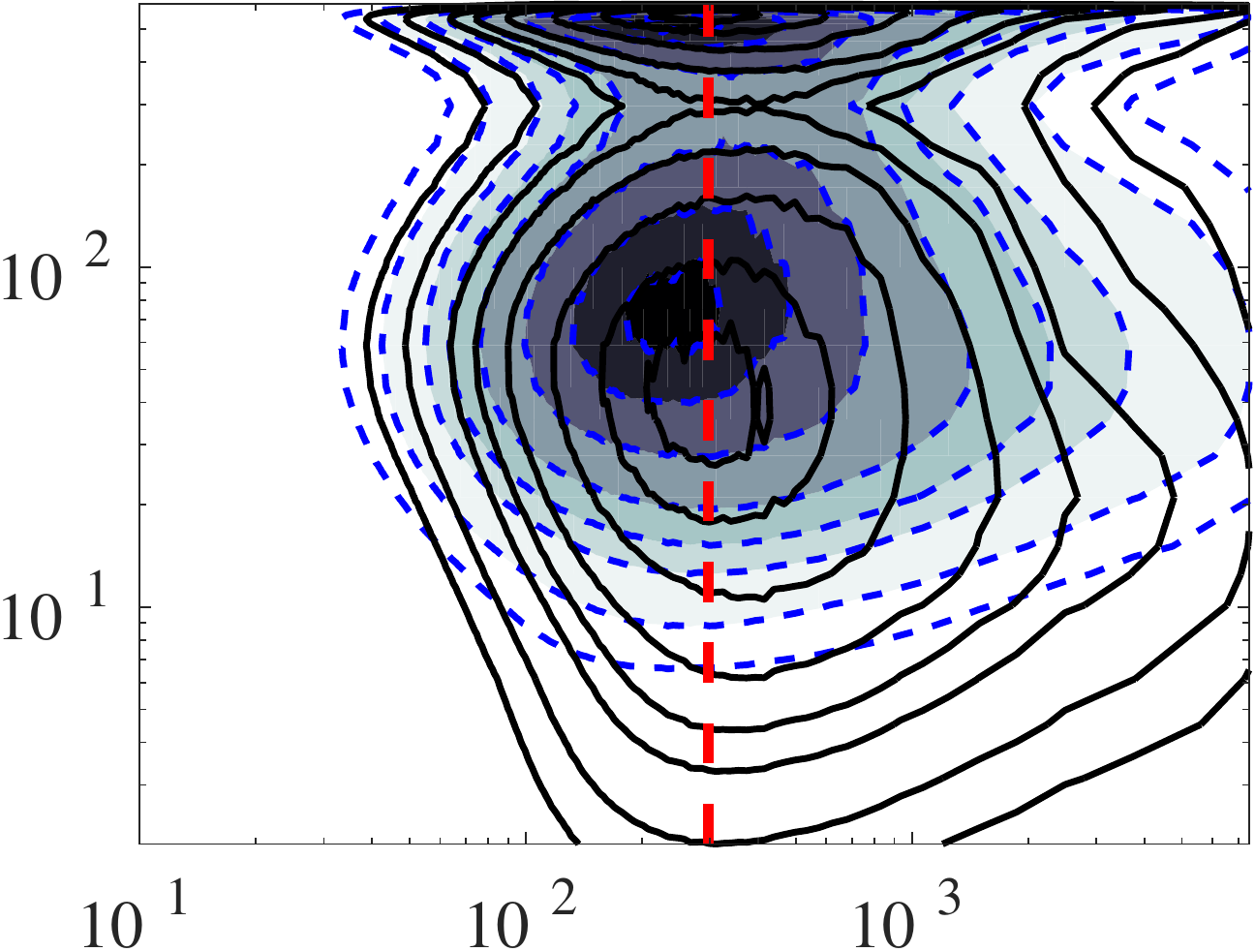}
\put(-0.1,3.9){$\displaystyle (b)$}
\put(-2.9,-0.3){$\lambda^*_x$}
\put(-6.0,2.1){$y^*$}

\includegraphics[width=55mm]{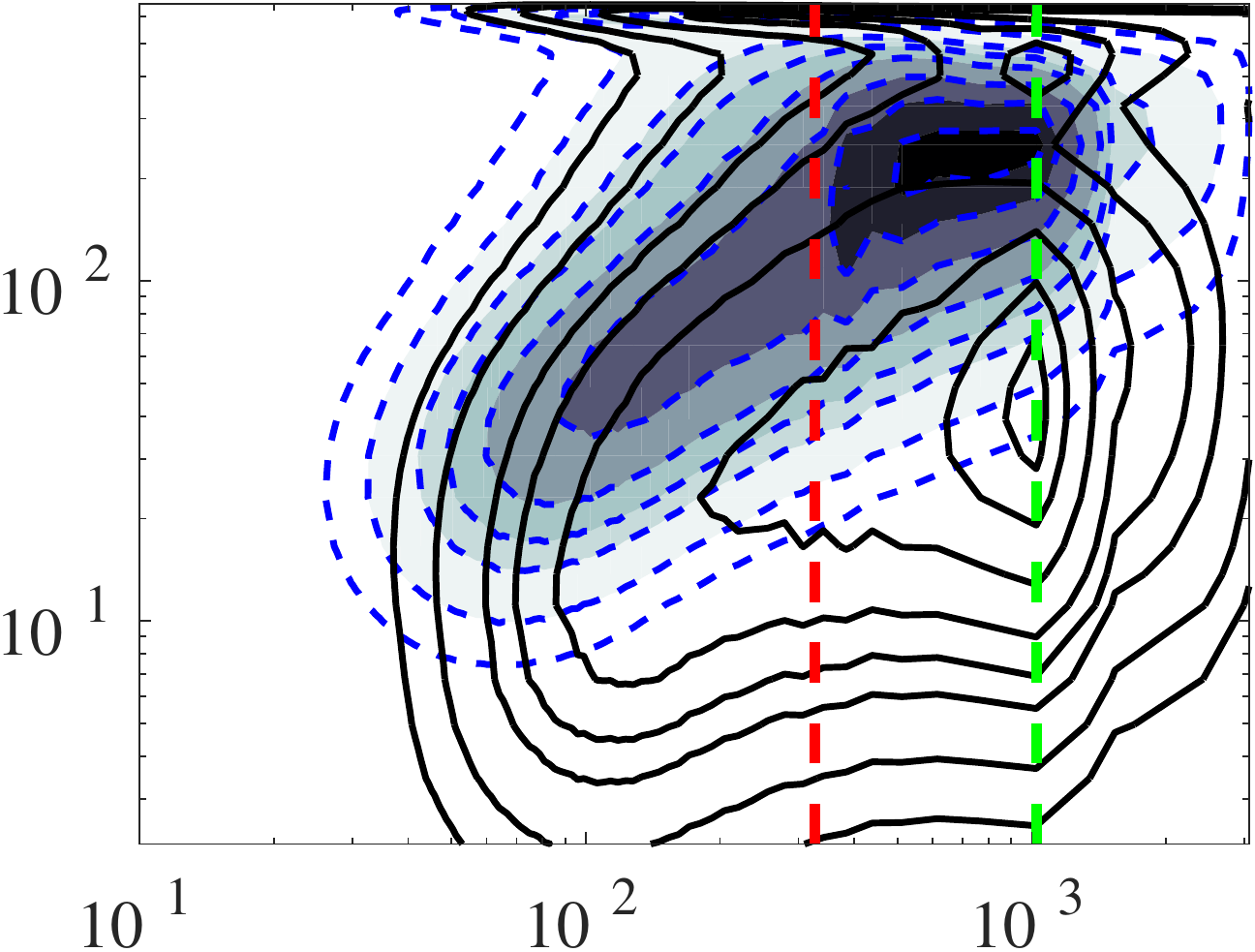}
\put(-0.1,3.9){$\displaystyle (c)$}
\put(-2.9,-0.3){$\lambda^*_z$}
\put(-6.0,2.1){$y^*$}
\hskip9mm
\includegraphics[width=55mm]{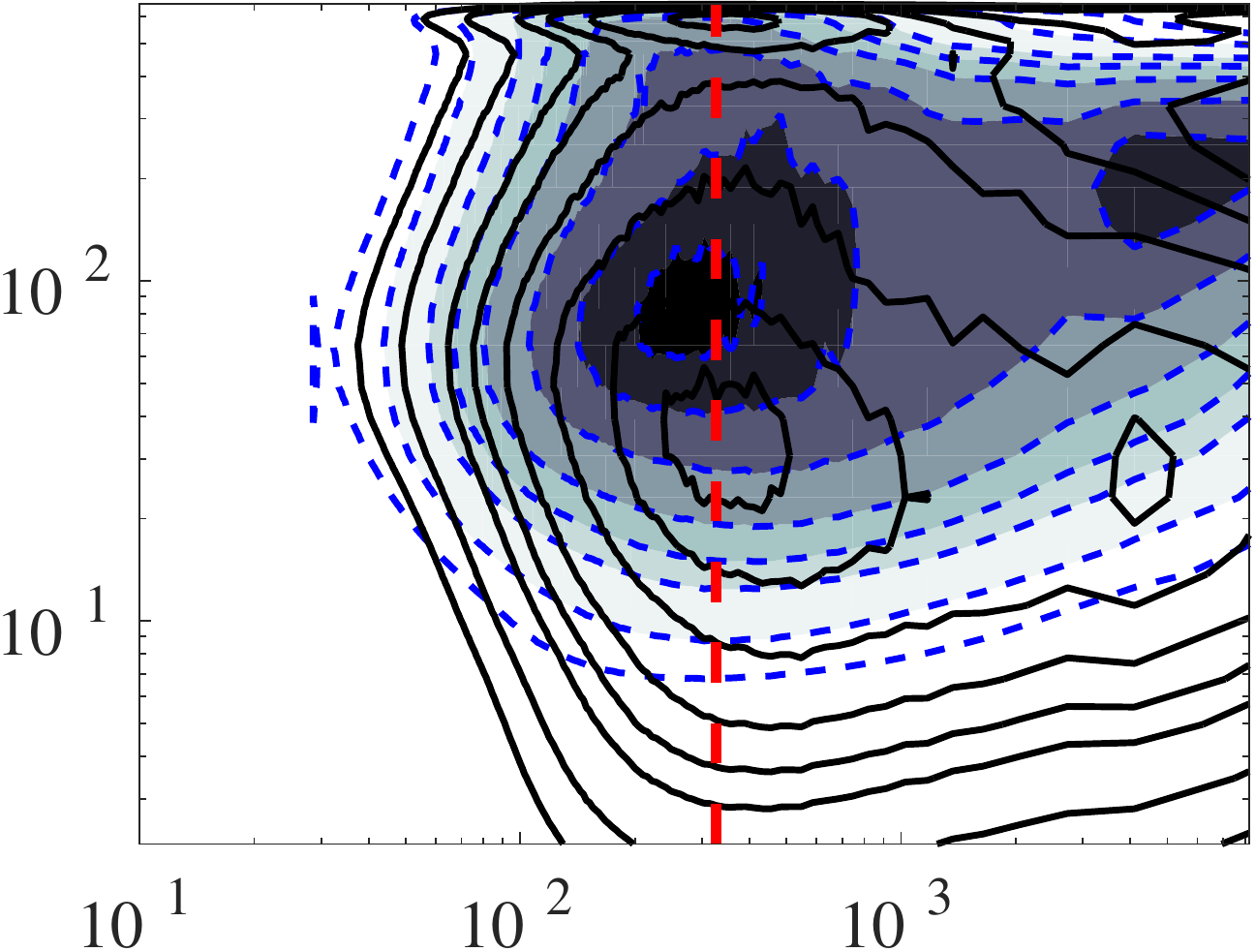}
\put(-0.1,3.9){$\displaystyle (d)$}
\put(-2.9,-0.3){$\lambda^*_x$}
\put(-6.0,2.1){$y^*$}

\includegraphics[width=55mm]{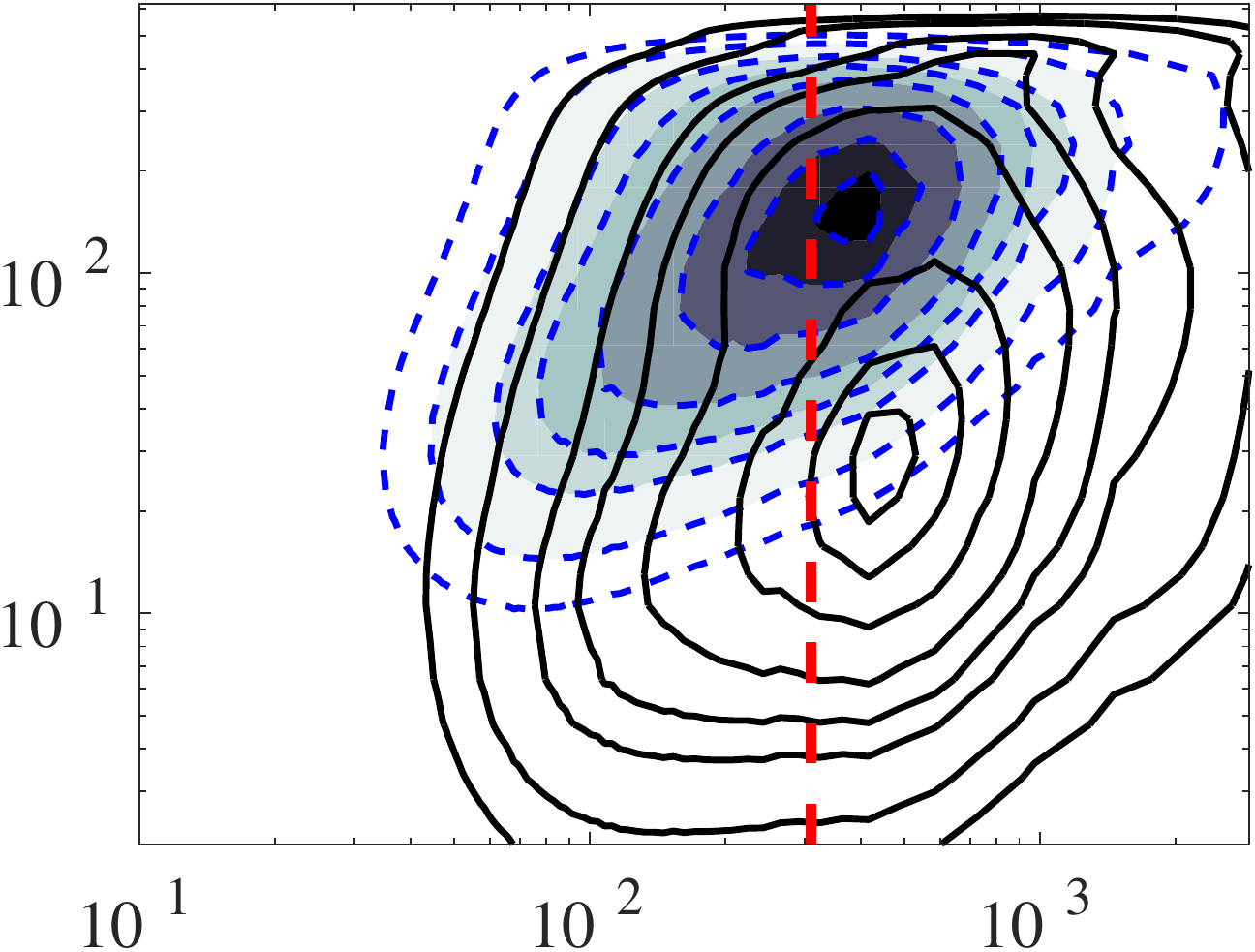}
\put(-0.1,3.9){$\displaystyle (e)$}
\put(-2.9,-0.3){$\lambda^*_z$}
\put(-6.0,2.1){$y^*$}
\hskip9mm
\includegraphics[width=55mm]{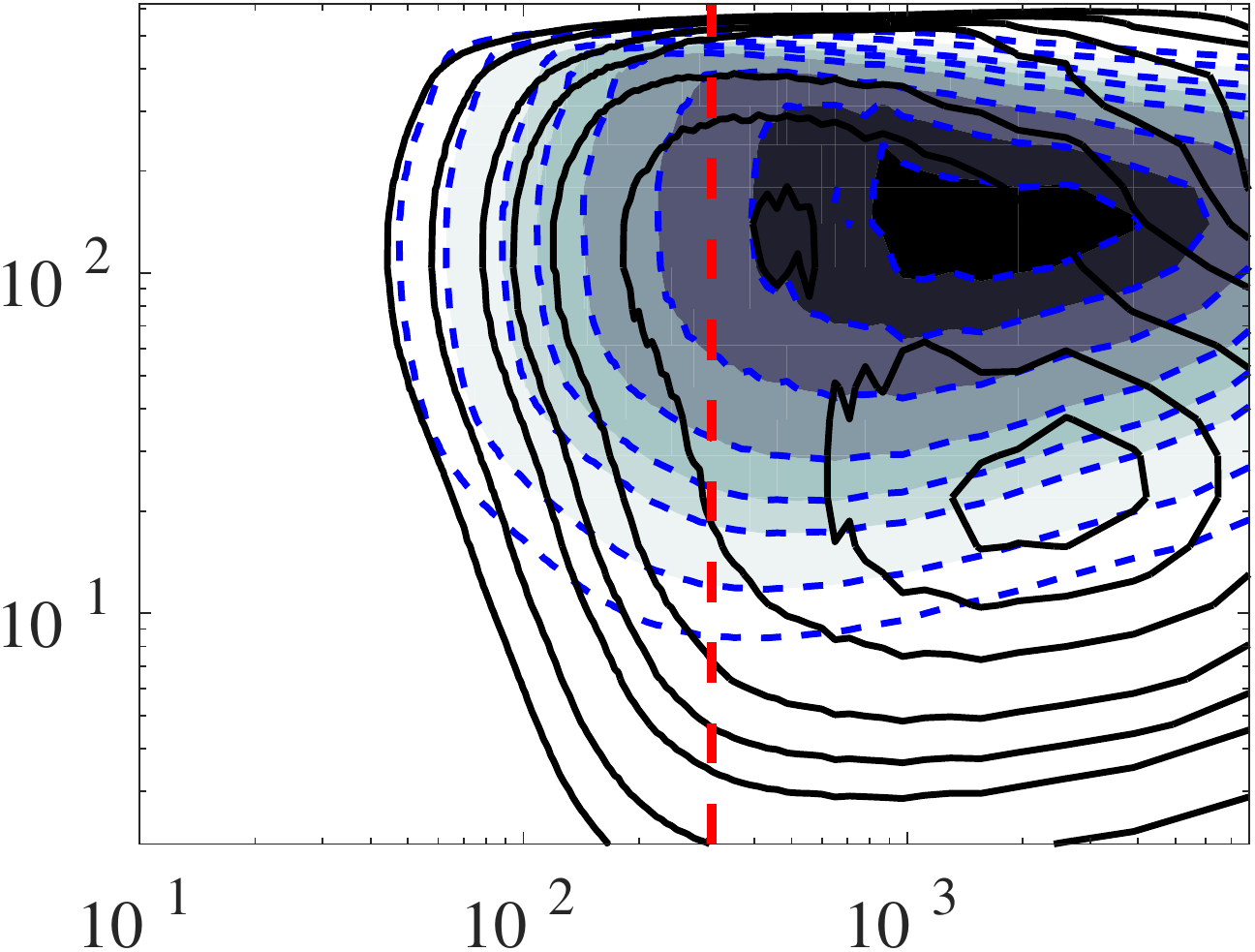}
\put(-0.1,3.9){$\displaystyle (f)$}
\put(-2.9,-0.3){$\lambda^*_x$}
\put(-6.0,2.1){$y^*$}

\includegraphics[width=55mm]{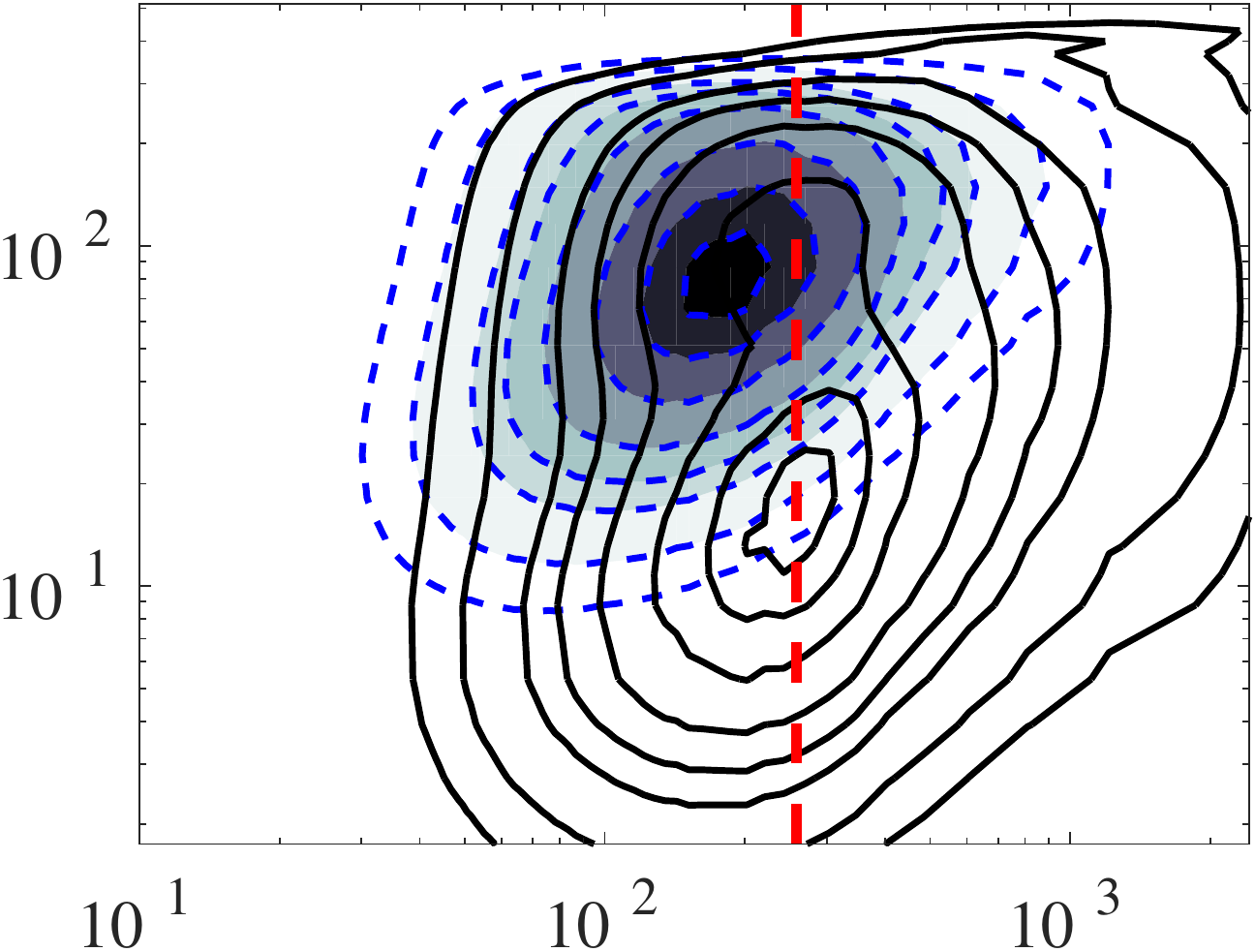}
\put(-0.1,3.9){$\displaystyle (g)$}
\put(-2.9,-0.3){$\lambda^*_z$}
\put(-6.0,2.1){$y^*$}
\hskip9mm
\includegraphics[width=55mm]{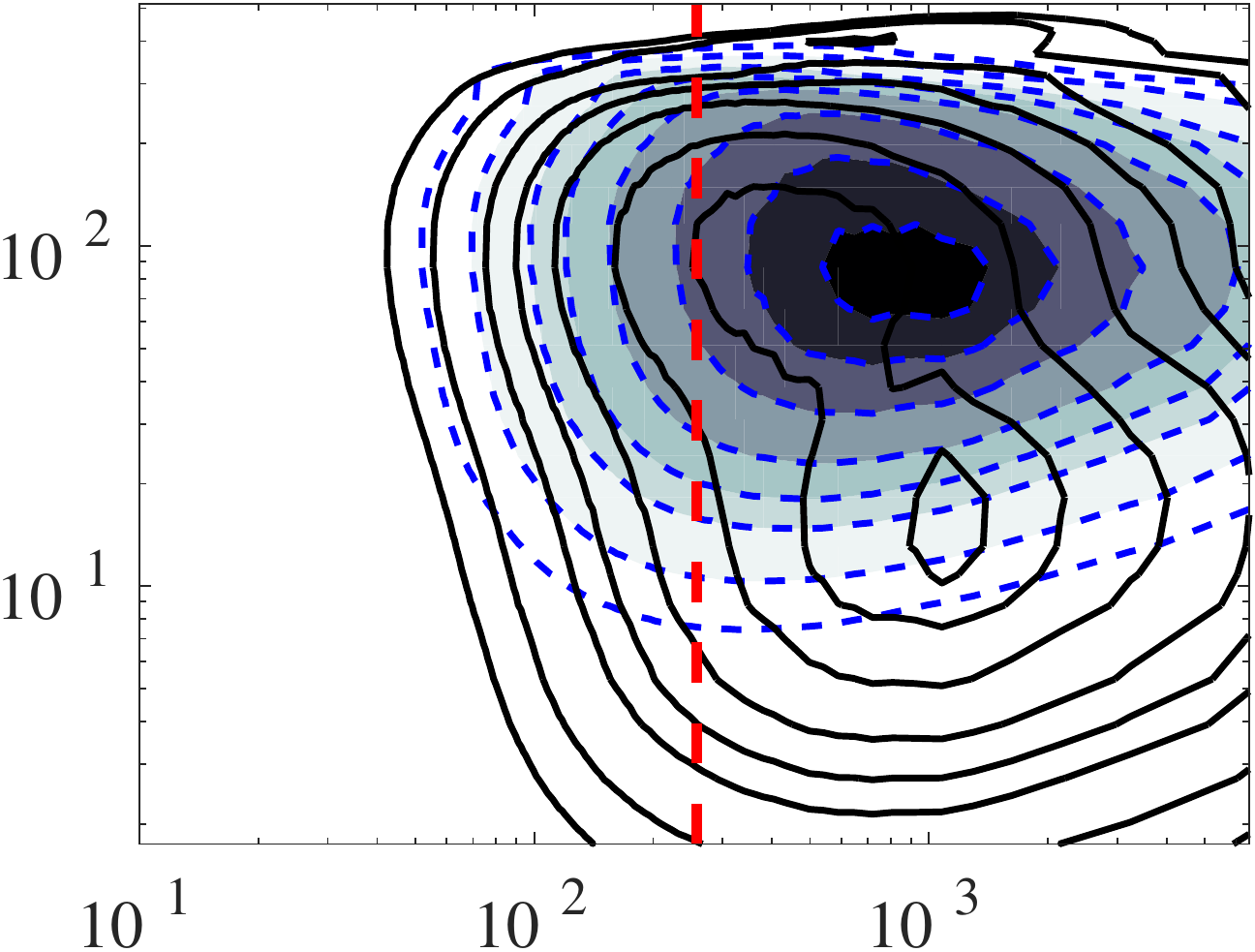}
\put(-0.1,3.9){$\displaystyle (h)$}
\put(-2.9,-0.3){$\lambda^*_x$}
\put(-6.0,2.1){$y^*$}
\end{center}
\caption{(Colour online) Maps of premultiplied one-dimensional spanwise 
and streamwise energy spectra 
at $Re=5000$ for $Ro=0$ ({\it a, b}), $Ro=0.15$ ({\it c, d}),
$Ro=0.45$ ({\it e, f}) and $Ro=0.9 $ ({\it g, h}).
Left column shows spanwise spectra $k_z \Phi_{vv}$ 
(blue dashed lines and colours) and $k_z \Phi_{ww}$ (black lines).
Right column shows streamwise spectra
$k_x \Phi_{vv}$ (blue dashed lines and colours) and $k_x \Phi_{ww}$ (black lines).
Contour levels from innermost to outermost are 0.94, 0.8, 0.6, 0.4, 0.28, 0.2, 0.1 and 0.05.
Spectra are scaled with their maximum value.
}
\label{spec5000}
\end{figure}

To answer more unequivocally whether at low $Re$ roll cells exist in a wider $Ro$ range 
than at higher $Re$, I have computed premultiplied one-dimensional
spanwise and streamwise spectra of the wall-normal and spanwise velocity fluctuations 
at $Re=5000$ for $Ro=0$, 0.15, 0.45 and 0.9, see figure \ref{spec5000}.
The wavelength and the distance to the wall is, as in the previous spectra,
scaled by the viscous length scale of the unstable side, $l^* = \nu/u_{\tau u}$,
and the spectra are scaled with their maximum value, see table \ref{maxval}.
The red dashed line marks again scales with a wavelength $h$.
Spanwise two-point correlations of the wall-normal velocity fluctuations
$R_{vv}=\overline{v(\vetx+z)v(\vetx)}/\overline{vv}$ 
at $Re=3000$, 5000 and $20\,000$ and various $Ro$ have also been calculated.
If streamwise vortices are present $R_{vv}$ is expected to display 
negative correlations with a minimum at a separation distance that is about
equal to the mean vortex diameter.
Accordingly, in non-rotating channel flow,
$R_{vv}$ has a minimum at $z^* \approx 25$ near the wall
as a result of near-wall streamwise vortices
(Kim \etal 1987).
The two-point correlations are shown at the unstable channel side at the wall-normal
position where $R_{vv}$ approximately has the largest negative values
and thus the signs of the roll cells are most clear.
In all cases, the mean velocity profile is approximately linear at this position.
The domain size is the same
in all DNSs used to compute the spectra and two-point correlations.

At $Ro=0.15$ and $Re=5000$, the spanwise spectrum 
$k_z \Phi_{ww}$ has a peak at $\lambda_z = \pi h$ (indicated by the green
dashed line) while the peak of
$k_z \Phi_{vv}$ extends from about
$\lambda_z = \pi h/2$ to $\pi h$
(figure \ref{spec5000}.{\it c})
and the streamwise spectra reveal structures that span the whole streamwise domain
(figure \ref{spec5000}.{\it d}).
Spectra at $Re=3000$ are not shown for brevity but at $Ro=0.15$ they show a peak at spanwise
wave lengths $\lambda_z =  \pi h$ and a lesser one at $3 \pi h/5$.
This indicates, similarly as at $Re = 20\,000$, roll cells with a size of $\pi h/2$ and smaller ones.
By comparing figure \ref{spec5000}.({\it e}) and ({\it f}) with 
figure \ref{spec5000}.({\it a}) and ({\it b}),
it can be concluded that at $Ro=0.45$ wider and longer structures exist on the 
unstable channel side than at $Ro=0$,
implying that large roll cells are present.
The spectral peaks in the spanwise 
$k_z \Phi_{vv}$ 
and streamwise spectrum
$k_x \Phi_{vv}$ 
are at $\lambda_z \approx 3 \pi h /8$
and $\lambda_x \approx 2 \pi h $, respectively. At similar wavelengths peaks
are found in the spectra at $Re=20\,000$, as shown before, and $Re=3000$.

\begin{figure}
\begin{center}
\setlength{\unitlength}{1cm}
\includegraphics[width=50mm]{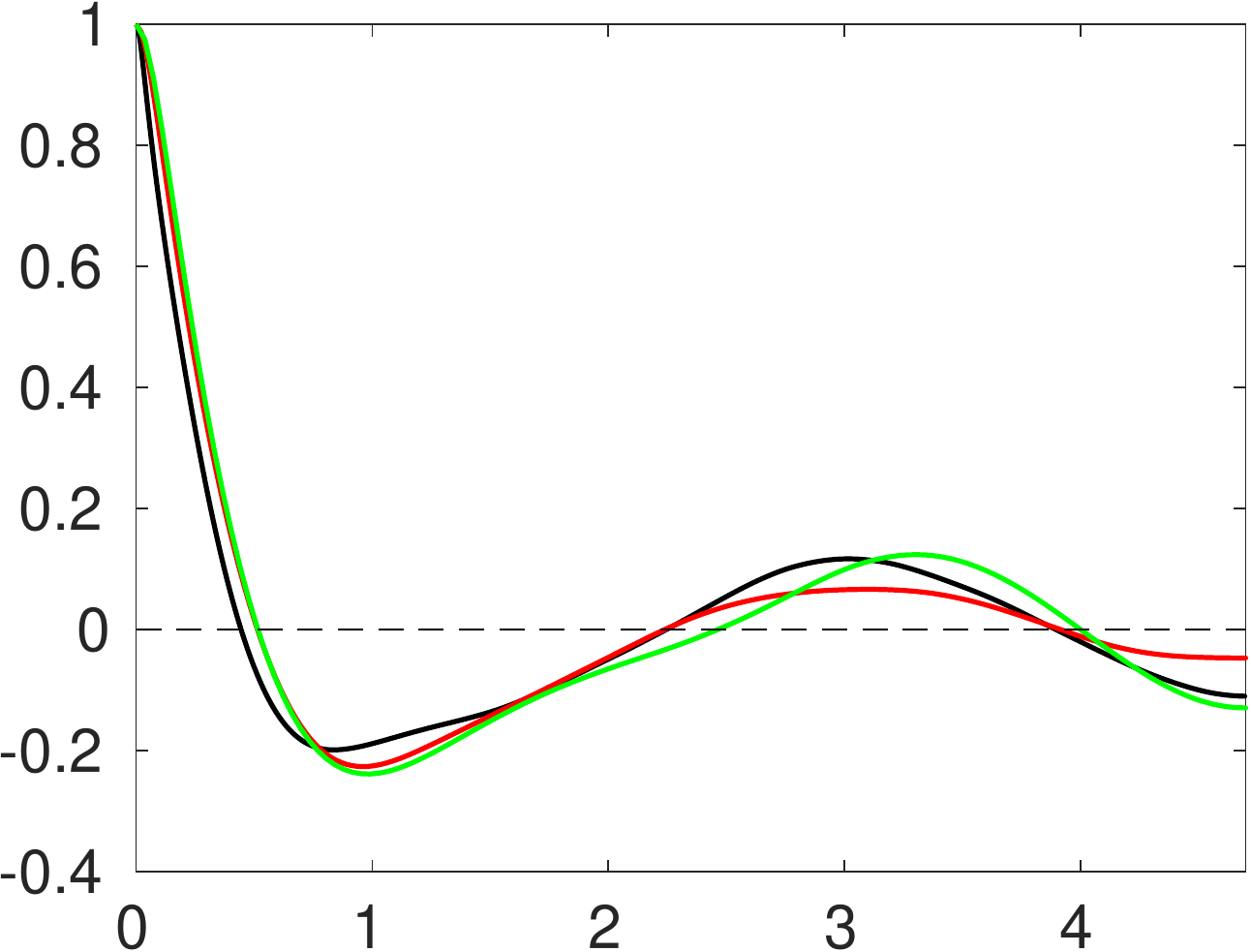}
\put(-0.1,3.5){$\displaystyle (a)$}
\put(-2.5,-0.3){$z$}
\put(-5.6,1.9){$R_{vv}$}
\hskip9mm
\includegraphics[width=50mm]{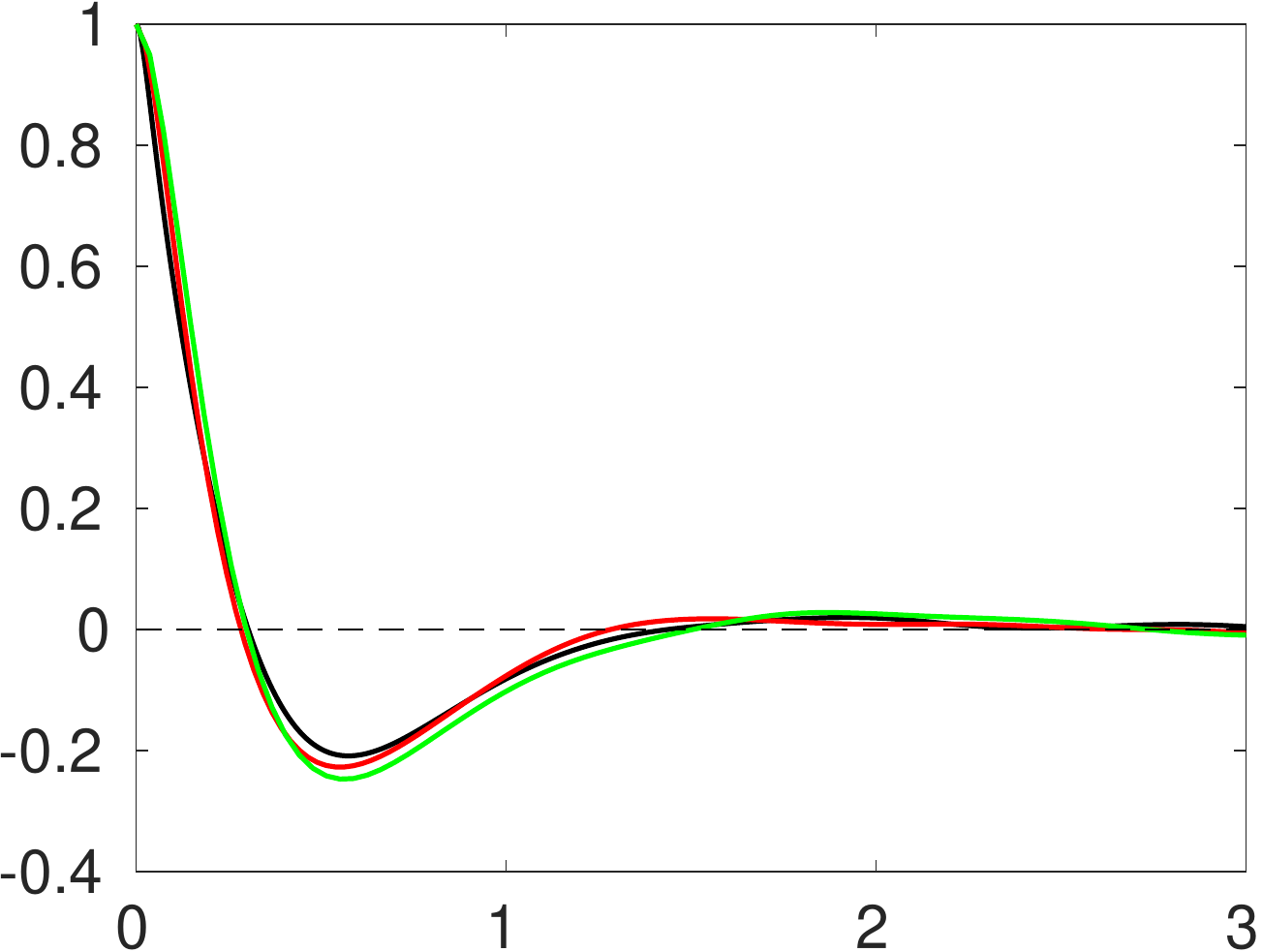}
\put(-0.1,3.5){$\displaystyle (b)$}
\put(-2.5,-0.3){$z$}
\put(-5.6,1.9){$R_{vv}$}

\includegraphics[width=50mm]{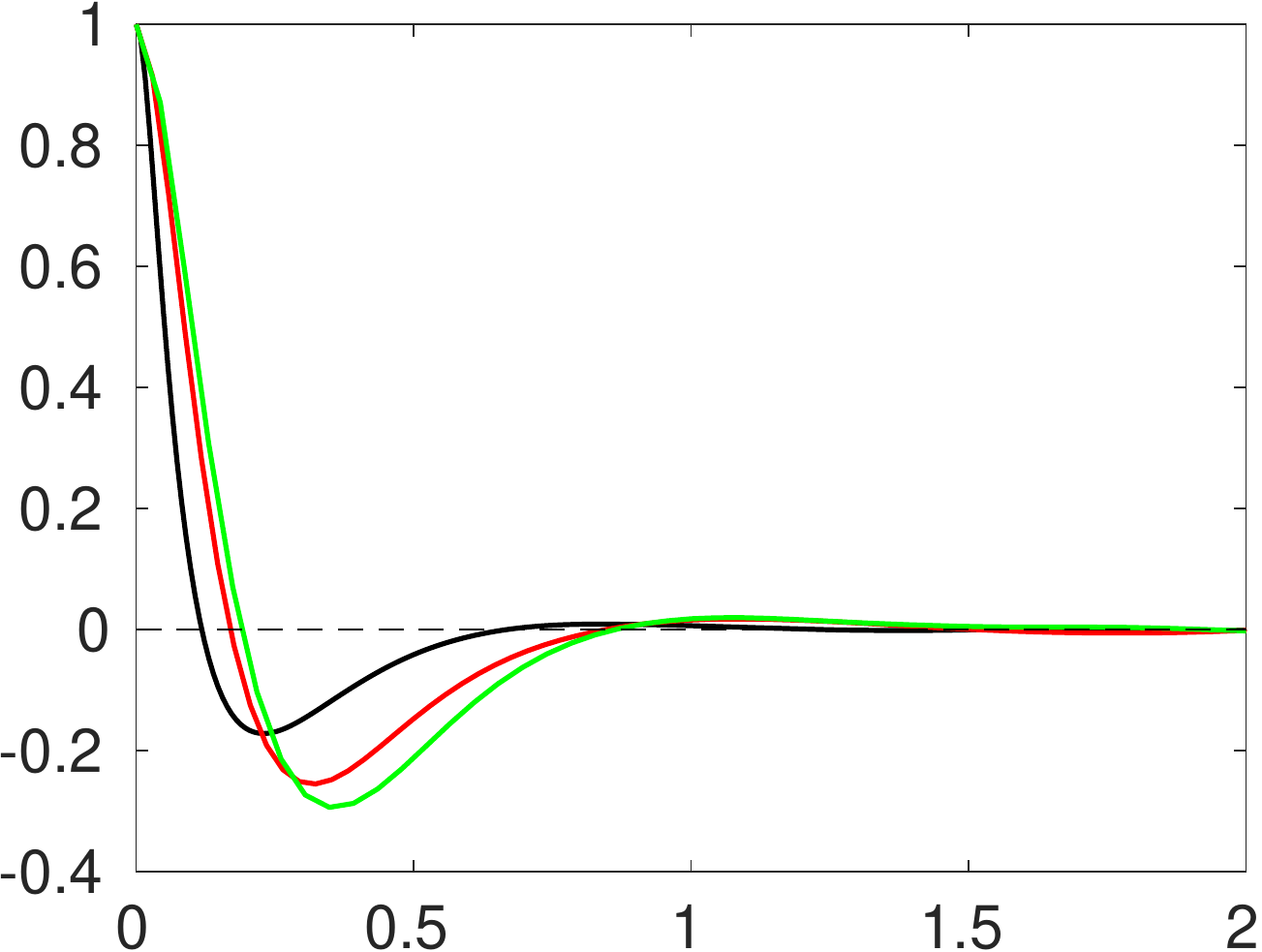}
\put(-0.1,3.5){$\displaystyle (c)$}
\put(-2.5,-0.3){$z$}
\put(-5.6,1.9){$R_{vv}$}
\hskip9mm
\includegraphics[width=50mm]{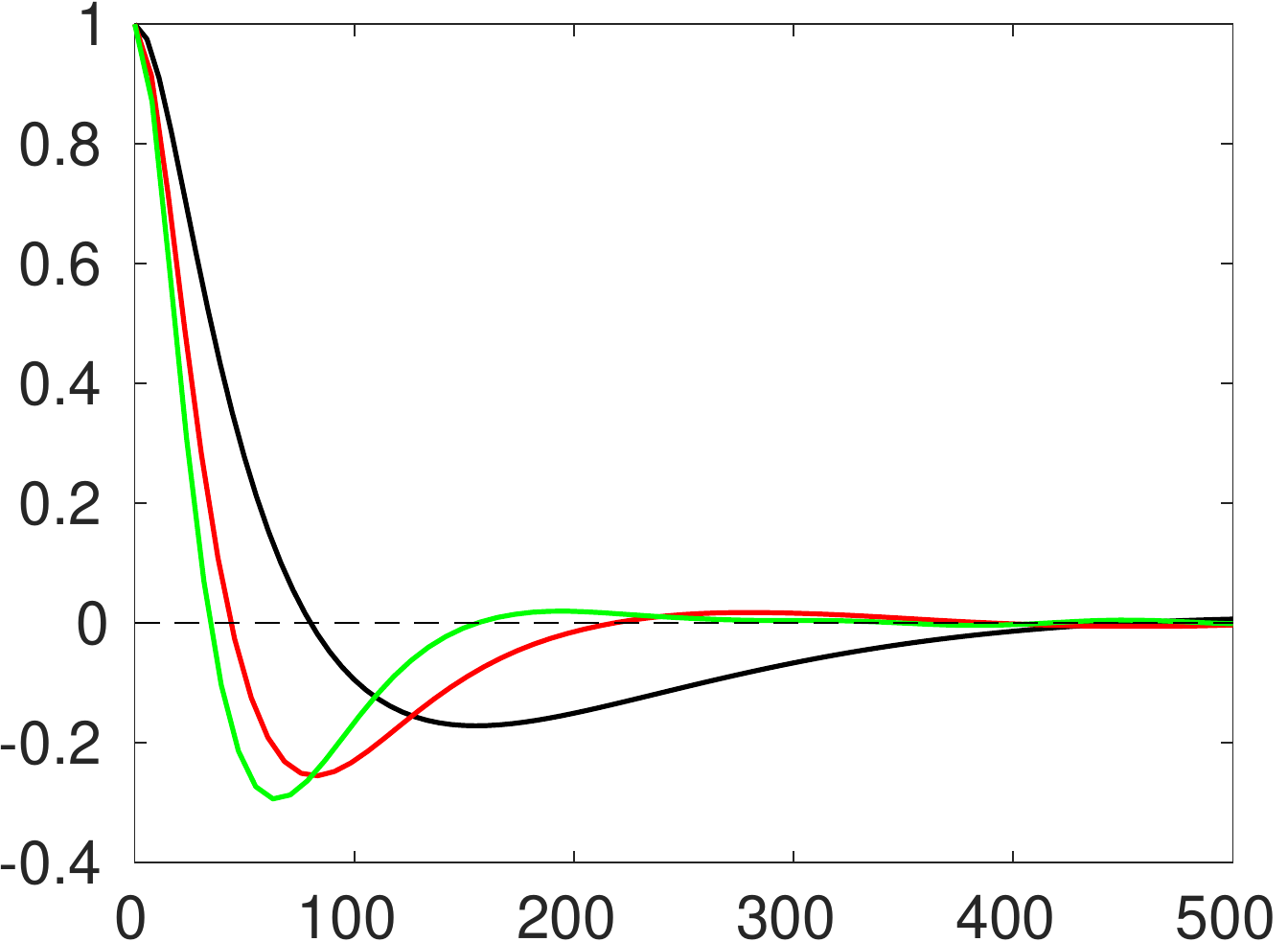}
\put(-0.1,3.5){$\displaystyle (d)$}
\put(-2.5,-0.3){$z^*$}
\put(-5.6,1.9){$R_{vv}$}

\includegraphics[width=50mm]{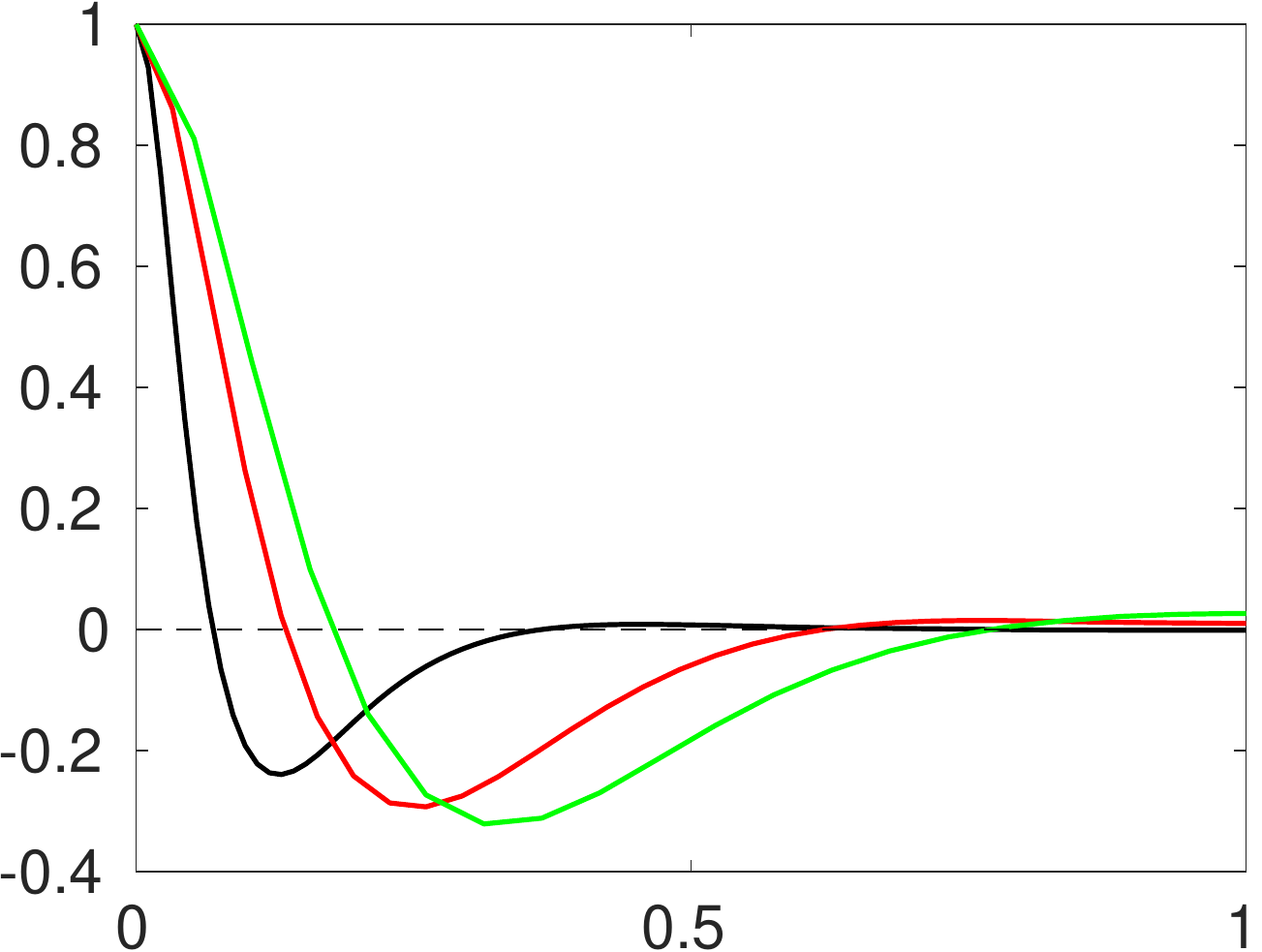}
\put(-0.1,3.5){$\displaystyle (e)$}
\put(-2.5,-0.3){$z$}
\put(-5.6,1.9){$R_{vv}$}
\hskip9mm
\includegraphics[width=50mm]{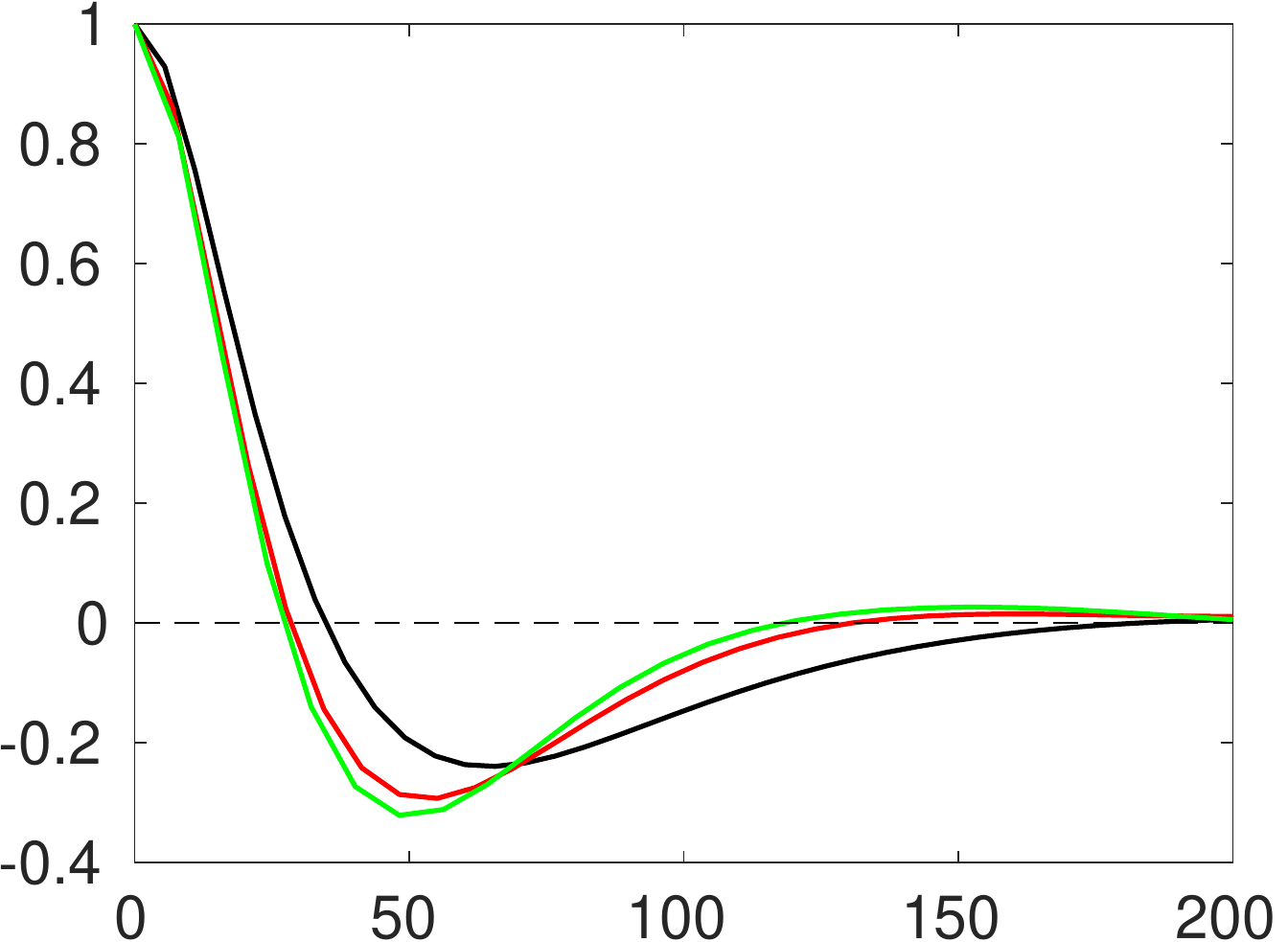}
\put(-0.1,3.5){$\displaystyle (f)$}
\put(-2.5,-0.3){$z^*$}
\put(-5.6,1.9){$R_{vv}$}
\end{center}
\caption{(Colour online) Spanwise two-point correlations of the wall-normal velocity fluctuations
at ({\it a}) $Ro=0.15$ and $Re=20\,000$ and $y=-0.41$,
$Re=5000$ and $y=0$, and $Re=3000$ and $y=0$;
({\it b}) $Ro=0.45$ and $Re=20\,000$ and $y=-0.54$,
$Re=5000$ and $y=-0.55$, and $Re=3000$ and $y=-0.5$;
({\it c}, {\it d}) $Ro=0.9$ and $Re=20\,000$ and $y=-0.83$,
$Re=5000$ and $y=-0.75$, and $Re=3000$ and $y=-0.75$;
({\it e},{\it f}) $Ro=1.2$ and $Re=20\,000$ and $y=-0.92$,
$Re=5000$ and $y=-0.83$, and $Re=3000$ and $y=-0.75$.
In ({\it a,b,c,e}) $z$ is scaled by $h$ and in ({\it d,f}) $z^*=z/l^*$.
($\textcolor{black}{~^{\line(1,0){20}}}$) $Re=20\,000$,
($\textcolor{red}{~^{\line(1,0){20}}}$) $Re=5000$ 
and ($\textcolor{green}{~^{\line(1,0){20}}}$) $Re=3000$. 
}
\label{twopoint}
\end{figure}

The spectra suggest that the roll cells at $Ro=0.15$ as well as at $Ro=0.45$ 
have a similar size for $Re=3000$ to $20\,000$.
This is confirmed by the two-point correlations. 
At $Ro=0.15$, $R_{vv}$ is clearly negative for $0.5 \lesssim z \lesssim 2 $ and its
minimum, which is a measure of the mean vortex diameter, is at about the same $z$
at all three Reynolds numbers (figure \ref{twopoint}.{\it a}).
The spectra suggest roll cells with a diameter $\pi h/2$ as well as smaller ones
and that appears to be consistent with the two-point correlations which suggest
roll cells with a mean diameter of about $h$.
Also $R_{vv}$ at $Ro=0.45$ has a minimum at about the same $z$ at
all three $Re$ (figure \ref{twopoint}.{\it b}). The spectra indicate that the roll
cells have a mean diameter of about $3 \pi h/16$, in agreement with the two-point correlations.

These results indicate that for $Re=3000$ to $20\,000$
the size of the roll cells is smaller at $Ro=0.45$ than at $Ro=0.15$ 
but approximately independent of $Re$.
The size $\pi h/2$ of the largest roll cells at $Ro=0.15$ indicated by the spanwise spectra
is consistent with the size observed in the DNSs by Kristoffersen \& Andersson (1993)
and Yang \& Wu (2012) at the same $Ro$.
Yang \& Wu also observed smaller roll cells
and found that the size of the roll cells becomes smaller with $Ro$ for $Ro\geq 0.15$, 
again consistent with the present results.
Also the DNSs by Dai \etal (2016) at $Re=2800$ show that the roll cells become
smaller when $Ro$ increases from 0.1 to 0.5.
On the other hand, the size $\pi h/2$ of the roll cells in the DNS by Hsieh \& Biringen (2016) 
at $Ro=0.5$ and $Re_\tau \approx 200$ is 
considerably larger than in the present DNSs at $Ro=0.45$. This difference 
might be related to the more restricted computational domain used by Hsieh \& Biringen.
Besides, they only show visualizations of the time-averaged velocity field and
these naturally emphasize the largest, most steady structures.

At a higher $Ro=0.9$ the peak in the spanwise spectrum 
$k_z \Phi_{vv}$ 
shifts towards a smaller $\lambda_z \approx 3 \pi h/13$ at $Re=5000$ 
(figure \ref{spec5000}.{\it g}) and
$\lambda_z \approx 3 \pi h/11$ at $Re=3000$, which is at a larger
wavelength than at $Re=20\,0000$  
(figure \ref{specz}.{\it h}). 
The large structures are also considerably longer in terms of outer units
at lower $Re$, see figure 
figure \ref{spec5000}.({\it h}) vs. figure \ref{specx}.({\it h}). 
The minimum in $R_{vv}$ shift towards smaller $z$ at higher $Re$
(figure \ref{twopoint}.{\it c}), confirming that at $Ro=0.9$ the large scales
become smaller at higher $Re$. 
The two-point correlations 
(figure \ref{twopoint}.{\it e}) and spectra (not shown here) show that 
at $Ro=1.2$ the differences are even more pronounced.
The size of the large flow structures in terms of outer units
shrinks monotonically with $Re$ and this trend continues at higher $Ro$.
On the other hand, if the separation distance $z$ is scaled by the viscous
length scale $l^*$ the two-point correlations differ considerably at $Ro=0.9$
(figure \ref{twopoint}.{\it d}) and lower $Ro$ but much less at $Ro=1.2$
(figure \ref{twopoint}.{\it f}) and higher $Ro$.
The two-point correlations at $Re=3000$ and $Re=5000$ show in fact an almost perfect collapse.
Thus, the size of the largest scales continues to decrease with $Ro$ but for
$Ro \geq 0.9$ it becomes dependent of $Re$ as well.
Whether these largest structures can be considered to be typical roll cells
is not clear since the roll cells and typical near-wall streamwise vortices become less
distinguishable at high $Ro$ and low $Re$.
The two-point correlations suggest that their size start
to display a viscous wall unit scaling
and therefore may no have the same physical origin as at lower $Ro$.

%
\section{Conclusions}

The present numerical study of fully developed plane turbulent channel flow subject
to system rotation about the spanwise direction covers a wide range of parameters
with $Re$ between 3000 and $31\,600$ and $Ro$ between 0 and 2.7.
At all $Re$ and for a wide range of $Ro$ the mean streamwise velocity profile
has a linear part with a slope $\diff U / \diff y \simeq 2 \Omega$ implying
that this mean zero-absolute-vorticity state is independent of the Reynolds number.
This zero-absolute-vorticity state has also been found in low-Reynolds number
and approximately in transitional rotating channel flow (Iida \etal 2010,
Wall \& Nagata 2013) and in laminar and turbulent plane Couette flow subject
to anticyclonic rotation (Kawata \& Alfredsson 2016, Gai \etal 2016).
In all these other cases, the flow is dominated by large streamwise roll cells. 
Hsieh \& Biringen (2016) found that when roll cells are suppressed in DNSs
of rotating turbulent channel flow by decreasing the spanwise domain size 
the slope of the mean velocity profile $\diff U / \diff y$ deviates from $2 \Omega$
pointing out that it is important to resolve the roll cells.
In the present study, the zero-absolute-vorticity state exists too at high $Ro$
when the flow is turbulent but roll cells are absent or small proving that
roll cells are not a prerequisite. Some possible explanations for the zero-absolute-vorticity
state have been proposed (Hamba 2006, Kawata \& Alfredsson 2016) 
but none rigorous.
Because of the linear mean velocity slope,
profiles of the production and dissipation rate of turbulent kinetic energy and
some of the budget terms in the Reynolds stress equations show
a linear part as well.
In this part of the unstable channel side, basically all energy
is fed into the wall-normal Reynolds stress component when
the production and Coriolis term are considered together. 
The energy is then redistributed to the streamwise and especially 
the spanwise Reynolds stress component through pressure-strain correlations. 

Through visualizations, one-dimensional spectra and two-point correlations the
influence on rotation on turbulence structures is investigated. A distinct unstable side
with intense turbulence and vortical structures and a stable side
with much weaker turbulence develops in the channel with an apparent 
sharp border between the two sides. If $Ro$ approaches 0.45 
the flow at higher $Re$ partly relaminarizes on the stable side of the channel and
oblique turbulent-laminar patterns develop which resemble
the oblique band-like structures found in transitional Couette and channel flows
at low Reynolds numbers (Duguet \etal 2010, Tuckerman \etal 2014).
It is quite remarkable that such patterns exist in rotating
channel flow at a significantly higher $Re$.
The study by Brethouwer \etal (2012) suggests 
that turbulent-laminar patterns similar to
those found in my DNS at $Ro=0.45$ and $Re=20\,000$ to $31\,600$
also exist at lower $Re$ but at a lower $Ro$ and 
that the patterns most likely have a longer wave length
at lower $Re$. This implies that the patterns may appear at low $Re$ in DNSs
of rotating channel flow if very larger computational domains are used.

If $Ro$ is raised further in the present DNSs the turbulent fraction of the flow on
the stable side goes down and eventually the flow relaminarizes there.
The unstable part of the channel with strong turbulence diminishes in size with $Ro$ 
and when the rotation rate is sufficiently high the whole flow becomes
laminar. However, as shown by Brethouwer (2016), a linear instability
can develop in rapidly rotating channel flows causing a continuous
cycle of turbulent bursts. 

The influence of $Re$ is investigated and found to be significant. 
At fixed $Ro$ Reynolds stresses are noticeably stronger on the stable side
of the channel at higher $Re$. 
When $Ro \leq 0.9$, the flow is partly or fully turbulent on the stable channel
side at higher $Re$ whereas at lower $Re$
the turbulent fraction of the flow is significantly smaller 
and the flow tends to relaminarize faster.
Care has therefore has to be exercised when drawing general conclusions from
lower Reynolds number studies of rotating channel flow.
Attention should also be paid to the size of the computational domain since
that may have a significant influence on the large-scale structures like
roll cells and the relaminarization of the stable channel side.

On the unstable side of the channel, large counter-rotating streamwise
roll cells are observed at $Ro=0.15$ and 0.45. 
The roll cells become smaller and less noticeable and
eventually disappear if $Ro$ is raised. This trend is stronger
at higher $Re$ since at high $Ro$
the large-scale modes, which are possibly related to roll cells,
are larger at lower $Re$.
This suggests that at low $Re$ roll cells may exist in a wider $Ro$ range,
but note there is yet no way to 
unambiguously determine if roll cells exists.
The spectra also indicate that at lower $Ro$ the unstable channel side
is populated by attached eddies whereas these appear to be absent at higher $Ro$,
indicating that the interaction between the inner and outer layer is weak in
rapidly rotating channel flows.

The present case gives some general insights into the effect
of rotation on wall-bounded flows. It is valuable as well for turbulence modelling since 
the influence of rotation is turbulence is still difficult to model. Even for
large-eddy simulation it could be a demanding case because
correctly predicting the relaminarization of rapidly rotating channel flow
may pose a challenge.

\begin{acknowledgments}
PRACE is acknowledged for the allocation of computing time
at the J{\"u}lich Supercomputing Centre in Germany for the REFIT project.
Computational resources at PDC were made available by SNIC.
The author further acknowledges financial support
by the Swedish Research Council (grant numbers 621-2013-5784 and 621-2016-03533).
\end{acknowledgments}

\bibliographystyle{jfm}


\end{document}